\documentclass[fleqn,usenatbib]{mnras}

\usepackage{newtxtext,newtxmath}

\usepackage[T1]{fontenc}

\DeclareRobustCommand{\VAN}[3]{#2}
\let\VANthebibliography\thebibliography
\def\thebibliography{\DeclareRobustCommand{\VAN}[3]{##3}\VANthebibliography}

\usepackage{graphicx}	
\usepackage{amsmath}	
\usepackage{pifont}
\usepackage{multirow}

\newcommand{\cmark}{\text{\ding{51}}}
\newcommand{\xmark}{\text{\ding{55}}}

\DeclareRobustCommand{\ion}[2]{%
\relax\ifmmode
\ifx\testbx\f@series
{\mathbf{#1\,\mathsc{#2}}}\else
{\mathrm{#1\,\mathsc{#2}}}\fi
\else\textup{#1\,{\mdseries\textsc{#2}}}%
\fi}

\title[AGN jet-driven outflows in Seyferts]{Blandford-Znajek jets in galaxy formation simulations: exploring the diversity of outflows produced by spin-driven AGN jets in Seyfert galaxies}

\author[R. Y. Talbot et al.]{
Rosie Y. Talbot$^{1,2}$\thanks{E-mail: rt421@cam.ac.uk (RYT)}, Debora Sijacki$^{1,2}$ and Martin A. Bourne$^{1,2}$
\\
$^{1}$Institute of Astronomy, University of Cambridge, Madingley Road, Cambridge, CB3 0HA, UK\\
$^{2}$Kavli Institute for Cosmology, University of Cambridge, Madingley Road, Cambridge, CB3 0HA, UK
}

\date{Submitted to MNRAS}
\pubyear{2021}

\begin{document}
\label{firstpage}
\pagerange{\pageref{firstpage}--\pageref{lastpage}}
\maketitle

\begin{abstract}
Recent observations of Seyfert galaxies indicate that low power, misaligned jets can undergo significant interaction with the gas in the galactic disc and may be able to drive large-scale, multiphase outflows. We apply our novel sub-grid model for Blandford-Znajek jets to simulations of the central regions of Seyferts, in which a black hole is embedded in a dense, sub-kpc circumnuclear disc (CND) and surrounded by a dilute circumgalactic medium (CGM). We find that the variability of the accretion flow is highly sensitive both to the jet power and to the CND thermodynamics and, ultimately, is determined by the complex interplay between jet-driven outflows and backflows. Even at moderate Eddington ratios, AGN jets are able to significantly alter the thermodynamics and kinematics of CNDs and entrain up to $10\%$ of their mass in the outflow. Mass outflow rates and kinetic powers of the warm outflowing component are in agreement with recent observations for black holes with similar bolometric luminosities, with outflow velocities that are able to reach $500 \,{\rm km \, s^{-1}}$. Depending on their power and direction, jets are able to drive a wide variety of large-scale outflows, ranging from light, hot and collimated structures to highly mass-loaded, multiphase, bipolar winds. This diversity of jet-driven outflows highlights the importance of applying physically motivated models of AGN feedback to realistic galaxy formation contexts. Such simulations will play a crucial role in accurately interpreting the wealth of data that next generation facilities such as JWST, SKA and Athena will provide.
\end{abstract}




\section{Introduction}

It is now widely thought that feedback from supermassive black holes plays a crucial role in shaping the properties of their host galaxies \citep[see e.g.][]{2013KormendyHo, 2012Fabian, 2007McNamaraNulsen}. Whilst the clearest observational signs of black hole feedback in the form of active galactic nuclei (AGN) jets are found in galaxy clusters, there is growing evidence that these jets are, in fact, ubiquitous \citep[for a recent review see][]{2019Blandford+}. Indeed, AGN jets have now been detected in a wide variety of galaxy contexts including dwarfs \citep{2020Yang+,2019Mezcua+,2012ReinesDeller}, Seyferts \citep{2017Williams+,2012Mingo+,2011Mingo+,2006Gallimore+,2006HotaSaikia,1999Morganti+} and ellipticals \citep{2018Tremblay+,2012Birzan,2009Mittal+,2007Best+}.

Recent observations indicate that it is not just high power jets and those found in massive systems that are able to significantly impact their host galaxy. These observations suggest that ionised outflows may be more prevalent and have a greater effect on the nuclear kinematics in low-excitation galaxies and in cases where the radio emission is compact (kpc scale) \citep{2021Speranza+,2019Molyneux+,2015Harrison+,2008Holt+,2006Das+}.

Jets propagating away from the black hole can also undergo significant interaction with the gas in the interstellar medium (ISM) and circum-galactic medium (CGM) \citep{2021Ruschel-Dutra,2021Morganti+,2018Morganti+,2021Roy+,2019Jarvis+}. Such interactions could be responsible for launching massive molecular outflows both on cluster scales \citep{2018Tremblay+,2018Simionescu+, 2017Russell+} and on galaxy scales \citep{2015Morganti+,2014Tadhunter+,2013Combes+}. 

Jets launched by systems that host a circumnuclear disc (CND) are also able to affect the kinematics of this central reservoir of cold, dense gas \citep{2014GarciaBurillo+,2013Couto+}. Indeed, there is now growing evidence that indicates that jets are able to significantly perturb the kinematics of the gas in the direction perpendicular to that of their propagation and may even drive outflows in this direction \citep{2021Venturi+,2021Ruschel-Dutra,2020Congiu+,2015Riffel+,2014Riffel+}. Such scenarios are particularly plausible in Seyfert galaxies, where observations suggest that the jet axis is randomly oriented with respect to the normal to the galactic plane \citep{2000Kinney+,1998Clarke+}.

AGN jet feedback is an inherently multi-physics, multi-scale phenomenon and, thus, is highly suited to exploration via numerical simulations. To investigate the interactions between AGN and their surrounding environment, it is vital that simulations are able to follow the long-time, large-scale evolution of the jet. To do so, however, means that the scales on which these AGN jets are launched fall well below that which can be resolved and the resulting lack of ab initio jet formation means that the launching mechanisms are necessarily encoded in sub-grid models. 

Such sub-grid models have been widely and effectively used in fully cosmological galaxy formation simulations \citep[see e.g.][]{2019Dave+,2018Weinberger+,2018Henden+,2015Sijacki+,2015Crain+,2012Dubois+, 2007Sijacki+}, which all indicate that the feedback from supermassive black holes is required to produce galaxy populations that are realistic and broadly consistent with observations. Simulations of AGN jets launched into galaxy cluster environments explore the ways in which these jets interact and transfer energy to the intracluster medium (ICM) \citep[see e.g.][]{2021Su+,2021BourneSijacki,2019Bourne+,2019Beckmann+,2018Ehlert+,2017BourneSijacki,2017Weinberger+,2014LiBryan,2012Krause+}, while simulations that focus on individual galaxies have been used to explore the effects of jets on the galactic star formation rate as well as the interaction between the jet and the ISM \citep{2018Cielo+,2018MMukherjee+,2016Mukherjee+,2016Wagner+,2012Gaibler+,2011Gaibler+,2011WagnerBicknell,2008Antonuccio-DeloguSilk}. Indeed, some observations of molecular outflows \citep{2017Osterloo+,2015Morganti+,2014Tadhunter+} have been shown to have properties that are consistent with simulations of jets expanding into a clumpy medium \citep{2011WagnerBicknell,2002Mellema+}.

In recent years, sub-grid models of black hole accretion have begun to track the evolution of the spin of the black hole \citep[see e.g.][]{2019BustamanteSpringel,2018Fiacconi+,2014Dubois+}. Doing so is particularly important as general relativistic magnetohydrodynamic (GRMHD) simulations of accreting, spinning black holes (which {\it are} able to produce ab initio jets) indicate that the Blandford-Znajek jet-launching mechanism \citep{1977BlandfordZnajek} is likely to be at play \citep[see e.g.][]{2010Tchekhovskoy+,2012Tchekhovskoy,2018Liska+} wherein the jet is powered by the spin-energy of the black hole. The modelling of spin-driven jets in galaxy-scale simulations, however, is still in it's infancy and the majority of jet models do not evolve the black hole spin, precluding the launching of self-consistent Blandford-Znajek jets via sub-grid modelling. 

In \cite{2021Talbot+} we presented a novel sub-grid model for black hole accretion via a thin (warped) $\alpha$-disc and feedback in the form of a Blandford-Znajek jet and described how we implemented it into the moving mesh code \textsc{arepo} \citep{2010Springel,2016Pakmor+,2020Weinberger+}. \textsc{arepo} solves the non-relativistic equations of hydrodynamics on a quasi-Lagrangian, moving, unstructured mesh based on the Voronoi tessellation of a set of points which move with the fluid. The gravitational forces are calculated using a hierarchical oct-tree method \citep{2005Springel, 1986BarnesHut} along with the PM method for long-range forces. We also benchmarked the spin-driven jet model using idealised simulations of the central regions of Seyfert galaxies and performing detailed comparisons between these spin-driven jets and analytic models of jet propagation as well as simulations of jets with fixed direction and power.

Here, we significantly extend the simulation suite presented in \cite{2021Talbot+} to explore how varying the initial black hole spin magnitude and direction, the density of the medium into which the jet propagates and the rate at which gas is funnelled towards the black hole affect the evolution and properties of the jet-driven outflows. 

The structure of this paper is as follows. In Section~\ref{Sec: BZ method} we provide a brief overview of the key details our spin-driven jet model. Then in Section~\ref{Sec: Simulation set-up} we describe the initial conditions of our simulations as well as the different black hole parameters and environmental properties that are explored in this work. In Section~\ref{Sec: Results} we present our results, examining the inflows and outflows produced in our simulations as well as providing comparisons to recent observations of AGN-driven outflows in Seyfert galaxies. We end with our conclusions which are presented in Section~\ref{Sec: Conclusions}.

\section{Summary of the Blandford-Znajek jet model}
\label{Sec: BZ method}

In \cite{2021Talbot+} we presented a sub-grid model for black hole accretion and feedback in the form of a Blandford-Znajek jet and described how we incorporated into the moving mesh code \textsc{arepo} \citep{2010Springel, 2016Pakmor+, 2020Weinberger+}. 

In this model, the black hole accretion prescription is based on the work of \cite{2018Fiacconi+}, wherein the black hole is assumed to be surrounded by a sub-grid, thin (potentially warped) $\alpha$-disc \citep{1973ShakuraSunyaev}. In our model, the black hole-$\alpha$-disc system evolves due to accretion from the surroundings, mutual Bardeen-Petterson torquing, the launching of the Blandford-Znajek jet and the accretion of material by the black hole from the ISCO of the $\alpha$-disc. The $\alpha$-disc modulates the flux of mass and angular momentum across the black hole horizon which allows us to accurately follow the mass and spin evolution of the black hole. This information is then used to self-consistently calculate the power and direction of the Blandford-Znajek jet \citep{1977BlandfordZnajek} that is launched by the black hole. 

We make use of the predictions from GRMHD simulations and analytic arguments in our calculations of the jet power, both of which are needed to fully parameterise the model. The sub-grid accretion prescription interfaces with the wider simulation via mass and angular momentum fluxes onto the $\alpha$-disc which are estimated from the properties of inflowing gas cells within the black hole smoothing length which is divided into four sectors (for a full description of how these quantities are calculated, see section~3.2 of \cite{2021Talbot+}). The numerical procedure through which the jet is injected is based on the kinetic energy conserving jet model presented in \cite{2017BourneSijacki}. To ensure that the spatial resolution of these gas cells are sufficiently high (i.e. comparable to the radius of the sub-grid $\alpha$-disc) we apply super-Lagrangian refinement techniques to the cells in the vicinity of the black hole, using procedures based on those presented in \cite{2015CurtisSijacki}. We also ensure that jet injection and lobe inflation is adequately resolved by employing additional, jet-specific refinement schemes that target the jet cylinder region as well as the jet lobes \citep{2017BourneSijacki}. Specifically, each half of the jet cylinder contains at least $10$ cells, however these refinement techniques typically lead to the cylinder being resolved by at least $100$ cells.

We now provide a brief overview of the physical processes described by our model and highlight some of the key details regarding the evolution of the sub-grid system. For a complete description of the model, however, we refer the reader to \cite{2021Talbot+}. The black hole is modelled as a sink particle that accretes from a sub-grid accretion disc. This disc is assumed to be a steady-state, geometrically-thin $\alpha$-disc \citep{1973ShakuraSunyaev}. The black hole is fully described by its mass, $M_{\rm BH}$, and angular momentum, $\boldsymbol{J}_{\rm BH}$, and similarly, the properties of the $\alpha$-disc are completely determined by its mass, $M_{\rm d}$, and total angular momentum, $\boldsymbol{J}_{\rm d}$. Hereafter $\boldsymbol{j}_{\rm BH}$ and $\boldsymbol{j}_{\rm d}$ correspond to unit vectors in the direction of the black hole and $\alpha$-disc angular momentum, respectively. $\dot{M}$ is the steady-state rest mass flux through the $\alpha$-disc and, therefore, is equal to the mass-loss rate at the innermost stable circular orbit (ISCO). During times where the jet is active our model assumes that a fraction, $1/(1 + \eta_{\rm J})$, of this mass flux is able to reach the black hole while the rest is drawn up into the jet. Throughout this work we assume $\eta_{\rm J} = 1$ \citep{2021Talbot+, 2017BourneSijacki}.

The mass of the black hole evolves due to accretion from the inner edge of the $\alpha$-disc and due to the launching of the Blandford-Znajek jet\footnote{The Blandford-Znajek jet is powered by the spin-energy of the black hole, meaning that with no other processes acting, the launching of the jet leads to a decrease in the mass-energy of the black hole \citep{1977BlandfordZnajek}.}
\begin{equation}
\label{eq: MdotBH}
    \dot{M}_{\rm BH} =\frac{(1-\epsilon_{\rm r}-\epsilon_{\rm BZ})}{1 + \eta_{\rm J}}\dot{M} \, ,
\end{equation}
where $\epsilon_{\rm r}$ is the spin-dependent radiative efficiency\footnote{For the full expression for $\epsilon_{\rm r}$, see appendix A of \cite{2021Talbot+}.} and $\epsilon_{\rm BZ}$ is our `Blandford-Znajek efficiency' which is given by
\begin{equation}
    \epsilon_{\rm BZ} = \frac{\kappa}{4\pi}\,\phi_{\rm BH}^2\, f(a) \, ,
\end{equation}
where $\kappa$ is a constant that depends (weakly) on the magnetic field structure threading the black hole horizon. In this work we choose $\kappa = 1/(6\pi)$ which corresponds to a split-monopole field. $\phi_{\rm BH}$ is the dimensionless magnetic flux threading one hemisphere of the black hole horizon. In our model, we use the spin dependent values for $\phi_{\rm BH}$ presented in \cite{2012Tchekhovskoy}. $f(a)$ is a dimensionless function of the black hole spin, $a$, \citep[see][]{2021Talbot+, 2010Tchekhovskoy+}.

The mass of the $\alpha$-disc evolves due to inflow from the wider simulation and the loss of material from the ISCO
\begin{equation}
\label{eq: MdotDisc}
    \dot{M}_{\rm d} = \dot{M}_{\rm in} -\dot{M} \,,
\end{equation}
where $\dot{M}_{\rm in}$ is the mass inflow rate from the simulation. $\dot{M}_{\rm in}$ is calculated using a smooth particle hydrodynamic interpolation of the local mass flux onto black hole, based on the properties of the gas cells within the black hole smoothing length. For a full description of this procedure, see section~$3.2$ of \cite{2021Talbot+}. 

The angular momentum of the black hole will evolve due to the accretion of material from the $\alpha$-disc and the launching of the jet. In addition, if the black hole angular momentum is misaligned with that of the $\alpha$-disc then the $\alpha$-disc will warp which imposes mutual Bardeen-Petterson torques on the black hole and disc \citep{1975BardeenPetterson}. These torques cause the black hole and $\alpha$-disc angular momenta to precess and align with their total angular momentum. 

As shown in \cite{2021Talbot+}, the angular momentum evolution of the black hole due to these processes can be written as
\begin{align}
\label{eq: jdotbh}
    \boldsymbol{\dot{J}}_{\rm BH} &= \,(L_{\rm ISCO}- L_{\rm BZ})\,\frac{\dot{M}}{1+\eta_{\rm J}}\, \boldsymbol{j}_{\rm BH}\nonumber \\ 
    &-J_{\rm BH}\bigg\{\frac{\sin(\pi/7)}{\tau_{\rm GM}}( \boldsymbol{j}_{\rm BH} \times \boldsymbol{j}_{\rm d}) + \frac{\cos(\pi/7)}{\tau_{\rm GM}} \big[\boldsymbol{j}_{\rm BH} \times(\boldsymbol{j}_{\rm BH} \times \boldsymbol{j}_{\rm d})\big]\bigg\} \, ,
\end{align}
where $\tau_{\rm GM}$ is the gravitomagnetic timescale
\begin{equation}
\label{eq: taugm}
    \tau_{\rm GM} \approx 0.17\,\bigg(\frac{M_{\rm BH}}{10^6{\rm M_\odot}}\bigg)^{-2/35}\bigg(\frac{f_{\rm Edd}}{\epsilon_{\rm r}/0.1}\bigg)^{-32/35}a^{5/7} \; {\rm Myrs} \, ,
\end{equation}
and $f_{\rm Edd}$ is the Eddington fraction. $L_{\rm ISCO}$ is the spin-dependent specific angular momentum of material at the ISCO\footnote{For the full expression for $L_{\rm ISCO}$, see appendix A of \cite{2021Talbot+}.} and $L_{\rm BZ}$ is our `Blandford-Znajek specific angular momentum'
\begin{align}
    \boldsymbol{L}_{\rm BZ} &= L_{\rm BZ}\,\boldsymbol{j}_{\rm BH}\; , \nonumber \\
    &=\frac{\kappa}{2\pi}\,\phi_{\rm BH}^2\,\Bigg(\frac{a}{2\,\big(1+ \sqrt{1-a^2}\big)}\Bigg)\,\frac{GM_{\rm BH}}{c}\,\boldsymbol{j}_{\rm BH} \, .
\end{align}

The angular momentum of the $\alpha$-disc evolves due to the inflow of material from the surroundings, the loss of material from the ISCO and Bardeen-Petterson torques
\begin{align}
\label{eq: jdotd}
    \boldsymbol{\dot{J}}_{\rm d} =& \,\dot{M}_{\rm in}\,\boldsymbol{L}_{\rm in} - L_{\rm ISCO}\,\dot{M} \,\boldsymbol{j}_{\rm BH}\nonumber\\
    &+J_{\rm BH}\bigg\{\frac{\sin(\pi/7)}{\tau_{\rm GM}}( \boldsymbol{j}_{\rm BH} \times \boldsymbol{j}_{\rm d}) + \frac{\cos(\pi/7)}{\tau_{\rm GM}} \big[\boldsymbol{j}_{\rm BH} \times(\boldsymbol{j}_{\rm BH} \times \boldsymbol{j}_{\rm d})\big]\bigg\} \, ,
\end{align}
where $\boldsymbol{L}_{\rm in}$ is the specific angular momentum of inflowing material.

The Blandford-Znajek jet is powered by the spin-energy of the black hole. The energy flux into the jet is given by
\begin{equation}
\label{eq: EdotJet}
    \dot{E}_{\rm J} = \frac{\epsilon_{\rm BZ}}{1 + \eta_{\rm J}}\dot{M} c^2 \, ,
\end{equation}
\citep{2021Talbot+,1977BlandfordZnajek}.

The jet is launched by giving momentum kicks to gas cells within the `jet cylinder' \citep[for full details of the numerical procedure used to launch the jet see][]{2021Talbot+,2017BourneSijacki}. Cells in the `northern half'\footnote{A gas cell with position $\boldsymbol{r}_i$ is in defined to be in the `north' if $(\boldsymbol{r}_i - \boldsymbol{r}_{\rm BH}) \cdot \boldsymbol{j}_{\rm BH} >0$, where $\boldsymbol{r}_{\rm BH}$ is the position of the black hole.} of the cylinder are kicked in the direction of $\boldsymbol{j}_{\rm BH}$ and cells in the `southern half' are kicked in the direction of $-\boldsymbol{j}_{\rm BH}$. The momentum kicks are chosen to ensure that the change in kinetic energy of the jet cylinder is described by equation~(\ref{eq: EdotJet}).

In addition to momentum, mass is also injected into the jet cylinder, assumed to have come from the inner-edge of the ISCO. The mass flux into the jet is
\begin{equation}
\label{eq: MdotJet}
    \dot{M}_{\rm J} = \frac{\eta_{\rm J}}{1+\eta_{\rm J}}\dot{M} \, .
\end{equation}

We should also point out that the assumption that the mass that loads the jet comes from the ISCO of the $\alpha$-disc means that, in addition to kicking the cells in the jet cylinder along the jet axis, exact angular momentum conservation requires that these cells should also be given a toroidal kick to account for the angular momentum of the material at the ISCO. We calculated order of magnitude estimates of these toroidal kicks and found them to be negligible in comparison to the typical magnitude of the vertical kicks due to jet launching. We, therefore, do not include these toroidal kicks in our model.

\section{Simulation set-up}
\label{Sec: Simulation set-up}

We set-up our simulations to model the central regions of a typical Seyfert galaxy: a $10^6\, {\rm M_\odot}$ black hole is placed at the centre of a CND which is embedded in a stellar bulge and surrounded by a warm CGM. For a full description of the initial conditions, we refer the reader to section~$4.1$ of \cite{2021Talbot+}, however we do wish to emphasise that all components are self-gravitating and that no radiative cooling processes (other than the cooled CND, see Section~\ref{Sec: Circumnuclear discs} below) are modelled in our simulations. The typical scale-length of the CNDs in our simulation is $\sim 70$~pc and the scale-height is $\sim 5$ to $\sim 9$~pc with the actual value depending on the simulation setup. Initially, scale-heights and scale-lengths of the CNDs are resolved by $\sim10$ and $\sim100$ gas cells respectively, although our targeted refinement criteria tend to increase this somewhat during the simulation.

Having created a set of initial conditions, they are then relaxed for $200$~Myrs to allow any transients in the CND to dissipate and for the initially static CGM to come into equilibrium with the rotating CND. During this time we do not apply black hole refinement techniques, nor do we use the $\alpha$-disc-BZ jet model as this would result in jets with unphysical properties. The final snapshots from the relaxation are then used as initial conditions for the jet simulations. During this relaxation period the black holes grow by $\sim10^4\;{\rm M_\odot}$ which is small in comparison to their initial mass and also leads to negligible changes in the masses of the CNDs (which are initially $10^8 \; {\rm M_\odot}$). 

Upon starting the jet simulations we switch on the $\alpha$-disc accretion model and the black hole refinement but we allow $2$~Myrs before launching the jet\footnote{All the equations describing the $\alpha$-disc-BZ jet model in Section~\ref{Sec: BZ method} are valid when the jet is inactive. In this case $\epsilon_{\rm BZ} = 0$ and $\eta_{\rm J} = 0$.}. Over the course of these $2$~Myrs the mass of the $\alpha$-disc comes into equilibrium with inflowing gas from the wider simulation and the black hole refinement increases the spatial resolution close to the black hole \citep[see][]{2021Talbot+, 2015CurtisSijacki}. At $2$~Myrs, we reset the black hole mass and spin as well as the direction of the $\alpha$-disc angular momentum due to the fact that these properties evolve somewhat during the $2$~Myrs. Throughout the rest of the work, we will discuss the `initial' properties of the black hole/jet/$\alpha$-disc and this should be understood as the properties immediately after the jet turns on at $2$~Myrs. When we refer to times `before the jet turns on' we are referring to this initial $2$~Myrs and not to the previous relaxation period.

Before the jet launches, typical radii of gas cells in the CGM are $10-100$~pc and those of gas cells in the CND lie in the range $1-10$~pc. With the additional jet and black hole-specific refinement procedures, gas cells close to the black hole can have radii of the order $0.01$~pc\footnote{For the purposes of comparison, the size of the Broad Line Region (BLR) in local Seyferts is typically $\sim0.01$~pc which is, therefore, largely spatially unresolved in our simulations.}.

Expanding on the work presented in \cite{2021Talbot+}, we carry out a suite of simulations that vary the initial black hole spin magnitude and direction, the density of the CGM and the CND structure. For a full list of the simulations discussed in this work, see Table~\ref{tab: all runs} which details the labels that we use to refer to the simulations as well as the parameter choices that differentiate between the runs. We now briefly describe how we vary these parameters and discuss their physical interpretations.

\subsection{Circumnuclear discs}
\label{Sec: Circumnuclear discs}
In this work we consider the effects of altering the structure of the CND as this allows some control over the rate at which gas is funnelled towards the black hole. Specifically, we consider a `thick' CND (which is the CND used in \cite{2021Talbot+}) as well as a `thin' CND and a `thin-cooled' CND. The `thick' and `thin' cases are realised by initialising the gas discs with temperatures of $2\times10^4$~K and $10^4$~K, respectively, when generating the initial conditions. The higher gas temperatures in the `thick' CND case gives extra pressure support and thus the scale-height\footnote{The scale-heights of the CNDs are calculated by fitting a Gaussian to the vertical density profile of the gas at the scale-length of the CND ($\sim70$~pc).} of the `thin' disc ($5.40$~pc) is smaller than that of the `thick' CND ($9.26$~pc).

The `thin-cooled' CND has the same initial disc structure as the `thin' CND but is cooled using a simple beta-cooling prescription \citep{2007Nayakshin+,2005Rice+, 2003Rice+, 2001Gammie}. The cooling enhances the mass flux towards the centre of the CND, providing a setup in which we can assess how the black hole spin direction (and thus the jet direction) evolves due to higher, more sustained rates of mass inflow. The cooling also leads to a much denser disc which is likely to provide more resistance to the propagation of the jet. The cooling is applied to the gas cells that are in the CND and are not in the jet which we identify using passive tracers. For a gas cell with specific internal energy $u_i$, the cooling function is given by
\begin{equation}
    \frac{{\rm d}u_i}{{\rm d}t} = -\frac{1}{\beta}\, u_i\,\Omega_{{\rm k},i} \exp\bigg(-\frac{|z_i|}{\gamma (R_i+R_{\rm c})}\bigg) \, ,
\end{equation}
where $z_i$ and $R_i$ are the vertical and cylindrical radial coordinates of the gas cell relative to the black hole, $R_{\rm c}$ softens the cooling function and ensures that it does not diverge at the origin. $\Omega_{{\rm k},i} = \sqrt{GM_{\rm BH}/ (R_i+R_\mathrm{c})^3}$ is the (softened) Keplerian rotation associated with the gas cell and $\beta$ is a free parameter, calibrated to get the desired amount of cooling. Since we cool the disc only with the intention of enhancing inflow onto the black hole, we apply an exponential suppression term, with scale-height fixed by the free parameter $\gamma$ to ensure that disc cells that are entrained by the jet undergo negligible cooling. In all simulations presented here, we use $\beta = 10$, $\gamma = 1$ and $R_{\rm c} = 5$~pc which gives sufficiently strong cooling without causing the disc to fragment. The $\beta$-cooling is then active during the initial $2$~Myr relaxation period, so as to generate a thinner, denser CND which has a scale-height of $5.00$~pc at the end of the relaxation period. For the jet simulations that have these `thin-cooled' CNDs (see Table~\ref{tab: all runs}), the $\beta$-cooling is also active throughout the entirety of the simulation.

\subsection{Spin magnitude}
Current observations indicate that black holes can have a wide range spins \citep{2021Reynolds} and these show no clear trend with cosmic time nor with black hole mass. Hence, in this study we consider two different initial black hole spins: $a_0 = 0.2$ and $a_0 = 0.9$ which represent useful brackets to the realistic range of values. Throughout this work we also refer to these as the `low' power and `high' power cases due to the fact the black hole spin has a significant impact on the magnitude of the jet power, as will be explored in Section~\ref{Sec: Jet power}. We also refer to these runs as the spin $0.2$ and spin $0.9$ cases and this should be understood as referring to the {\it initial} black hole spin as they are free to evolve over the course of the simulation. 

\subsection{Spin direction}
The initial jet direction is described by the inclination of the black hole spin to the vertical $\theta_{\rm BH} \equiv \cos^{-1}(\boldsymbol{j}_{\rm BH}\cdot \hat{\boldsymbol{z}})$. Our analysis will largely be focused on three different jet inclinations: $\theta_{\rm BH, 0} \in \{0^\circ,\, 45^\circ, \, 90^\circ\}$ which we will also refer to as the `vertical', `inclined' and `horizontal' cases. These should be understood as referring to the {\it initial} direction of the black hole spin (and thus the direction of the jet) as the spin direction evolves with time. For a subset of our analyses, we will additionally consider two highly inclined jets with $\theta_{\rm BH, 0} = 70^\circ,\, 80^\circ$.

Our choice to investigate misaligned jets is motivated by the fact that some studies find that, in Seyferts, there is little evidence for correlation between the jet axis and the major axis of the host galaxy \citep{1999NagarWilson, 1998Clarke+,1997Schmitt+}.

\subsection{CGM density}
In \cite{2021Talbot+} we showed that the the morphology and evolution of the jets was particularly sensitive to the density of the surrounding CGM. We follow on from this work by considering the `standard' CGM ($\rho_{\rm CGM} = 10^{-27}\,{\rm g\, cm^{-3}}$) and `dense' CGM ($\rho_{\rm CGM} = 10^{-25}\,{\rm g\, cm^{-3}}$) cases, allowing us to fully investigate how the properties of self-consistent spin-driven jets are affected by the density of the ambient medium.

\begin{table*}
    \caption{This table details the properties of all simulations that are discussed in this paper. The first $5$ columns give the labels which we use when referring to a simulation in the text. Columns $6-11$ then list the physical quantities to which these labels correspond. Specifically, column $6$ gives the target cell volume of pure jet material, as described in Section~\ref{Sec: Resolution}, column $7$ indicates whether the CND is cooled, as described in Section~\ref{Sec: Circumnuclear discs}, column $8$ gives the temperature of gas cells in the CND in the initial conditions, column $9$ gives the density of the CGM, column $10$ gives the initial black hole spin magnitude and column $11$ gives the initial inclination of the black hole spin to the vertical. \textsuperscript{*} These runs will have the jet inclination stated specifically when being referred to in the text.}
    \label{tab: all runs}
    \begin{tabular}{ccccccccccc}
        \hline 
        \hline
        Resolution & CND & CGM density & Spin/Jet power  & Jet direction & $V_{\rm J}^{\rm max}$& $\beta$-cooling & $T_{\rm CND,0}$ & $\rho_{\rm CGM}$ & $a_0$ & $\theta_{\rm BH,0}$  \\
        label & label & label & label & label & [${\rm pc^3}$] & & [K] & [${\rm g\, cm^{-3}}$] & & [degrees]\\
        \hline
        \hline
        High & Thick & Standard & Low & Vertical & $10^2$ & $\xmark$ & $2\times10^4$ & $10^{-27}$ & $0.2$ & $0$ \\
        High & Thick & Standard & Low & Inclined & $10^2$ & $\xmark$ & $2\times10^4$ & $10^{-27}$ & $0.2$ & $45$ \\
        High & Thick & Standard & Low & Inclined\textsuperscript{*} & $10^2$ & $\xmark$ & $2\times10^4$ & $10^{-27}$ & $0.2$ & $70$ \\
        High & Thick & Standard & Low & Inclined\textsuperscript{*} & $10^2$ & $\xmark$ & $2\times10^4$ & $10^{-27}$ & $0.2$ & $80$ \\
        High & Thick & Standard & Low & Horizontal & $10^2$ & $\xmark$ & $2\times10^4$ & $10^{-27}$ & $0.2$ & $90$ \\
        High & Thick & Standard & High & Vertical & $10^2$ & $\xmark$ & $2\times10^4$ & $10^{-27}$ & $0.9$ & $0$ \\
        High & Thick & Standard & High & Inclined & $10^2$ & $\xmark$ & $2\times10^4$ & $10^{-27}$ & $0.9$ & $45$ \\
        High & Thick & Standard & High & Horizontal & $10^2$ & $\xmark$ & $2\times10^4$ & $10^{-27}$ & $0.9$ & $90$ \\
        Low & Thick & Standard & Low & Vertical & $10^3$ & $\xmark$ & $2\times10^4$ & $10^{-27}$ & $0.2$ & $0$ \\
        Low & Thick & Standard & Low & Horizontal & $10^3$ & $\xmark$ & $2\times10^4$ & $10^{-27}$ & $0.2$ & $90$ \\
        Low & Thick & Standard & High & Vertical & $10^3$ & $\xmark$ & $2\times10^4$ & $10^{-27}$ & $0.9$ & $0$ \\
        Low & Thick & Standard & High & Horizontal & $10^3$ & $\xmark$ & $2\times10^4$ & $10^{-27}$ & $0.9$ & $90$ \\
        \hline 
        High & Thick & Dense & High & Vertical & $10^2$ & $\xmark$ & $2\times10^4$ & $10^{-25}$ & $0.9$ & $0$ \\
        High & Thick & Dense & High & Horizontal & $10^2$ & $\xmark$ & $2\times10^4$ & $10^{-25}$ & $0.9$ & $90$ \\
        \hline
        High & Thin & Standard & High & Vertical & $10^2$  & $\xmark$ & $10^4$ & $10^{-27}$ & $0.9$ & $0$ \\
        \hline
        High & Thin-cooled & Standard & Low & Vertical & $10^2$ & $\cmark$ & $10^4$ & $10^{-27}$ & $0.2$ & $0$ \\
        High & Thin-cooled & Standard & Low & Horizontal & $10^2$ & $\cmark$ & $10^4$ & $10^{-27}$ & $0.2$ & $90$ \\
        High & Thin-cooled & Standard & High & Vertical & $10^2$ & $\cmark$ & $10^4$ & $10^{-27}$ & $0.9$ & $0$ \\
        High & Thin-cooled & Standard & High & Horizontal & $10^2$ & $\cmark$ & $10^4$ & $10^{-27}$ & $0.9$ & $90$ \\
        Low & Thin-cooled & Standard & Low & Vertical & $10^3$ & $\cmark$ & $10^4$ & $10^{-27}$ & $0.2$ & $0$ \\
        \hline 
        \hline 

    \end{tabular}
\end{table*}

\subsection{Resolution}
\label{Sec: Resolution}
In this work we carried out simulations where the jet lobes were evolved with two different spatial resolutions. The `high' resolution simulations are better able to resolve the instabilities in the jet lobes, meaning that these simulations are ideal for exploring the interaction of the jet lobes with the CND and CGM. These `high' resolution simulations, however, are particularly computationally expensive due to the large number of cells in the jet lobes, making it hard to follow the long-time evolution of the black hole and jet. For this reason, we carry out a set of `low' resolution simulations wherein the lobes are followed at a lower resolution.

We implement this by altering the jet lobe volume refinement criterion, described in detail in section $3.7.3$ of \cite{2021Talbot+}. Briefly, this refinement procedure identifies cells using a passive jet tracer, $ m_{{\rm J},i}$ by calculating the jet fraction of the cell ($f_{{\rm J},i}\equiv m_{{\rm J},i}/m_i$, where $m_i$ is the cell mass) and selecting those cells that satisfy $f_{{\rm J},i} > 10^{-5}$. Such cells are then split if their volumes, $V_i$, satisfy
\begin{equation}
    V_i > \big[1 - \log_{10}(f_{{\rm J},i})\big]\,V_{\rm J}^{\rm max} \; .
\end{equation}
$V_{\rm J}^{\rm max}$ is a parameter that should be understood as being the maximum volume of a cell that contains pure jet material. In this work, we consider two values: $V_{\rm J}^{\rm max} = 10^2\,{\rm pc^3}$ and $V_{\rm J}^{\rm max}= 10^3\,{\rm pc^3}$, which correspond to `high' and `low' resolution simulations respectively. The effects of lowering the resolution of the jet lobes are explored in more detail in Appendix~\ref{app: res test}.

\section{Results}
\label{Sec: Results}

\subsection{Qualitative exploration of jet-driven outflows}
\label{Sec: Qualitative exploration}
\begin{figure*}
    \centering
    \includegraphics[width=\textwidth]{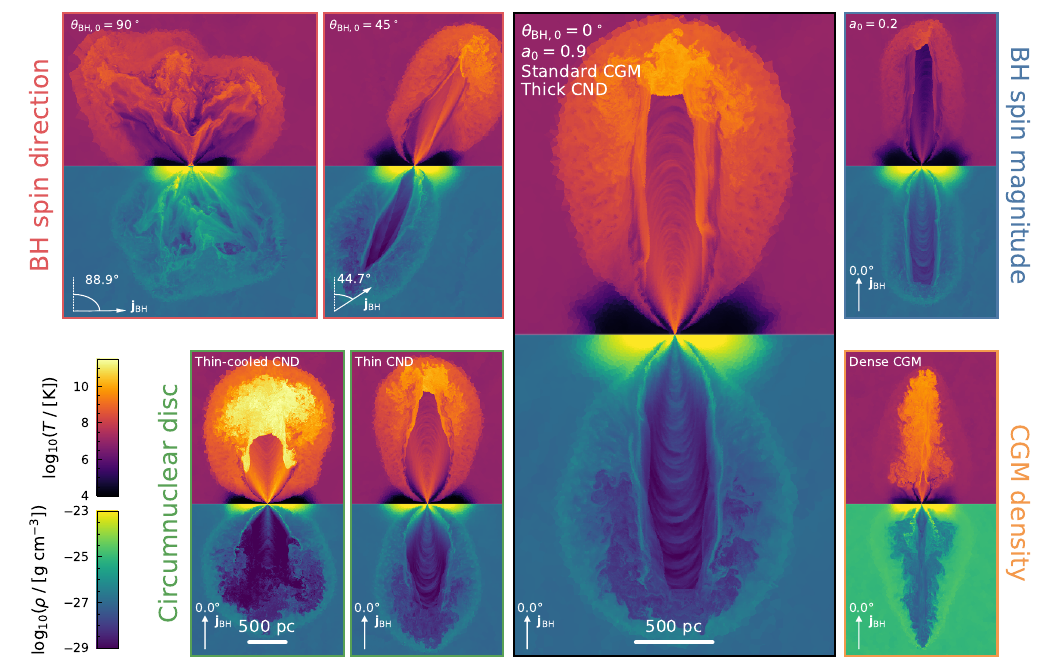}
    \caption{Slices in the $x$-$z$ plane, centred on the black hole, for a representative sample of the `high' resolution simulations. The top half of each panel shows the temperature field and the bottom half shows the density field. The main panel shows a `reference' run and it's initial spin direction and magnitude, the CGM density and the CND structure are indicated in the top left corner. Each of the other panels then show slices from a simulation in which we vary one of these parameters whilst keeping the others fixed. The parameter being varied is indicated by the colour of the border around the slice and its value is shown in the top left corner of the panel. Each slice corresponds to the time when the $z$-extent of the outflow is closest to $2$~kpc (so the spatial scales of the smaller panels are all the same). The inclination of the jet is shown in the bottom left of each panel. It is worth noting that changing these parameters within the realistic range of values leads to very different jet morphologies.}
    \label{fig: slices}
\end{figure*}

We begin our analysis of the simulation suite by providing a qualitative overview of the different morphologies of the jet-driven outflows. Fig.~\ref{fig: slices} shows slices in the $x$-$z$ plane, centred on the black hole, of a representative selection of the `high' resolution simulations. The top half of each panel shows the temperature field and the bottom half shows the density field. Each slice corresponds to the time when the $z$-extent of the outflow\footnote{The `$z$-extent of the outflow' is defined to be the maximal displacement, relative to the black hole, of a gas cell defined to be in the jet ($f_{\rm J, i}> 10^{-3}$). In the `vertical' jet cases, this is analogous to the length of the largest jet lobe.} is closest to $2$~kpc. The main panel shows the `high' spin `vertical' jet in the `thick' CND launched into the `standard' CGM. Each of the other panels then shows a simulation in which we vary one of these parameters whilst keeping the others fixed. The parameter being varied is indicated by the colour of the border around the slice and its value is shown in the top left corner of the panel. Specifically, the two panels in the top left show simulations where we change the initial black hole spin direction, the two in the bottom left show simulations with different CND structures. The panel in the top right shows the `low' spin case and the bottom right shows the `dense' CGM case. 

It is immediately obvious that a wide variety of outflow morphologies are produced just by changes to these four parameters. The jet launched directly into the CND drives a multiphase, quasi-bipolar wind as it blows off material from the surface of the CND. Inclining the jet at $45^\circ$ produces an outflow that has a greater resemblance to the `vertical' jet case but the jet channel is more susceptible to disruption and there is clear evidence for asymmetric entrainment of CND material. The jet launched from the `thin-cooled' CND is much more powerful (this is explored quantitatively in Section~\ref{Sec: Jet power}) which leads to a significantly hotter and faster jet with a broader, shorter jet channel and a more extended cocoon. The `low' spin jet is comparatively cold and slow. The shock at the head of the jet is weaker and the surrounding cocoon is cooler, with less backflowing gas than is seen in other runs. The increased pressure of the ambient medium in the `dense' CGM run leads to a much narrower jet channel and a series of recollimation shocks form along the jet axis which divert material off-axis and form a hot cocoon (for a more extensive discussion of the jet morphologies in `standard' vs. `dense' CGM cases, see \cite{2021Talbot+}). We now proceed to explore these simulations in more detail.

\subsection{Quantitative exploration of the black hole and jet properties}
\label{Sec:Quantitative exploration}

\begin{figure*}
   \centering
 \includegraphics[width=\textwidth]{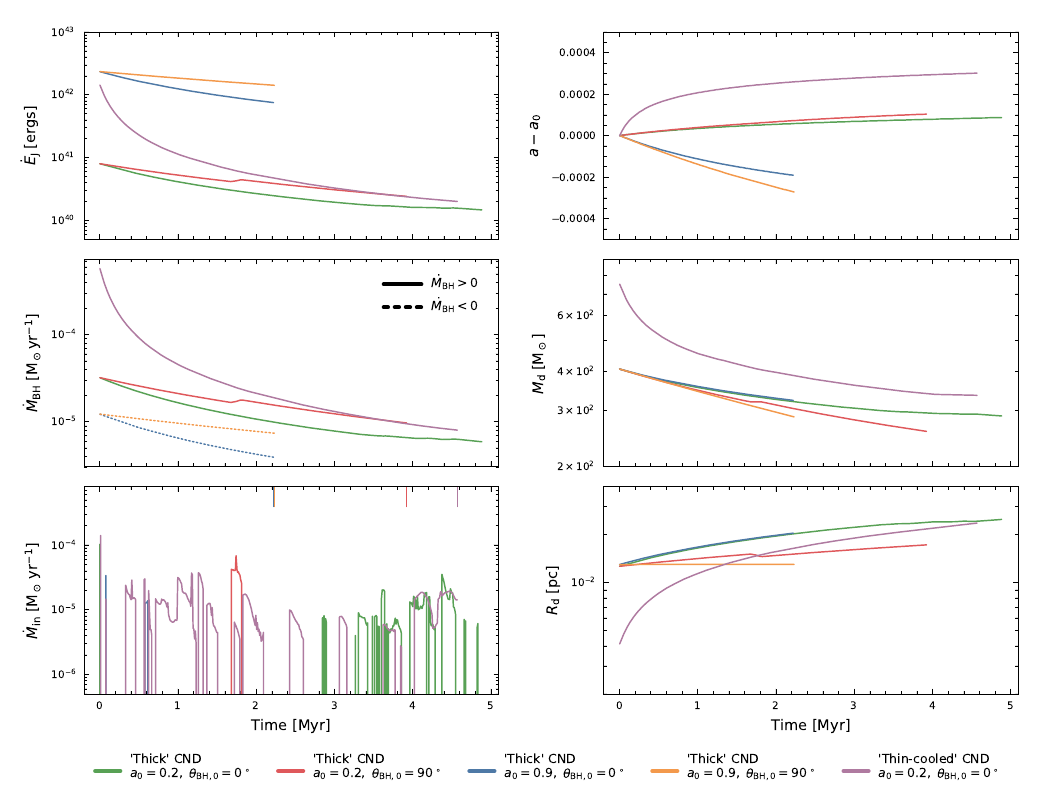}
  \caption{Time evolution of various diagnostic properties for a representative sample of the `low' resolution runs. In the first row, the time evolution of the jet power is shown on the left and that of the black hole spin magnitude (relative to the initial spin) is shown on the right. In the second row, the evolution of the black hole mass growth rate is shown on the left and that of the accretion disc mass is shown on the right. In the third row, the panel on the left shows the evolution of the mass inflow rate onto the $\alpha$-disc and the evolution of its radius is shown on the right. The legend below the panels indicates which simulation each line colour corresponds to. In the panel that shows the mass inflow rate, for clarity, the coloured tick marks on the upper $x$-axis indicate the final time of those simulations that, due to computational limitations, did not reach $5$~Myrs.}
  \label{fig: subgrid}
\end{figure*}

The initial jet configurations in our simulations correspond to bolometric luminosities in the range $10^{41}-10^{43} \,{\rm erg \, s^{-1}}$ which are comparable to typical Seyfert nuclear luminosities for black holes of mass $10^6 \; {\rm M_\odot}$ which span the range $10^{40}-10^{43} \,{\rm erg \, s^{-1}}$ \citep[see e.g.][for further details regarding bolometric luminosities see Section~\ref{Sec: outflow obs comparison}]{2021Ruschel-Dutra, 2021Venturi+, 2020Chen+, 2018Komossa, 2015Foschini+, 2013Combes+}. Typical accretion rates through the $\alpha$-disc range from $\sim 9\times 10^{-4}$ to $\sim 6\times 10^{-2}$ of the Eddington rate which are on the lower side of those inferred for radio-loud Narrow-line Seyferts \citep{2018Komossa, 2012Xu+, 2010Grupe+, 2002Boroson}. This is to be expected given that our simulations probe the lower ranges of black hole masses and jet powers that are observed in these systems.

In this and the following section we use Fig.~\ref{fig: subgrid} to aid our discussion of the $\alpha$-disc, black hole and jet properties. This figure shows the time evolution of various diagnostic properties for a representative sample of the `low' resolution runs\footnote{In Appendix~\ref{app: res test} we show that lowering the jet lobe resolution has negligible impact on the evolution of the black hole and $\alpha$-disc properties.}. Specifically, the panels show the time evolution of: the jet power (top-left), the black hole spin magnitude (relative to the initial spin; top-right), the black hole mass growth rate (centre-left), the accretion disc mass (centre-right), the mass inflow rate onto the $\alpha$-disc (bottom-left), and the $\alpha$-disc radius (bottom-right). In Fig.~\ref{fig: subgrid} we do not show any of the runs in the `dense' CGM as the evolution of their $\alpha$-disc, black hole properties and jet power is largely similar to those of the `standard' CGM runs.

In all simulations, the evolution of the jet power is non-negligible, this is particularly obvious in the case of the `low' power `vertical' jet in the `thin-cooled' CND in which the jet power drops by over an order of magnitude within $2$~Myrs. Unsurprisingly, higher spin black holes launch more powerful jets. Similarly, black holes in the `thin-cooled' CND launch more powerful jets relative to those in the `thick' CND (the jet power evolution will be discussed in more detail in Section~\ref{Sec: Jet power}).

Interestingly, from inspecting the black hole mass growth rates it is clear that the masses of the `high' spin black holes are actually decreasing (the growth rate is slightly negative), whereas the `low' spin black holes are able to grow. Equation~(\ref{eq: MdotBH}) shows that the mass of a black hole will increase if $1 - \epsilon_{\rm r} - \epsilon_{\rm BZ}>0$, and will decrease otherwise. To understand this, recall that the black hole mass evolution is driven by fluxes of {\it energy} across its horizon. The $1 - \epsilon_{\rm r} - \epsilon_{\rm BZ}$ factor accounts for the fact that the binding energy of accreted gas is reduced below that of it's rest-mass energy and for the fact that power of a Blandford-Znajek jet is driven by the {\it spin-energy} of the black hole (this is explained in more detail in sections~$2.1$~and~$2.3$ of \cite{2021Talbot+}). In the framework of our model, $1 - \epsilon_{\rm r} - \epsilon_{\rm BZ}$ is a function of spin alone that is positive for $a<0.81$ and negative otherwise.

We now turn to the evolution of the mass inflow rate onto the $\alpha$-disc. Focusing on the runs in the `thick' CND, we see that in the $\sim 3$~Myrs after the jet turns on there is no significant inflow onto the $\alpha$-disc. After $3$~Myrs, inflow is able to resume in the `low' power `vertical' jet case, however this inflow is fairly bursty in comparison to that which is found in the absence of a jet (not shown in Fig.~\ref{fig: subgrid}; although see Fig.~\ref{fig: mdot in slices}). For the `low' power `vertical' jet in the `thin-cooled' CND, inflow is able to resume earlier and is somewhat less bursty than the corresponding run in the `thick' CND. The fact that modest differences in the cooling of the CND gas can either result in no inflows (for $\sim 3$~Myrs) or largely sustained inflows highlights the fact that the feeding of the $\alpha$-disc and black hole is particularly sensitive to changes in the thermodynamics of the surrounding gas. It should also be noted that the jets themselves can significantly influence gas kinematics in the vicinity of the black hole and can cause significant disruption (see Section~\ref{Sec: Close to the black hole} for further details) which is why the runs in which the jets are initially launched directly into the CND take longer for inflow to resume.

In contrast to the jet power evolution, the black hole spins all show very modest changes over the course of the simulations. Perhaps the most obvious trend, however, is that all `high' spin ($a_0 = 0.9$) runs spin down while all `low' spin ($a_0 = 0.2$) runs spin up. The evolution of the black hole spin is determined by 
\begin{equation}
    \label{eq: adot}
    \frac{\dot{a}}{a} = \frac{\dot{J}_{\rm BH}}{J_{\rm BH}} - 2\frac{\dot{M}_{\rm BH}}{M_{\rm BH}}  \approx \frac{\dot{J}_{\rm BH}}{J_{\rm BH}} \; .
\end{equation}
In all runs, the contribution to spin evolution from black hole growth is subdominant relative to the contribution from changes to the black hole angular momentum. From equation~(\ref{eq: jdotbh}), recall that the magnitude of the black hole angular momentum evolves only due to accretion from the $\alpha$-disc and launching of the jet
\begin{equation}
    \label{eq: jdot mag}
    \dot{J}_{\rm BH} = (L_{\rm ISCO} - L_{\rm BZ}) \frac{\dot{M}}{1+\eta_{\rm J}} \; .
\end{equation}
Whether the black hole spins up or down, therefore, depends on the sign of $L_{\rm ISCO} - L_{\rm BZ}$ which is a function of spin (with normalisation set by the black hole mass). The behaviour of $L_{\rm ISCO} - L_{\rm BZ}$ is such that black holes with $a<0.38$ will spin up and those with $a>0.38$ will spin down. This is consistent with the spin evolution of our simulations in Fig.~\ref{fig: subgrid}. It is also clear that the rate of increase of the black hole spin in the `low' power `vertical' jet launched from the `thin-cooled' CND evolves faster than that of the black hole in the corresponding run in the `thick' CND. This is due to the fact that the rate of change of the magnitude of the black hole angular momentum is proportional to $\dot{M}$ (see equation~(\ref{eq: jdot mag})) which is higher in the `thin-cooled' CND than in the `thick' CND.

The initial $\alpha$-disc mass and radius in each simulation are determined by the behaviour of the mass flows through the CND and accretion onto the black hole during the $2$~Myrs before the jet turns on. For this reason, the initial values of the $\alpha$-disc mass and radius are the same for all runs with the `thick' CND, while all runs with the `thin-cooled' CND host higher initial $\alpha$-disc masses (and lower radii) due to enhanced inflows facilitated by the rapidly cooling gas. Once the jet is launched, it temporarily suppresses the inflow onto the $\alpha$-disc. This causes the mass of the $\alpha$-disc to drop as the mass that is lost at the ISCO, which either feeds the black hole or is drawn up into the jet, is not able to be replenished. The specific angular momentum of material at ISCO is significantly lower than that of the $\alpha$-disc and so, in the absence of inflows, the specific angular momentum of the $\alpha$-disc increases, causing the $\alpha$-disc radius to increase.

Fresh inflows of mass onto the $\alpha$-disc can lead to changes in its radius, as is seen at $t \approx 1.8$~Myr in the `low' power `horizontal' jet in the `thick' CND case (red line). At this time there is a brief burst of inflowing gas, which slightly increases the $\alpha$-disc mass (and consequently, the jet power) but leads to a small reduction in the radius of the $\alpha$-disc. Whether the radius increases or decreases in response to inflows onto the $\alpha$-disc depends on the angular momentum of the inflowing gas relative to that of the disc. In the framework of our sub-grid model, we only allow inflows of gas that would be able to circularise and settle in the $\alpha$-disc\footnote{The criteria we use to enforce this is described in detail in section $3.2$ of \cite{2021Talbot+}.} (i.e. the gas needs to have specific angular momentum smaller than that of the outer edge of the $\alpha$-disc). In these simulations, the specific angular momentum of such inflowing gas is typically lower than that of the $\alpha$-disc, which is why the disc radius tends to decrease in response to inflows. In Section~\ref{Sec: Close to the black hole} we will explore the origin of this low angular momentum gas.

In this section we do not discuss the black hole and $\alpha$-disc angular momentum directions; the evolution of these quantities is briefly explored in Appendix~\ref{App: bh dir evolution} in which we also analyse a simulation in which the black hole spin direction is initially `horizontal' and the angular momentum of the $\alpha$-disc is not aligned with the vertical, but rather is slightly misaligned with that of the black hole spin.

\subsection{Jet power}
\label{Sec: Jet power}

We now explore the underlying physical processes that lead to evolution of the powers of these jets. First, recall that the jet power is given by equation~(\ref{eq: EdotJet}). From this, it is clear that the jet power depends on the mass flux through the $\alpha$-disc and the Blandford-Znajek efficiency $\epsilon_{\rm BZ}$ which is a function of spin alone. 

The mass flux through the $\alpha$-disc is
\begin{equation}
    \dot{M} = {\rm min}(f_{\rm Edd}, 1) \; \dot{M}_{\rm Edd}\,,
\end{equation}
where
\begin{equation}
\label{eq: fedd}
    f_{\rm Edd} \approx 0.76 \bigg(\frac{\epsilon_{\rm r}}{0.1}\bigg)\bigg(\frac{M_{\rm d}}{10^4 \,{\rm M_\odot}}\bigg)^5\bigg(\frac{M_{\rm BH}}{10^6 \,{\rm M_\odot}}\bigg)^{-47/7}\bigg(\frac{a \; J_{\rm d} / J_{\rm BH}}{3}\bigg)^{-25/7}\, .
\end{equation}
For the derivation of this expression, see Appendix A of \cite{2018Fiacconi+}. The jet power therefore scales as
\begin{equation}
\label{eq: edot scaling}
    \dot{E}_{\rm J} \propto\epsilon_{\rm BZ}\,M_{\rm d}^{5}\,M_{\rm BH}^{-40/7}\,\bigg(\frac{aJ_{\rm d}}{J_{\rm BH}}\bigg)^{-25/7} \,.
\end{equation}

As discussed in Section~\ref{Sec:Quantitative exploration}, the jet powers all evolve over the course of the simulations, but from Fig.~\ref{fig: subgrid} it is also evident that the initial powers of these jets differ as well. We now briefly explore what determines this initial jet power.

\subsubsection{Initial jet power}
\label{Sec: Initial jet power}
Two clear trends can be seen in the initial jet powers: (i) The `high' spin runs in the `thick' CND have initial powers that are higher than the corresponding `low' spin runs and (ii) the `low' spin runs in the `thin-cooled' CND have higher initial jet powers than the corresponding runs in the `thick' CND\footnote{Whilst, for clarity, only one run in the `thin-cooled' CND is shown in Fig.~\ref{fig: subgrid}, we have verified that this is the case for all jet launched from the `thin-cooled' CND.}.

To understand these differences, we return to equation~(\ref{eq: edot scaling}) which shows that the initial jet power depends on $a$, $M_{\rm d}$, $M_{\rm BH}$ and the ratio $aJ_{\rm d}/J_{\rm BH}$. For all simulations, initial differences in black hole masses are negligible, as are initial variations in $aJ_{\rm d}/J_{\rm BH}$. As discussed in the previous section, the masses of the $\alpha$-discs are such that accretion onto the black hole is balanced by inflow from the surroundings. The higher inflow rates in the `thin-cooled' CND leads to a higher initial $\alpha$-disc mass ($\sim750 \; {\rm M_\odot}$) relative to that found in the `thick' CND runs ($\sim400 \; {\rm M_\odot}$).

Initial differences in jet power between the `high' and `low' spin cases arise due to the fact that the efficiency of black hole energy extraction into the jet, $\epsilon_{\rm BZ}(a)$, increases with spin. Differences in jet power between the `thick' and `thin-cooled' CND runs, however, are due to the fact that higher mass $\alpha$-discs are able to feed the black hole at higher rates.

\subsubsection{Jet power evolution}
\label{Sec: Jet power evolution}

Having discussed the factors affecting the initial powers of the jets, we now explore the processes that are driving the evolution of the power of these jets that is seen in Fig.~\ref{fig: subgrid}. 

Differences in the initial power due to the black hole spin dependence of the jet efficiency persist throughout the simulations (recall that the black hole spins do not evolve appreciably in all cases). The initial difference in power between the `low' spin runs in the `thick' and `thin-cooled' CNDs, however, diminishes over time. This is due to the more rapid decrease of the $\alpha$-disc masses in the `thin-cooled' CND runs which brings their masses in line with those found in the `thick' CND runs (see the middle-right panel of Fig.~\ref{fig: subgrid}).

Before the jet turns on, inflow and accretion are in an equilibrium state. Gas inflow is then initially cut off when the jet is launched and, during the time where gas is unable to reach the $\alpha$-disc, the mass of the disc decreases as it is drained by the black hole. When inflow resumes, it is still insufficient to balance accretion onto the black hole but it somewhat slows the rate at which the $\alpha$-disc mass decreases and, thus, reduces the rate of decline of jet power (see eg. the `low' power `vertical' jet in the `thick' CND). 

Having explored the most obvious features of the jet power evolution, we now turn to some `higher-order' effects which have yet to be discussed. First, it is clear that the power of `vertical' jets initially drops faster than `horizontal' jets\footnote{By `initially' we mean at times before mass inflow onto the $\alpha$-disc is able to resume.}. This trend is due to the way the $\alpha$-disc angular momentum evolves in these two scenarios: in the `vertical' jet case, $J_{\rm d}$ stays approximately constant and in the `horizontal' jet case $J_{\rm d}$ drops fairly rapidly (recall that the jet power is proportional to $J_{\rm d}^{-25/7}$).

To understand this, we rewrite the evolutionary equation for $\boldsymbol{J}_{\rm d}$ (equation~(\ref{eq: jdotd})) in the form
\begin{equation}
\label{eq: jdotd simple}
    \boldsymbol{\dot{J}}_{\rm d} =  -\dot{J}_{\rm acc, d}\,\boldsymbol{j}_{\rm BH}  + \dot{J}_{\rm prec}\frac{\boldsymbol{j}_{\rm BH} \times \boldsymbol{j}_{\rm d}}{|\boldsymbol{j}_{\rm BH} \times \boldsymbol{j}_{\rm d}|} + \dot{J}_{\rm align}\frac{\boldsymbol{j}_{\rm BH} \times (\boldsymbol{j}_{\rm BH} \times \boldsymbol{j}_{\rm d})}{|\boldsymbol{j}_{\rm BH} \times (\boldsymbol{j}_{\rm BH} \times \boldsymbol{j}_{\rm d})|} \, ,
\end{equation}
where we have neglected $\boldsymbol{\dot{J}}_{\rm in}$ as we are considering times before inflow restarts.

Initially, $\boldsymbol{j}_{\rm BH} = \boldsymbol{j}_{\rm d} = \hat{\boldsymbol{z}}$ in the vertical jet case and $\boldsymbol{j}_{\rm BH} = \hat{\boldsymbol{x}}$ and $\boldsymbol{j}_{\rm d} = \hat{\boldsymbol{z}}$ in the horizontal jet case. Thus, for the vertical jet
\begin{equation}
    \boldsymbol{\dot{J}}_{\rm d, v} = -\dot{J}_{\rm acc, d}\,\hat{\boldsymbol{z}}\,,
\end{equation}
and for the horizontal jet
\begin{equation}
    \boldsymbol{\dot{J}}_{\rm d, h} = -\dot{J}_{\rm acc, d}\,\hat{\boldsymbol{x}} - \dot{J}_{\rm prec}\,\hat{\boldsymbol{y}} - \dot{J}_{\rm align}\,\hat{\boldsymbol{z}}\,.
\end{equation}
If the initial disc angular momentum is $J_{\rm d,0}\, \hat{\boldsymbol{z}}$ then after time $\Delta t$
\begin{equation}
    J_{\rm d}^2 = |J_{\rm d,0}\hat{\boldsymbol{z}} + \boldsymbol{\dot{J}}_{\rm d} \Delta t|^2\,,
\end{equation}
which to $\mathcal{O}(\Delta t)$ is 
\begin{equation}
    J_{\rm d, v}^2 =J_{\rm d,0}^2-2J_{\rm d,0}\dot{J}_{\rm acc, d}\Delta t\,,
\end{equation}
for the vertical jet and
\begin{equation}
    J_{\rm d, h}^2 =J_{\rm d,0}^2-2J_{\rm d,0}\dot{J}_{\rm align}\Delta t\,,
\end{equation}
for the horizontal jet. 

We find that Bardeen-Petterson torques ($\dot{J}_{\rm align}$ and $\dot{J}_{\rm prec}$) are orders of magnitude larger than accretion torques ($\dot{J}_{\rm acc, d}$) which is unsurprising given the low mass fluxes through the $\alpha$-disc. This is why the magnitude of the $\alpha$-disc angular momentum in the `horizontal' case decreases faster than in the `vertical' case. 

Another feature in the evolution of the jet powers is that, for black holes of a given spin, the power of the jets launched from the `thin-cooled' CND drops faster than those launched from the `thick' CND. This comes about due to the initially higher accretion rates in the `thin-cooled' CND case. These higher accretion rates lead to faster draining of the $\alpha$-disc which, in turn, leads to a higher rate of change of $\dot{M}$ and, ultimately, a more rapid drop in jet power (the jet power is proportional to $\dot{M}$, see equation~(\ref{eq: EdotJet})).

\subsection{Jet interaction with the CND}
\begin{figure*}
    \centering
    \includegraphics[width=\textwidth]{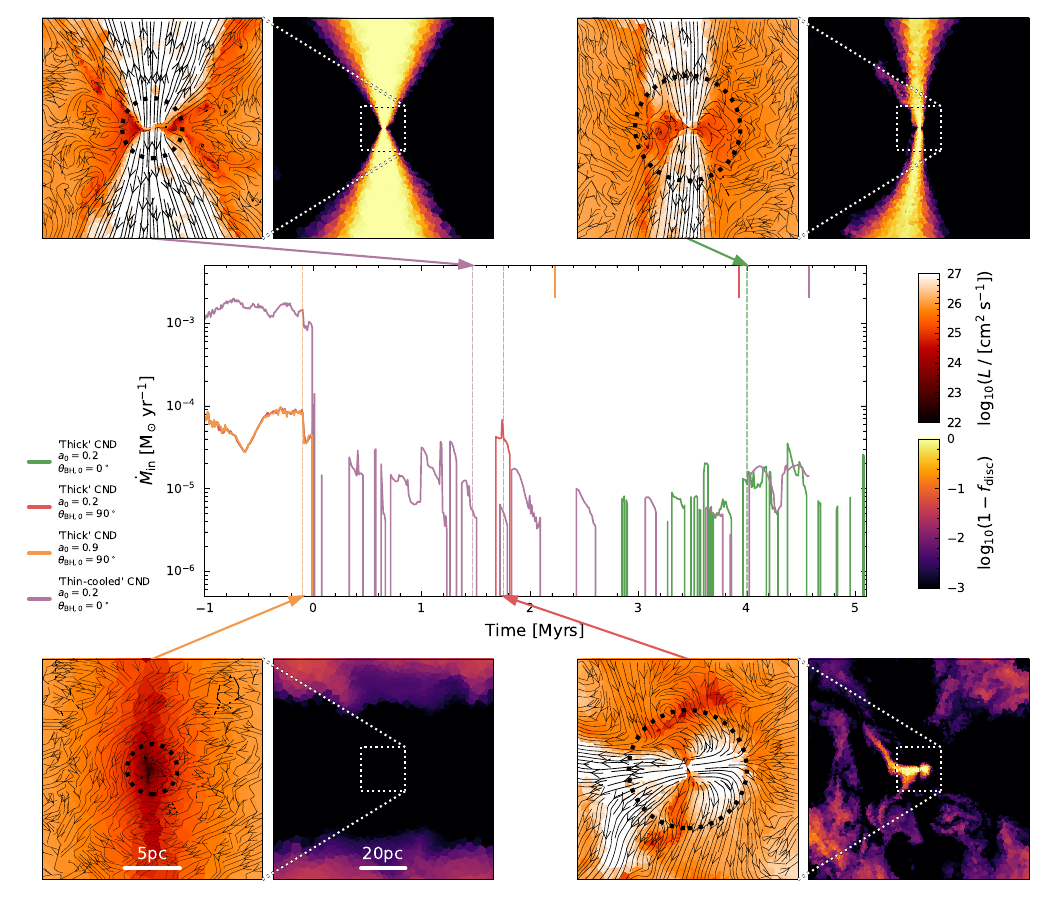}
    \caption{The main panel shows the mass inflow rate onto the $\alpha$-disc for three of the runs in the `thick' CND and one run in the `thin-cooled' CND (see legend). Negative times correspond to times before the jet is turned on. The four pairs of panels contain slices, centred on the black hole, in which the angular momentum of the $\alpha$-disc is aligned with the $y$ axis. In each pair of panels, the slice on the left shows the specific angular momentum of the gas with the streamlines indicating the velocity field and the slice on the right shows a passive tracer that identifies all material that does not originate from the CND. The arrow from each of the specific angular momentum plots points to the time at which this pair of slices was made, chosen such that they show the state of the gas at times when inflows of gas are able to reach the $\alpha$-disc. These times are also highlighted by the vertical dashed lines in the main panel with the corresponding colour. The coloured tick marks on the upper $x$-axis indicate the final time of those simulations that, due to computational limitations, did not reach $5$~Myrs. The dotted circle in each specific angular momentum plot indicates the smoothing length of the black hole. Whilst accretion onto the $\alpha$-disc is largely sustained before the jet turns on, after jet launching it becomes bursty, with peaks corresponding to the times when jet-driven backflows deliver low angular momentum material to the black hole, including entrained CND gas.}
    \label{fig: mdot in slices}
\end{figure*}

In this section, we explore the processes affecting the behaviour of the mass inflow rate onto the sub-grid $\alpha$-disc, $\dot{M}_{\rm in}$. As we saw in the previous section, $\dot{M}_{\rm in}$, plays an important role in regulating the mass of the $\alpha$-disc which, in turn, affects the power of the jet via changes to the black hole accretion rate.

We saw that mass inflow rate is significantly altered by the launching of the jet. But also, by design, the exact value of $\dot{M}_{\rm in}$ also depends on the properties of the gas in the vicinity of the black hole as only radially inflowing gas that is able to circularise is allowed to settle on the $\alpha$-disc. Hence, we first look at how the CND gas close to the black hole responds to the jet launching as this is what determines the instantaneous accretion rate. Afterwards we will consider the effect of the jets on the CND on larger scales as this determines what gas may be available for future accretion.

\subsubsection{Inflows close to the black hole}
\label{Sec: Close to the black hole}

The main panel in Fig.~\ref{fig: mdot in slices} shows the mass inflow rate onto the $\alpha$-disc for three of the runs in the `thick' CND and one run in the `thin-cooled' CND. This panel is almost identical to the left-hand panel in the bottom row of Fig.~\ref{fig: subgrid} except for the fact that inflow rates are additionally shown for a period of time before the jets turn on. The four pairs of panels contain slices, centred on the black hole in which the angular momentum of the $\alpha$-disc is aligned with the $y$ axis. In each pair of panels, the slice on the left shows the specific angular momentum of the gas with streamlines that indicate the velocity field and the slice on the right is of a passive tracer that identifies all gas that does not originate from the CND.

At times before the jet is launched, the inflow is coherent and axisymmetric and the gas has a clear (cylindrical) radial gradient in specific angular momentum (see the slices on the bottom-left of Fig.~\ref{fig: mdot in slices}). This corresponds to a sustained mass inflow rate, indicating that the sub-grid $\alpha$-disc systems are all in approximate equilibrium with inflow from the surroundings. From the main panel, it is clear that before the jet turns on, the run in the `thin-cooled' CND hosts inflow rates that are approximately an order of magnitude larger than those found in the `thick' CND.

In all cases, when the jet is launched, the gas local to the black hole is disrupted, with considerable changes to the velocity field indicating that the kinematics of this gas has been significantly affected. The regions occupied by the jet itself are clearly seen in the tracer maps (primarily identified by yellow colours) and these jet regions are also comprised of gas with comparatively high specific angular momentum. 

The slices of the `horizontal' jet case (shown in the bottom-right of Fig.~\ref{fig: mdot in slices}) clearly show that, on small-scales, the jet is propagating directly into the CND. The large-scale outflow, however, has a quasi-bipolar morphology, as can be seen in the top-left panel of Fig.~\ref{fig: slices}. The fact that outflows tend to escape along the rotation axis of the CND, regardless of the initial direction, was highlighted in \cite{2021Talbot+} (see also \cite{2019Nelson+,2016CurtisSijacki,2014Costa+}). 

\begin{figure*}
    \centering
    \includegraphics[width=\textwidth]{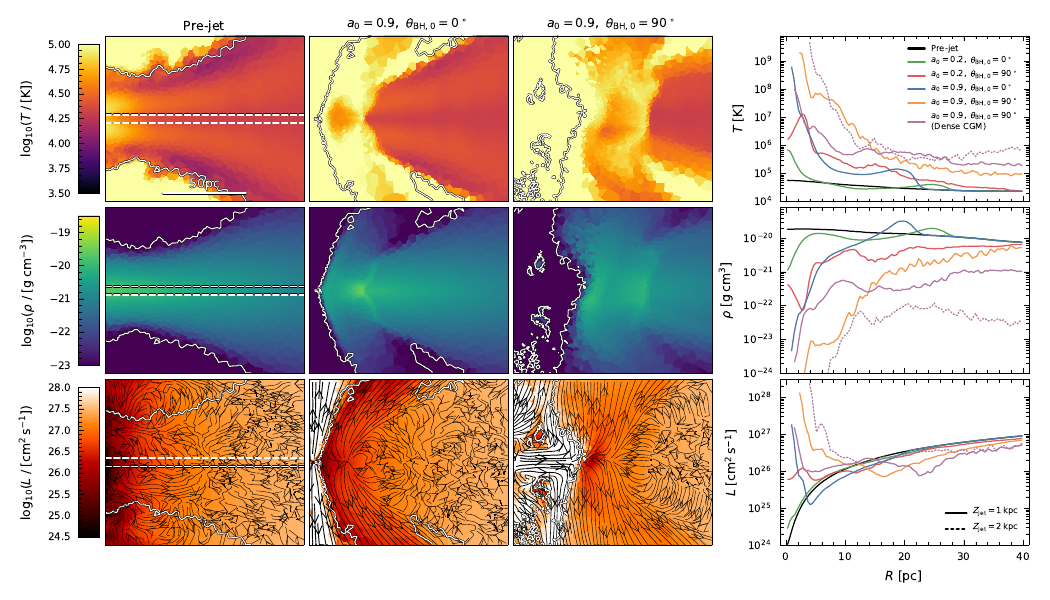}
    \caption{The first three columns show slices (with the angular momentum of the CND aligned with the $y$ axis and the centre of mass of the CND in the centre-left) of the temperature (first row), density (second row) and specific angular momentum (third row) for the `high' spin runs in the `thick' CND. Streamlines of the gas velocity are overlaid on the specific angular momentum slices. The first column corresponds to the state of the gas prior to jet launching, the second column shows the `vertical' jet case and the third column shows the `horizontal' jet case. The slices in the second and third columns were made using the simulation output in which the extent of the outflow in the $z$-direction is closest to $2$~kpc. The solid line in each slice shows a contour of a passive tracer that identifies CND material. The panels in the final column show cylindrical radial profiles of gas in the mid-plane of the CND. The mass-weighted temperature is shown in the top, the volume-weighted density in the middle and the mass-weighted specific angular momentum is shown in the bottom. The colour of the line indicates the initial jet configuration (see the legend in the upper left panel). In addition to the four runs in the `thick' CND with the `standard' CGM, we also show the `high' power `horizontal' jet in the `thick' CND that is launched into the `dense' CGM. All profiles correspond to the time when the $z$-extent of the outflow is closest to $1$~kpc. We additionally show the profile for the run in the `dense' CGM when its $z$-extent is closest to $2$~kpc as this corresponds to the case in which the CND is maximally affected. The initial CND profile is shown in black. In the first column of the slices, the region enclosed by the dotted line indicates the gas which was used to calculate the profiles.}
    \label{fig: cnd profiles slices thick}
\end{figure*}

In simulations in which inflow is able to resume after the jet turns on, the presence of vortices near the base of the jet indicates that jet-driven backflows become important for feeding the black hole and, ultimately, regulating the power of the jet. These backflows draw inflowing gas with sufficiently low angular momentum to within the black hole smoothing length, hence resulting in a burst of inflow. This behaviour is seen in both the `vertical' and the `horizontal' jet cases. In the `vertical' jet cases, these vortices predominantly draw in gas from the CND (indicated by dark colours in the tracer slice). This is also observed in the `horizontal' jet case, however, we additionally see that the backflows driven by these jets are able to draw in material that does not originate from the CND (i.e. pure jet or CGM gas). 

From this discussion, it is clear that the bursty nature of gas inflow after jet launching comes about as a result of the interplay between two processes: the jets cut off inflow by generating funnels of high angular momentum, outflowing material which shock heat the local gas and alter its kinematics. But these jets additionally drive backflows that act to draw low angular momentum gas towards the black hole which, ultimately, refuels the jet activity. 

These small-scale, jet-driven backflows close to the base of the jet clearly play an important role in the feeding of the black hole. Other works have also shown that large-scale backflows within the jet cocoon can draw low angular momentum gas back towards the black hole \citep{2014Cielo+,2010Antonuccio-DeloguSilk}. Understanding the way jets alter the angular momentum structure and kinematics of the gas on both small and large scales is, therefore, vital if we are to understand how self-regulating jet feedback cycles are established and maintained.

\subsubsection{The impact of the jet on the thermodynamics of the CND}
\label{Sec: CNDs}

\begin{figure*}
    \centering
    \includegraphics[width=\textwidth]{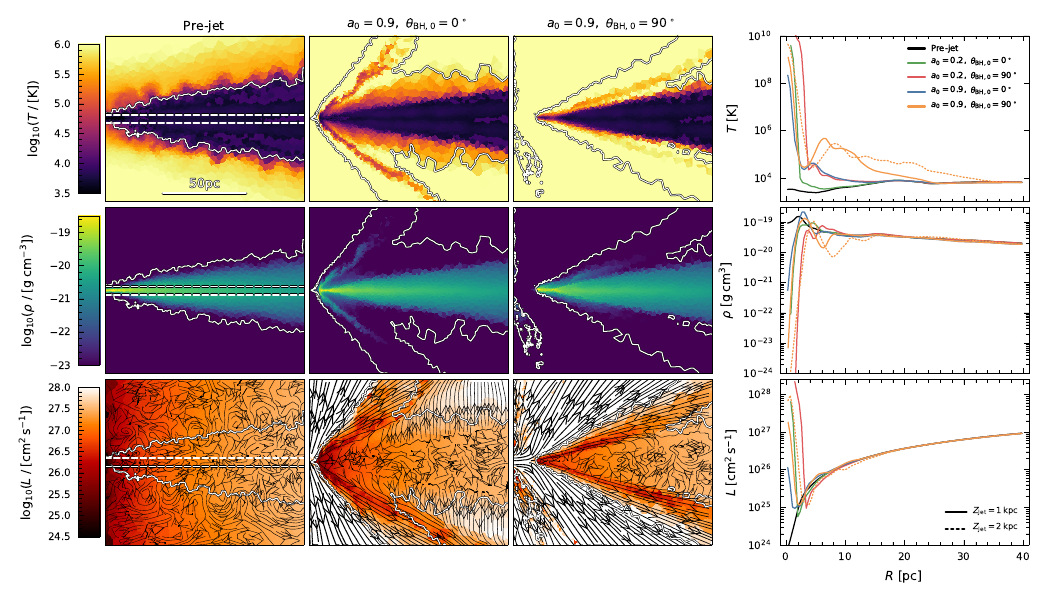}
    \caption{The first three columns show slices (with the angular momentum of the CND aligned with the $y$ axis and the centre of mass of the CND in the centre-left) of the temperature (first row), density (second row) and specific angular momentum (third row) for the `high' spin runs in the `thin-cooled' CND. Streamlines of the gas velocity are overlaid on the specific angular momentum slices. The first column corresponds to the state of the gas prior to jet launching, the second column shows the `vertical' jet case and the third column shows the `horizontal' jet case. The slices in the second and third columns were made using the simulation output in which the extent of the outflow in the $z$-direction is closest to $2$~kpc. The solid line in each slice shows a contour of a passive tracer that identifies CND material. The panels in the final column show cylindrical radial profiles of gas in the mid-plane of the CND. The mass-weighted temperature is shown in the top, the volume-weighted density in the middle and the mass-weighted specific angular momentum is shown in the bottom. The colour of the line indicates the initial jet configuration (see the legend in the upper left panel). All profiles correspond to the time when the $z$-extent of the outflow is closest to $1$~kpc. We additionally show the profile for the `high' power `horizontal' run at the time when its $z$-extent is closest to $2$~kpc as this corresponds to the case in which the CND is maximally affected. The initial CND profile is shown in black. In the first column of the slices, the region enclosed by the dotted line indicates the gas which was used to calculate the profiles.}
    \label{fig: cnd profiles slices thincooled}
\end{figure*}

Having considered how the gas in the immediate vicinity of the black hole responds to jet launching, we now examine the extent to which the jet is able to alter the CND gas on larger scales. Specifically, we explore the way in which the thermodynamics of different CNDs respond to the launching of the jet as well as how this response differs depending on the initial jet configuration. This is important to understand as changes to the structure of the CND will influence the long-term availability of gas for accretion which, in turn, will play a role in regulating the evolution of the jet properties on these longer time-scales. In this section and those that follow, we analyse simulations after $z$-extent of the jet-driven outflows have reached fixed lengths (calculated as described in Section~\ref{Sec: Qualitative exploration}), rather than at a fixed time. This allows us to compare the simulations on a more equal footing, as differences in the power and direction of the jets affect their propagation speeds.

The first three columns of Figs.~\ref{fig: cnd profiles slices thick}~and~\ref{fig: cnd profiles slices thincooled} show slices of the temperature (first row), density (second row) and specific angular momentum (third row) that focus on the CND gas (with the angular momentum of the CND aligned with the $y$ axis and the centre of mass of the CND in the centre-left). The fourth column shows cylindrical radial profiles of the mid-plane gas for each of these thermodynamic quantities. Fig.~\ref{fig: cnd profiles slices thick} shows simulations in which jets are launched from the `thick' CND, while Fig.~\ref{fig: cnd profiles slices thincooled} shows those in which jets are launched from the `thin-cooled' CND. 

Comparison of the pre-jet slices in Figs.~\ref{fig: cnd profiles slices thick}~and~\ref{fig: cnd profiles slices thincooled} indicates that, before the jet turns on, the `thin-cooled' CND has a much smaller scale-height (as discussed in Section~\ref{Sec: Circumnuclear discs}), lower temperature, is less pressure supported due to the enhanced cooling and also hosts a more complex velocity structure. The comparatively small scale-height of the `thin-cooled' CND means that the surface area for interaction with the jet is significantly smaller than in the `thick' CND case. Additionally, the higher densities in the `thin-cooled' disc mean that it is more resistant to disruption. 

The `high' power, `horizontal' jet has the greatest effect on the CND structure in both the `thick' and `thin-cooled' CND cases, whilst the `low' power, `vertical' jet has the least impact on the CND. Of the jets launched from the `thick' CND, those that propagate into the `dense' CGM have larger impacts on the CND thermodynamics compared to those launched into the `standard' CGM. To highlight this, in Fig.~\ref{fig: cnd profiles slices thick} we additionally show the CND mid-plane profiles in the case where the `high' power, `horizontal' jet is launched into the `dense' CGM. These profiles indicate that this jet is able to substantially affect the temperature, density and angular momentum of the gas throughout the entire radial extent of the CND\footnote{Fig.~\ref{fig: cnd profiles slices thick} only shows the CND profiles out to $40$~pc to highlight the main changes (note that the scale-length of the CND is approximately $70$~pc), but we have confirmed that the CND profiles are perturbed beyond this point.}. The CND thermodynamics undergo the most extreme changes in the `dense' CGM case because the enhanced resistance that this CGM provides means that less jet energy can be diverted vertically and so more is deposited in the vicinity of the CND. Additionally, significant fluid instabilities develop throughout the outflow in the `dense' CGM case (see Fig.~\ref{fig: slice mass flux}) and the flow structure close to the CND interface is much more complex. Whilst this does not necessarily affect the mid-plane profiles shown in Fig.~\ref{fig: cnd profiles slices thick}, it will, however, lead to considerable changes to the structure of the CND as a whole.

The launching of the jet drives shocks into the surrounding CND which are particularly visible as sharp features in the temperature and density maps in Fig.~\ref{fig: cnd profiles slices thick} as well as in the respective mid-plane profiles. Due to the cocoon geometry, shocks in the `thick' CND are not purely radial but, rather, interfere with structure in the disc and reflect of the surface layers. The higher initial (mid-plane) densities in the `thin-cooled' case mean, not only that this disc is less susceptible to disruption by the jet, but also that shock energy is dissipated faster meaning that the majority of the CND remains largely unaffected by jet-launching. Interestingly, the `thin-cooled' CND structure is not changed significantly even by the `high' power, `horizontal' jet. Indeed, this jet is unable to propagate to large distances in the CND and is, instead, rapidly redirected towards the galactic polar regions which offer the least resistance. 

Focusing on changes to the specific angular momentum of the gas in the `thick' CND, the profiles in Fig.~\ref{fig: cnd profiles slices thick} show that the launching of the jets raises the specific angular momentum of the gas in the central regions considerably, as was also highlighted in Section~\ref{Sec: Close to the black hole}. At radial distances of $5$-$10$~pc, however, the profiles indicate that jet launching is able to lower the specific angular momentum of the gas below that of the pre-jet profile by radially perturbing gas orbital motions. These changes to the gas orbits are qualitatively similar to those explored by \cite{2013DehnenKing} who, by means of analytical arguments, analysed such effects in the context of momentum-driven winds and found that, once feedback ceases, material launched on largely parabolic orbits can be re-accreted by the black hole. 

Turning now to the specific angular momentum of the `thin-cooled' CND, the profiles in Fig.~\ref{fig: cnd profiles slices thincooled} indicate that, as in the `thick' CND case, jet launching considerably raises the specific angular momentum in the central regions. These changes, however, are confined to the central $\sim 10$~pc and there is no significant lowering of the specific angular momentum of the gas beyond this point, contrary to what is observed in the `thick' CND case. The gradient of the specific angular momentum profile in the inner regions of the `thin-cooled' CND, however, is steeper than that found in the `thick' CND. This, along with the lack of considerable perturbations to the CND gas beyond $\sim 10$~pc, explains why inflows of low specific angular momentum material are able to resume on shorter timescales in the `low' power `vertical' jet launched from this `thin-cooled' CND, as was observed in Sections~\ref{Sec:Quantitative exploration}~and~\ref{Sec: Close to the black hole}. 

In all runs, the jet-driven outflows cause considerable changes to the gas close to the interface between the CND and the CGM. Indeed, the velocity streamlines in Figs.~\ref{fig: cnd profiles slices thick}~and~\ref{fig: cnd profiles slices thincooled} indicate that the flow structure and gas kinematics are substantially perturbed by jet launching. In the `thick' CND case, the `vertical' jet clearly drives a high velocity, bipolar outflow, while leaving the the wider disc kinematics largely unchanged. The small-scale jet is clearly identifiable in the `horizontal' jet case, as it advances laterally into the disc. The propagation of this jet, however, is rapidly halted by the CND which leads to the formation of a quasi-bipolar outflow that emanates from the stagnation point of the flow. Turning now to the `thin-cooled' CND case, the streamlines shown in Fig.~\ref{fig: cnd profiles slices thincooled} indicate that the flow structure in the `vertical' and `horizontal' jet cases are fairly similar, with neither jet able to significantly perturb the CND kinematics. 

One final feature of interest regarding the `high' power `vertical' jet launched from the `thin-cooled' CND is the vertical streamlines that emanate from the surface of the CND. These indicate that the fast jet is able efficiently to draw material vertically, unbinding the upper strata of the disc. This is more pronounced in the `thin-cooled' CND case than the `thick' CND case primarily due to the wider jet opening angles that the `thin-cooled' CND is able to accommodate, which allow for more interaction between the outflow and the surface of the disc. 

In summary, we find that the CND kinematics are perturbed the most by `high' power 'horizontal' jets when they are launched into our lowest density CND and CGM setup. Conversely, our highest density CND setups are able to redirect `horizontal' jets, leading to vertical, bipolar outflows, and kinematic perturbations to the CND that are largely confined to its innermost regions. We find that the jets are able to have considerable effects on the specific angular momentum profiles of the CND gas, but that efficient radiative gas cooling in the CND is the primary reason why gas with low specific angular momentum can reach the $\alpha$-disc.

\subsection{Outflow composition and interaction with the CGM}
\label{Sec: Outflow composition}
In this section, we explore the properties of the jet-driven outflows themselves and how they interact with and are affected by the surrounding CGM. To analyse the composition of the outflows we separate the gas into the cold phase ($T<2\times10^4\, {\rm K}$), the warm phase ($2\times10^4\,{\rm K}<T<5\times10^5\,{\rm K}$) and the hot phase ($T>5\times10^5{\rm K}$). Additionally, we explore the origin of the gas in the outflow using passive tracers that are placed in the CND material and injected with the jet. Since these jet and disc tracers are independent, any remaining mass in a cell can be unambiguously identified as CGM material. 

Before beginning this analysis we, again, caution the reader that these simulations are idealised and do not model the wider galaxy, nor any large-scale cosmological inflows. The interaction of the jet with the ISM and CGM in self-consistent cosmological environments may be more complex than what we find in our simulations. It should also be noted that we do not include radiative cooling of the gas in the outflows\footnote{Recall that the $\beta$-cooling prescription, which is used in a subset of our simulations, is solely applied to CND gas that has not been entrained by the outflow.}. This means cooling of outflow gas is purely adiabatic and that formation of cold gas in the outflow via thermal instabilities cannot occur.  In addition, one would expect that the inclusion of radiative cooling would lead to higher fractions of cold gas in the outflow. This could be due to direct condensation of the warm, outflowing gas and also due to the rapid cooling that would occur in the mixing layers that separate the outflowing gas and the hot, ambient medium \citep{2021Banda-Barragan+}. Ultimately this could affect the amount of gas that is entrained in the outflows as cold gas is more resistant to destruction.

\begin{figure*}
    \vspace{-0.3cm}
    \centering
    \includegraphics[width=0.95\textwidth]{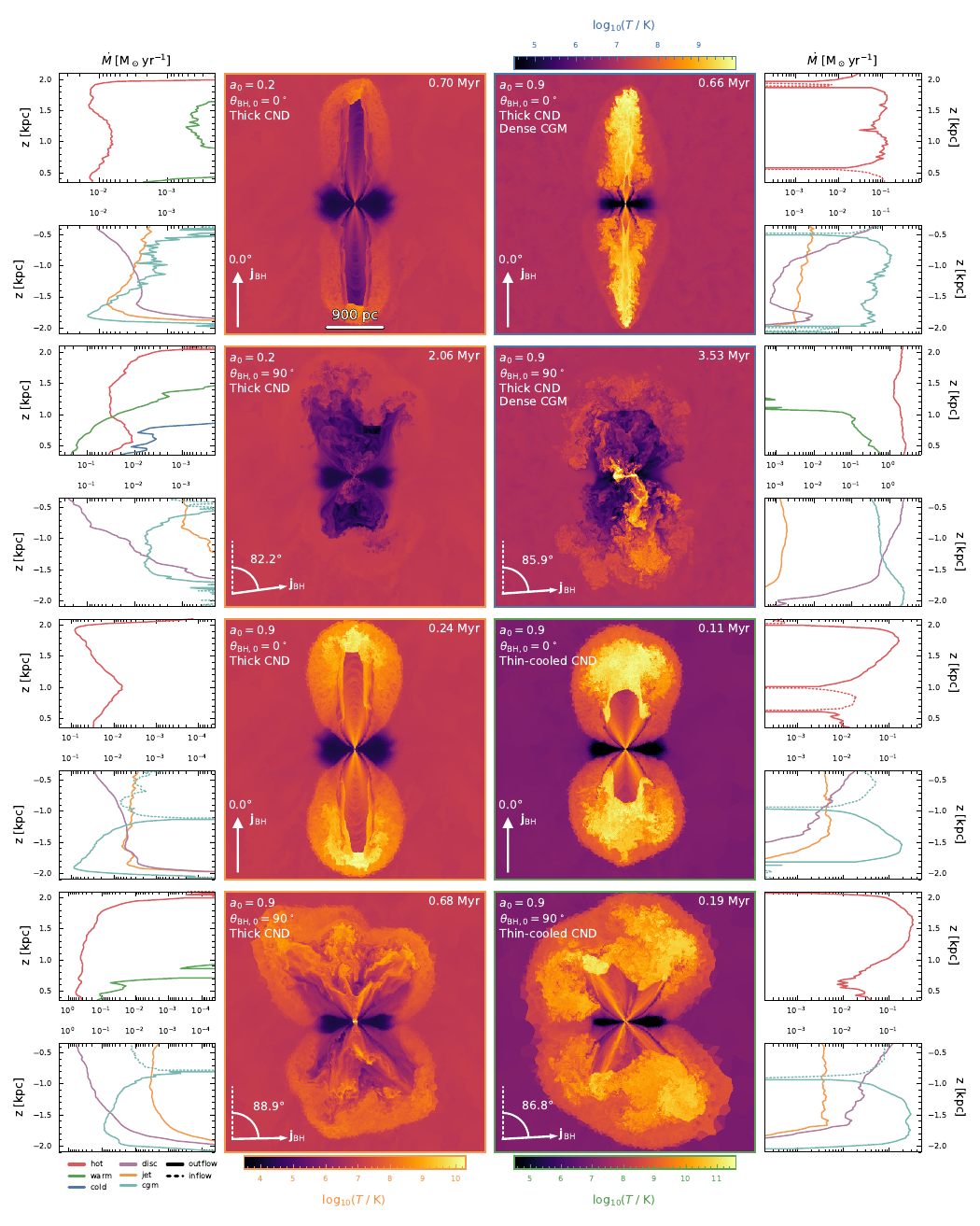}
    \vspace{-0.3cm}
    \caption{The central two columns show slices through the $x$-$z$ plane, centred on the black hole, for a selection of simulations at the time when the $z$-extent of the outflow is closest to $2$~kpc. The colour-scale in the bottom-left is used for jets launched from `thick' CND (left-hand column, indicated by orange borders). The colour-scale in top-right corresponds to the jet launched from the `thick' CND into the `dense' CGM (top two rows of the right-hand column, indicated by blue borders). Finally, the colour-scale in the bottom-right corresponds to jets launched from the `thin-cooled' CND (bottom two rows of the right-hand column, indicated by green borders). The arrow in the bottom left indicates the inclination of the jet relative to the vertical. Each slice has two smaller panels associated with it, which show corresponding vertical mass flux profiles. The $y$-axis of each of these flux plots is coincident with that of the corresponding slice. The mass flux profile in the upper profile is split into contributions from hot, warm and cold gas. The mass flux profile in the lower panel is split into contributions from CND gas, jet gas and CGM gas. A solid/dashed line indicates net outflow/inflow.}
    \label{fig: slice mass flux}
\end{figure*}

The main panels in Fig.~\ref{fig: slice mass flux} show slices through the $x$-$z$ plane, centred on the black hole, for a representative sample of our `high' resolution simulations, where each slice was made at the time when the $z$-extent of the outflow is closest to $2$~kpc. The left-hand column shows outflows launched by jets in the `thick' CND, the top two panels in the right-hand column show outflows launched by jets in the `thick' CND that propagate into the `dense' CGM while the bottom two panels show outflows launched from the `thin-cooled' CND. Each slice has two smaller panels associated with it that show vertical mass flux profile associated with the outflow. The $y$-axis of these flux profiles are {\it spatially coincident} with that of the slice. In the upper profile the mass flux is split into contributions from hot, warm and cold gas while in the lower panel, the flux is split into contributions from CND, jet\footnote{ Throughout this discussion, when we discuss `jet' material, we are referring to gas associated with the passive jet tracer.} and CGM gas, with a solid/dashed line indicating net outflow/inflow.

For a given $z$, the vertical mass flux is calculated via 
\begin{equation}
\label{eq: mdot out}
    \dot{M}_{\rm out}(z,t) = \sum_{|z_i-z| < {\rm d}z/2} {\rm sign}(z_i)\,\frac{m_i v_{z,i}}{{\rm d}z}\; ,
\end{equation}
where the sum is taken over all cells that satisfy $|z_i-z| < {\rm d}z/2$, ${\rm d}z$ is the width of the slab through which the fluxes are measured and $m_i$, $z_i$ and $v_{z,i}$ are the mass, z-coordinate and $z$-velocity of the gas cell relative to the black hole, respectively.

From Fig.~\ref{fig: slice mass flux} it is clear that the `high' power `vertical' jets are all completely dominated by hot gas. The one `low' power, `vertical' jet that is shown also has a significant hot component but additionally contains a non-negligible amount of warm gas. This warm component is largely associated with jet material and is not seen in the `high' power runs due to the stronger internal shocks in the jet channel which are more efficient at thermalising the jet material.

The outflows driven by `horizontal' jets, in comparison, all have a warm component, with the only exception being the `high' power jet in the `thin-cooled' CND whose outflow wholly consists of hot gas. To understand the lack of a warm component in this simulation recall that, for a given spin magnitude and direction, jets launched from the `thin-cooled' CND are more powerful due to the higher inflow rates before the jet turns on (see Section~\ref{Sec: Jet power}). Thus, the `high' power `horizontal' jet launched from the `thin-cooled' CND is sufficiently powerful that the shocks it drives rapidly thermalise a non-negligible fraction of the jet material as well as any entrained CGM and CND gas. In addition, the small scale-height of the `thin-cooled' CND as well as its enhanced resistance to lateral jet propagation together mean that these outflows undergo less interaction with the CND gas (see Section~\ref{Sec: CNDs}). The `high' power `horizontal' jet launched from the `thin-cooled' CND, therefore, has a comparatively low contribution from shock-heated CND material which, in other `horizontal' jet runs, can be significant. The fact that the contribution to the outflow mass flux from CND material is smaller when in cases where jets are launched from the `thin-cooled' CND compared to jets launched from the `thick' CND holds true both for `horizontal' and for `vertical' jet cases.

From the vertical flux profiles of the `horizontal' jet-driven outflows, it is apparent that the warm component, where it exists, is not present throughout the full vertical extent of the outflow but rather is found predominantly towards the base. This is due to the fact that mass fluxes in warm gas largely trace slower moving CND material that has been entrained and shock heated, although a smaller contribution can come from gas that has cooled via adiabatic expansion in the jet channel (the kinematics of this warm component is explored in Section~\ref{Sec: los velocities warm hot}). 

It is clear that these outflows are manifestly multiphase in nature and, additionally, show distinct signs of interaction between thermodynamically and kinematically distinct components (the outflow kinematics will be analysed in Section~\ref{Sec: los velocities}). For example, in the `high' spin case where the jet is launched from the `thick' CND (bottom-left panel) there is a clump of warm gas that is being accelerated by the hot outflow. This clump is sufficiently massive that it can also be identified in the warm gas flux profile.

The only outflow shown in Fig.~\ref{fig: slice mass flux} that hosts a non-negligible cold gas component is the `low' power `horizontal' jet that is launched into the `standard' CGM. Examination of the outflow composition in simulations that are not shown in Fig.~\ref{fig: slice mass flux} reveals that cold outflows are preferentially found when the jets have low powers (so that the they do not shock heat the gas appreciably), when they are highly inclined and when they are launched from the `thick' CND (so that they are able to interact significantly with the cold CND gas). In Section~\ref{Sec: los velocities cold}, we provide further analysis of these cold outflows and identify the black hole/CND configurations under which they arise.

Outflows driven by `horizontal' jets, in general, host higher mass fluxes when compared to the relevant `vertical' jet case. Looking at the mass flux contributions from disc, jet and CGM material, it is clear that these larger fluxes arise due to the greater contributions from CND material, which is more readily entrained in the `horizontal' jet cases. 

Whilst the majority of the jets drive a net flow of gas away from the black hole, the flux profiles indicate that net inflows of material can be present in individual components (primarily hot, CGM material). Such inflows can arise when outflowing material displaces gas in the CGM which then falls back towards the potential minimum. Inflows can also arise when there is a significant mass of backflowing material in the jet cocoon, as is the case in the `vertical' jet in the `dense' CGM case (top right) where there is a hot inflow close to the base of the jet and at the termination shock at the head of the jet axis. 

Since the morphological differences between jets launched into the `standard' and `dense' CGMs were examined in detail in \cite{2021Talbot+}, we will not revisit this here, except to briefly highlight the fact that the `dense' CGM runs all exhibit more vigorous instabilities. The efficient entrainment of CGM gas by these instabilities mean that both inflows and outflows in the `dense' CGM runs have comparatively higher contributions from CGM gas, although this is also due in part, simply, to the higher density of this CGM material.

It should also be noted that these two `dense' CGM runs both take longer to reach $2$~kpc than the analogous `high' power runs in the `standard' CGM. Again, this highlights the fact that these jets have to do more work to propagate and displace the denser material in the CGM. Indeed, the enhanced confinement provided by the `dense' CGM also contributes to the fact that the `horizontal' jet in the dense CGM has a greater disruptive effect on the CND structure, as was observed in Section~\ref{Sec: CNDs} and can very clearly be seen in the temperature slice of the `horizontal' jet in Fig.~\ref{fig: slice mass flux}. 

\subsection{Mass, momentum and energy loading of the outflows}
\label{Sec: Loadings}

The efficiencies by which mass, energy and momentum are communicated from the small-scale outflows that are launched in the vicinity of the black hole to those that are found on galactic scales are often quantified using so-called `loading factors'. These loading factors typically normalise the large-scale mass, energy and momentum fluxes by the black hole accretion rate, the AGN bolometric luminosity, $L_{\rm Bol}$, and the total radiative momentum output from the AGN, $L_{\rm Bol}/c$, respectively. 

Whilst such quantities are useful for comparing with observations (and indeed, we perform such analysis in Section~\ref{Sec: outflow obs comparison}), it is important to remember, however, that simulations impose numerical loadings when feedback is injected due to the fact that mass, momentum and energy are deposited into cells of finite mass, volume and often non-negligible thermal energy. These numerical loadings can play a significant role in the way feedback interacts with the wider environment as changes to the spatial or mass resolution of the injection volume will alter the morphology and evolution of the resulting outflow. Furthermore, it is worth stressing that the outflow properties will greatly depend on the local environment. For example, jets injected into the resolved, dense disc of the central galaxy will evolve very differently than if they were to be directly launched into the hot, dilute gas in the halo.

The use of loading factors to quantify the efficiency of mass, momentum and energy transfer requires the fluxes associated with the large-scale outflows to be normalised by quantities that correctly marginalise over these numerical effects. In this work we, therefore, choose to normalise by the fluxes at the base of the jet which we measure directly from the gas in the jet cylinder in our simulations. In this section, we calculate the spatial average of the loading factors, $\left\langle\eta_{\bullet}\right\rangle_{z}$, by vertically averaging the large-scale fluxes and, rather than using the instantaneous injection flux, we use time-averaged values, $\left\langle \bullet \right\rangle_{t}$, as the properties of the outflow will depend on the entire history of these injection fluxes, which are not constant in time.

Specifically, we define the mass loading factor at a given $z$ to be
\begin{equation}
\label{eq: mass loading factor}
    \eta_{\rm m}(z,t) = \frac{\dot{M}_{\rm out}(z,t)}{\left\langle\dot{M}_{\rm load}\right\rangle_{t}} \; ,
\end{equation}
where $\dot{M}_{\rm out}(z)$ is the mass flux given by equation~(\ref{eq: mdot out}) and $\left\langle\dot{M}_{\rm load}\right\rangle_{t}$ is time average of the mass loading rate at the base of the jet
\begin{equation}
\label{eq: mass loading rate}
    \dot{M}_{\rm load}(t) = \sum_{\rm cyl \,(north)} \frac{m_i\, \boldsymbol{v}_{i}\cdot\boldsymbol{j}_{\rm BH}}{h_{\rm J}}\; .
\end{equation}
Here, the sum is taken over all cells in the northern half of the jet cylinder, $\boldsymbol{v}_{i}\cdot\boldsymbol{j}_{\rm BH}$ is the velocity of the cell in the direction of the jet axis and $h_{\rm J}$ is the height of one half of the jet cylinder.

Similarly, we define the momentum loading factor
\begin{equation}
\label{eq: momentum loading factor}
    \eta_{\rm p}(z,t) = \frac{\dot{P}_{\rm out}(z,t)}{\left\langle\dot{P}_{\rm load}\right\rangle_{t}} \; ,
\end{equation}
where
\begin{equation}
\label{eq: pdot out}
    \dot{P}_{\rm out}(z,t) = \sum_{|z_i-z| < {\rm d}z/2} {\rm sign}(z_i)\frac{m_i v_{i} v_{z,i}}{{\rm d}z}\; , 
\end{equation}
and
\begin{equation}
\label{eq: momentum loading rate}
    \dot{P}_{\rm load}(t) = \sum_{\rm cyl \,(north)} \frac{m_i\, v_i \, \boldsymbol{v}_{i}\cdot\boldsymbol{j}_{\rm BH}}{h_{\rm J}}\; .
\end{equation}
Finally, the energy loading factor is
\begin{equation}
    \eta_{\rm e}(z,t) = \frac{\dot{E}_{\rm out}(z,t)}{\left\langle\dot{E}_{\rm load}\right\rangle_{t}} \; .
\end{equation}
where the energy flux is 
\begin{equation}
    \dot{E}_{\rm out}(z,t) = \sum_{|z_i-z| < {\rm d}z/2} {\rm sign}(z_i)\frac{e_i v_{z,i}}{{\rm d}z}\; ,
\end{equation}
with $e_i$ being the sum of the total kinetic and thermal energy of the cell and 
\begin{equation}
    \dot{E}_{\rm load}(t) = \sum_{\rm cyl \,(north)} \frac{e_i \, \boldsymbol{v}_{i}\cdot\boldsymbol{j}_{\rm BH}}{h_{\rm J}}\; .
\end{equation}

We calculated these vertically averaged mass, energy and momentum loading factors for the simulations shown in Fig.~\ref{fig: slice mass flux} and they are detailed in Table~\ref{tab: loadings}.

From inspection of the mass loading factors, it is clear that they largely follow the same trends as the absolute mass fluxes in Fig.~\ref{fig: slice mass flux}. In all cases, the outflows driven by the `horizontal' jets have higher mass loading factors than those driven by `vertical' jets. The difference between the mass loading factors in the `horizontal' and `vertical' jets launched from the `thin-cooled' CND is smaller, however, which is largely due to the fact that the `horizontal' jet undergoes less interaction with the CND relative to `horizontal' jets in the `thick' CND (as discussed in Section~\ref{Sec: CNDs}). 

The highest mass loading factor is found in the case of the `horizontal' jet launched into the `dense' CGM. As explored in Section~\ref{Sec: Outflow composition}, this jet-driven outflow entrains a significant mass of gas from the CND and vigorous instabilities are efficient at entraining the dense CGM gas in the outflow. 

The fact that all simulations have average mass loading factors greater than unity indicates that significant quantities of gas are {\it genuinely} entrained as the outflows propagate away from the black hole, irrespective of our resolution and a choice of jet cylinder mass.

The momentum loading factors all largely follow similar trends to those seen in the mass loading factors. It is apparent, however, that the differences between the `vertical' and `horizontal' jet-driven outflows are smaller which is due to the lower velocities found in the latter case. We will further discuss the outflow velocities in Section~\ref{Sec: los velocities}, however, the fact that the `horizontal' jet-driven outflows are slower can be seen simply by considering the time it takes for them to reach $2$~kpc (shown in the top-right of each slice in Fig.~\ref{fig: slice mass flux}).

Whilst the energy loadings are fairly similar across all runs, that of the `low' power `horizontal' jet is considerably lower. This is due to the fact that it is much more difficult for this jet to clear a channel for the outflow to escape. Additionally, the fact that this jet has lower power means that a comparatively larger amount of the jet energy will be used to lift CND material out of the gravitational potential. Such an effect is not seen in the energy loadings of the low power vertical jet due to the comparatively small opening angle of the outflow and the lower mass of CND material that it entrains. All energy loading factors are of order unity, as expected. Some simulations exhibit slightly higher values, which are largely due to fast moving, entrained material.

Whilst the partitioning of the energy loading factors into kinetic and thermal components is not shown in Table~\ref{tab: loadings}, we find that all outflows are kinetically dominated, with the only exception being the `horizontal' jet launched into the `dense' CGM which is thermally dominated. This is likely due to the fact that this outflow takes a longer time to reach $2$~kpc and, as we showed in \cite{2021Talbot+}, the energy of jet-driven outflows tends to thermalise at later times. We would, therefore, expect recent jet-driven outflows to be kinetically-dominated, especially closer to the base of the jet, while older outflows propagating to greater distances may transition to the thermally-dominated regime. The details of this transition will crucially depend on the jet power as well as the pressure confinement of the surrounding medium, as shown in \cite{2021Talbot+}.

\begin{table}
    \caption{Mass, momentum and energy loading factors for the `high' resolution simulations shown in Fig.~\ref{fig: slice mass flux} at the same time as the slices and mass fluxes were made (i.e. when the $z$-extent of the outflow is closest to $2$~kpc). The first column gives the name of the run using the labels described in Table~\ref{tab: all runs}. Unless stated, the jet is launched into the `thick' CND and `standard' CGM. The second, third and fourth column list the spatially averaged mass, momentum and energy loading factors, respectively. See Section~\ref{Sec: Loadings} for a full explanation of how these values were calculated.}
    \label{tab: loadings}
    \begin{tabular}{lccc}
        \hline 
        \hline
        \multicolumn{1}{c}{\small Run} & $\left\langle\eta_{\rm m}\right\rangle_{z}$ & $\left\langle\eta_{\rm p}\right\rangle_{z}$ & $\left\langle\eta_{\rm e}\right\rangle_{z}$ \\
        \hline
        \hline
        {\small \hspace{-0.3cm} `Low' power `vertical'} & $3.88$ & $3.14$ & $3.37$ \\
        {\small \hspace{-0.3cm} `Low' power `horizontal'} & $53.91$ & $8.08$ & $0.96$ \\
        {\small \hspace{-0.3cm} `High' power `vertical'} & $4.60$ & $3.21$ & $3.57$\\
        {\small \hspace{-0.3cm} `High' power `horizontal'} & $78.61$ & $16.12$ & $3.56$\\
        \hline 
        {\small \hspace{-0.3cm} `Dense' CGM, `high' power `vertical'} & $19.06$ & $2.29$ & $1.89$\\
        {\small \hspace{-0.3cm} `Dense' CGM, `high' power `horizontal'} & $1132.09$ & $33.65$ & $2.02$\\
        \hline
        {\small \hspace{-0.3cm} `Thin-cooled' CND, `high' power `vertical'} & $0.41$ & $0.72$ & $1.21$ \\
        {\small \hspace{-0.3cm} `Thin-cooled' CND, `high' power `horizontal' \hspace{-0.8cm}} & $0.81$ & $1.10$ & $1.07$  \\   
        \hline 
        \hline 
    \end{tabular}
\end{table}

\subsection{Mass outflow rates and outflow powers}
\label{Sec: outflow obs comparison}

\begin{figure*}
    \centering
    \includegraphics[width=\textwidth]{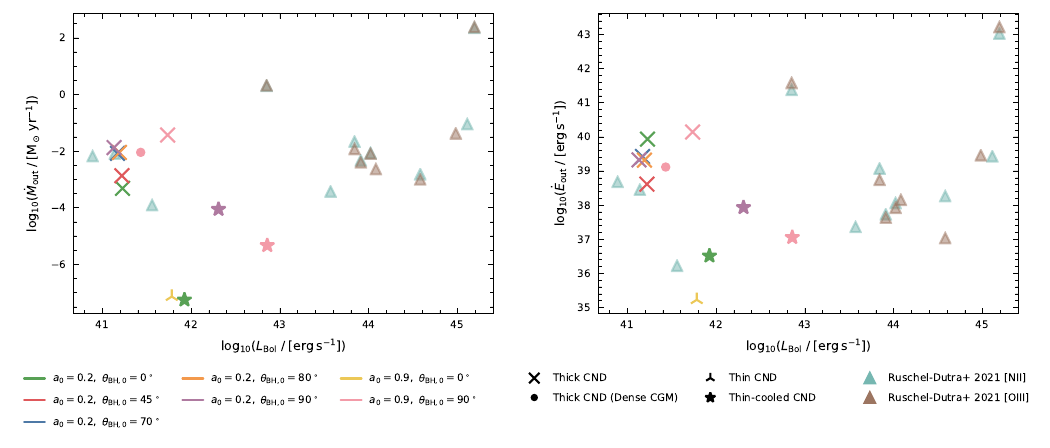}
    \vspace{-0.5cm}
    \caption{Mass outflow rate (left) and outflow kinetic power (right) of the warm gas, plotted as a function of bolometric luminosity for all `high' resolution simulations that host warm outflows. The colour of the marker indicates the initial black hole configuration, while the shape of the marker indicates the CND from which it was launched. Unless stated, the jet is launched into the `standard' CGM. We additionally plot outflow rates and kinetic powers derived from a sample of Seyferts and LINERS \citep{2021Ruschel-Dutra}. These points are indicated by triangular markers with turquoise points corresponding to values based on the [\ion{N}{ii}] line and brown points corresponding to values that use the [\ion{O}{iii}] line. On the whole, our mass outflow rates and kinetic powers are consistent with the observations and show a relatively large spread depending on the initial black hole properties and those of the surrounding environment.}
    \label{fig: lbol out}
\end{figure*}

Numerical simulations can provide vital information that can be used to constrain properties of galactic-scale outflows such as the distribution of mass and energy between the different phases. Observations of AGN-driven outflows that examine the mass outflow rates and energetics are available for a wide variety of systems \citep[see e.g.][]{2021Ruschel-Dutra,2021Fluetsch+,2020Santoro+,2020Davies+,2019BaronNetzer,2019Shimizu+,2017Fiore+}, including some in which jets are believed to be responsible for driving the outflow \citep{2017Osterloo+, 2015Morganti+}. 

In this section we compare mass outflow rates and kinetic powers of the outflows, measured directly from our simulations, to those found in a sample of Seyferts and LINERS that have bolometric luminosities and black hole masses comparable to ours \citep{2021Ruschel-Dutra}. 

\cite{2021Ruschel-Dutra} estimate the ionised gas mass outflow rates and kinetic powers using velocities derived from the kinematics of both the [\ion{O}{iii}] and the [\ion{N}{ii}] emission lines. The mass of outflowing, ionised gas is calculated using the fraction of the H$\alpha$ luminosity that corresponds to velocities above $600 \, {\rm km\,s^{-1}}$ from the line centre and using electron densities estimated from the [\ion{S}{ii}] $\lambda6716\,/\,\lambda6731$ diagnostic line ratio. 

For each of our `high' resolution simulations, we select the time when the $z$-extent of the outflow reaches $2$~kpc and consider only warm gas ($2\times10^4\,{\rm K} < T < 5\times10^5\,{\rm K}$) with radial velocities greater than $600\,{\rm km\, s^{-1}}$, in order that our calculations are consistent with \cite{2021Ruschel-Dutra}. From Fig.~\ref{fig: slice mass flux}, it is apparent that not all of the outflows in our simulations contain warm gas. Any of our simulations that do not host this warm, fast component are not included in our analysis. For each simulation, we calculate the radial mass outflow rate and the kinetic power of the outflow through $70$ spherical shells with radii in the range $0.5\,{\rm kpc} < r < 2\,{\rm kpc}$, where the lower spatial limit ensures that our calculations do not include fast, rotationally supported gas in the CND. For each simulation we take radial averages of the mass outflow rate and the kinetic power and present them, alongside those of \cite{2021Ruschel-Dutra}, in Fig.~\ref{fig: lbol out} as a function of the bolometric luminosity of the system.

At this time, our systems have warm gas outflow rates in the range $10^{-7}-10^{-1} \, {\rm M_\odot \, yr^{-1}}$, kinetic powers in the range $10^{35}-10^{41} \,{\rm erg \, s^{-1}}$ and bolometric luminosities in the range $10^{41}-10^{43} \,{\rm erg \, s^{-1}}$. From Fig.~\ref{fig: lbol out} it is clear that these values are broadly consistent with the observations. It is also apparent, however, that changes to the initial black hole properties and those of the surrounding environment can lead to a fairly large spread in measured outflow rate and kinetic power, which we also expect to be time dependent.

For a given black hole spin and CND structure, `vertical' jets drive the least massive outflows. As the inclination of the jet increases, so does the warm gas mass flux, as the jet is able to interact to a greater degree with the CND. With more recent observations, such as those of \cite{2021Ruschel-Dutra}, able to detect misalignments between the AGN ionisation axis and the wider galaxy, it will be interesting to see if these correlations between mass outflow rate and jet direction are also found in observed systems. 

For a given black hole configuration, the warm outflows launched from the `thin-cooled' CND are less massive that those launched from the `thick' CND. Again, this is due to differences in the CND geometry wherein the smaller scale-height of the `thin-cooled' CND means that jets launched from this disc are less efficient at entraining CND material. Additionally, jets in the `thin-cooled' disc are more powerful and are, therefore, able to shock heat gas to higher temperatures, meaning that the outflows they drive are dominated by hot gas (see the discussion in Section~\ref{Sec: Outflow composition}).

Differences between the outflow kinetic powers found in our simulations follow similar trends to those of the mass outflow rates. It is apparent, however, that `vertical' jets, where they are able to drive warm outflows, have comparatively high kinetic powers, as can be clearly seen in the case of the `low' power jet in the `thick' CND. This is due to the fact that, while the mass content of the outflows driven by these `vertical' jets is relatively low (see Fig.~\ref{fig: slice mass flux}), the outflowing gas in `vertical' jets has higher velocities than those in which the jet is misaligned. 

Interestingly, the outflows launched from the `thin-cooled' CND do not have the highest kinetic powers, despite having the highest bolometric luminosities. This is due to the fact that these outflows are largely dominated by hot gas.

Whilst we find good agreement between our simulations and the systems presented in \cite{2021Ruschel-Dutra} in terms of the mass outflow rates and kinetic powers, it is worth bearing in mind some of the caveats regarding our simulation methodology. The lack of radiative cooling of the outflowing gas in our simulations could affect the distribution of mass between the different phases. In addition, our idealised simulations of the central regions of a galaxy will not capture any interactions between the outflow and the wider galactic environment. Such interactions may lead to further entrainment of gas and this, along with other processes such as shock-heating and compression of the outflow as it collides with pre-existing gas structures, could lead to changes in the phase-distribution of the gas. Ultimately, these could both affect the warm gas mass outflow rates and kinetic powers measured in our simulations.

For our AGN-driven outflows, we find that the jets are able to kinematically perturb\footnote{Here we define `kinematically perturbed' gas to be that which has radial velocity $v_r > 100 \, {\rm km \, s^{-1}}$, which is an upper limit to the radial velocities present in the warm gas before the jets were launched.} $0.01-10\%$ of the ionised gas mass in the central $2$~kpc, the majority of which comes from the CND (which has an initial mass of $10^8\,{\rm M_\odot}$). Considering the fact that both the jet powers and the Eddington ratios in these simulations are relatively moderate, the kinematic perturbations from these jets are significant. The resulting masses of ionised outflowing gas lie in the range $10^4 - 10^7 \, {\rm M}_\odot$ which are similar to those measured in \citep{2021Venturi+} and towards the lower end of those measured in \cite{2021Ruschel-Dutra}. The densities found in our jet-driven outflows typically do not exceed $n_{\rm H} \sim 10 \, {\rm cm^{-3}}$, which is likely lower than the highest densities found in the outflows of \cite{2021Ruschel-Dutra}. These differences in density could be due to the lack of radiative cooling of the outflows in our simulations and also the fact that we do not model the wider cosmological environment, as mentioned previously. In order to shed further light on the origin of dense gas observed in AGN-driven outflows these effects need to be included in future work.

As highlighted in Section~\ref{Sec: Loadings}, AGN-driven outflows in observational studies are often quantified by normalising the outward fluxes of mass, momentum and kinetic energy by the black hole accretion rate, the total radiative momentum output from the AGN and the AGN bolometric luminosity, respectively. 

For the simulations shown in Fig.~\ref{fig: lbol out}, we calculate these quantities for the warm and hot component together, considering only gas that is outflowing (i.e. $v_r>100\, {\rm km \,s^{-1}}$) and taking radial averages between $0.5$ and $2$~kpc of the spherical fluxes, $\left\langle\bullet\right\rangle_r$, and normalising by the {\it instantaneous} black hole accretion rate, $\dot{M}_{\rm BH,0}$, radiative momentum output, $L_{\rm Bol}/c$ and bolometric luminosity $L_{\rm Bol}$. 

For all of our AGN jet-driven outflows that are shown in Fig.~\ref{fig: lbol out}, $\left\langle\dot{M}_{\rm out}\right\rangle_r \, /\,  \dot{M}_{\rm BH,0}$ lies in the range $3\times 10^2 - 3\times 10^5$. The higher values are found exclusively in the `dense' CGM runs, while others typically show values of $\sim1000$. The values of $\left\langle\dot{P}_{\rm out}\right\rangle_r \, / \, (L_{\rm Bol}/c)$ lie in the range $15 - 1500$ where, again, the highest values\footnote{Note that including radiative cooling in the outflows may decrease the amount of $P{\rm d}V$ work done by the expanding gas and hence could lead to lower values.} are found in the `dense' CGM runs and most show values of $\sim100$. The kinetic powers, on the other hand, are more modest and all simulations have values of $\left\langle\dot{E}_{\rm kin, out}\right\rangle_r \, / \, L_{\rm Bol}$ that lie in the range $0.05 - 3.60$. For the range of values predicted by AGN wind models, see \cite{2020Costa+}.

In this section, we have made clear the differences between the way we calculated $\left\langle\dot{M}_{\rm out}\right\rangle_r \, /\,  \dot{M}_{\rm BH,0}$, $\left\langle\dot{P}_{\rm out}\right\rangle_r \, / \, (L_{\rm Bol}/c)$ and $\left\langle\dot{E}_{\rm kin, out}\right\rangle_r \, / \, L_{\rm Bol}$ and the way we calculated the loading factors in Section~\ref{Sec: Loadings}. Indeed, these lead to very different quantities that serve different purposes: normalising fluxes by the instantaneous properties of the AGN allow simulated outflows to be directly compared with observations, whereas normalising by fluxes at the launch-point of the outflow allows us to explore the physical processes influencing mass, energy and momentum transfer in the outflows.

Moreover, we find that the mass and momentum loadings calculated by normalising fluxes by the instantaneous values of $\dot{M}_{\rm BH,0}$ and $L_{\rm Bol}/c$ can be orders of magnitude higher than when normalised by fluxes calculated directly in the simulation. 

Altogether, this reinforces the point we made in Section~\ref{Sec: Loadings} that using loading factors to make inferences about the efficiency by which mass, energy and momentum is communicated from small to large scales requires the loadings factors to be defined using physically meaningful quantities.

\subsection{Line-of-sight velocity maps}
\label{Sec: los velocities}

Some observations of AGN jets clearly show large-scale outflows of gas with properties that correlate with the radio emission from the jet. Spatially resolved kinematics of these outflowing components are now available for a wide range of such sources, both for the ionised gas \citep{2019Jarvis+,2017Vayner+,2015Cresci+, 2015Harrison+,2008Nevsbada+} and also for molecular gas \citep{2013Combes+,2014GarciaBurillo+,2014Tadhunter+,2017Osterloo+,2015Morganti+}. 

In this section we use line-of-sight (LoS) velocity maps to explore the kinematics of the cold, warm and hot gas phases that are present in our jet-driven outflows.

\subsubsection{Warm and hot phase LoS velocities}
\label{Sec: los velocities warm hot}

\begin{figure*}
    \vspace{-0.3cm}
    \centering
    \includegraphics[width=0.85\textwidth]{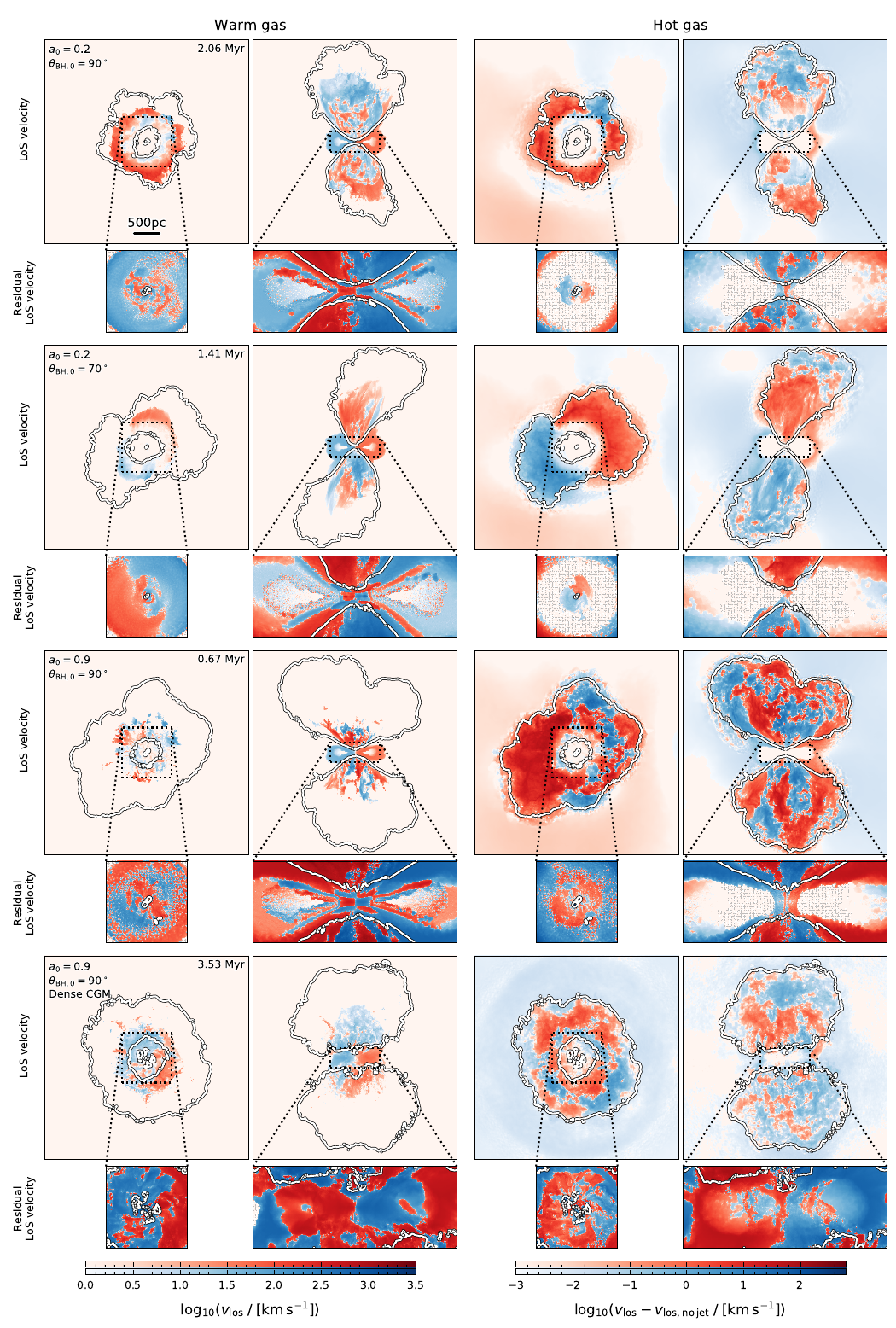}
    \vspace{-0.28cm}
    \caption{Each row shows line-of-sight velocity maps for a simulation in which a jet is launched from the `thick' CND. The relevant properties that identify each run are shown in the top-left of each panel. The first and second columns show the velocities in the warm ($2\times10^4\,{\rm K}<T<5\times10^5\,{\rm K}$) phase and the third and fourth columns show the hot ($T>5\times10^5{\rm K}$) phase, with projections down the $z$/$y$-axis shown on the left/right. Each projection corresponds to the time when the $z$-extent of the outflow is closest to $2$~kpc. The inset plots show residual line-of-sight velocity maps, relative to simulations without a jet. The white contours identify jet material. The colourscale used in the main panels is shown on the bottom-left and that used in the inset panels is shown on the bottom-right.}
    \label{fig: los velocity hot}
\end{figure*}

\begin{figure*}
    \centering
    \includegraphics[width=0.8\textwidth]{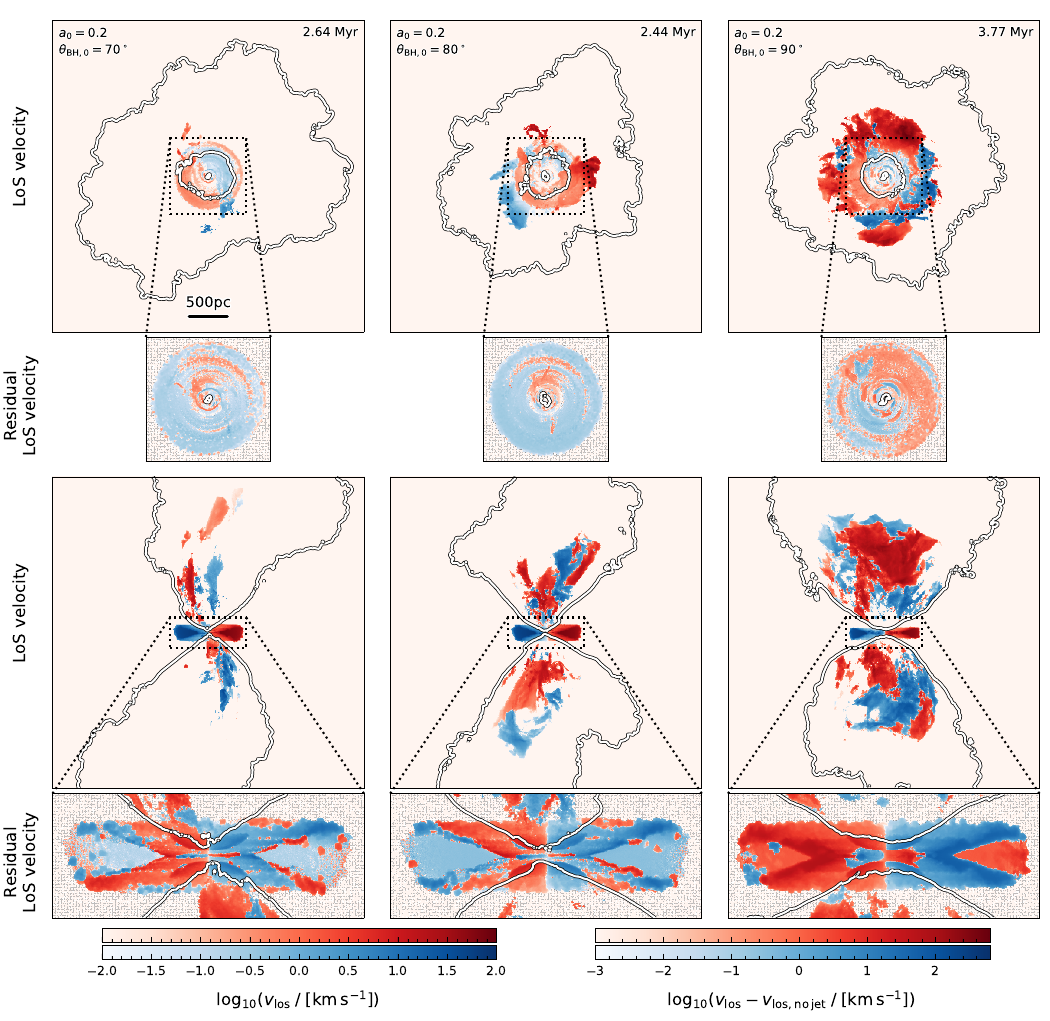}
    \caption{Each column shows line-of-sight velocity maps of the cold ($T<2\times10^4\,{\rm K}$) gas phase for a simulation in which a `low' power jet that is highly inclined ($\theta_{\rm BH,0} = 70^\circ, 80^\circ, 90^\circ$ from left to right) is launched from the `thick' CND. Each projection corresponds to the time when the $z$-extent of the outflow is closest to $3$~kpc. The first row shows projections down the $z$-axis while the second row shows projections down the $y$-axis. The inset plots show residual line-of-sight velocity maps, relative to simulations without a jet. The white countours identify jet material. The colourscale used in the main panels is shown on the bottom-left and that used in the inset panels is shown on the bottom-right.}
    \label{fig: los velocity cold}
\end{figure*}

We begin by examining the kinematics of the hot ($T>5\times10^5 \, {\rm K}$) and warm ($2\times10^4 \, {\rm K}<T<5\times10^5\,{\rm K}$) components of our jet-driven outflows.

Each row in Fig.~\ref{fig: los velocity hot} shows LoS velocity maps of both the warm and hot gas phases for an outflow driven by a significantly misaligned jet that has been launched from the `thick' CND. Each projection corresponds to the time when the $z$-extent of the outflow (as determined by the jet tracer) is closest to $2$~kpc. The inset plots associated with each panel show residual LoS velocity maps, relative to simulations without a jet, through the central $1\times1\times0.4\;{\rm kpc}$ and $1\times0.4\times1\;{\rm kpc}$ for the projections down the $z$ and $y$ axes, respectively. We should highlight the fact that, whilst the volume through which the velocities are projected in these smaller maps largely contains CND material, they can also contain material associated with the outflow and thus do not unambiguously indicate changes to the {\it CND} kinematics. Such an example can be seen in the projections down the $z$ axis in `low' power $70^\circ$ jet (second row), where the residual LoS velocity plot is clearly picking up the high velocity, bipolar outflow.

In all cases, the hot gas in the outflow is largely volume-filling whereas the warm gas does not extend to the outer edges of the jet dominated regions (which are outlined by the white contours in Fig.~\ref{fig: los velocity hot}). This is consistent with the fact that the velocities in the warm outflow are typically lower than those in the hot component. The lowest velocities in all of the simulations shown in Fig.~\ref{fig: los velocity hot} are found in the `high' spin, `horizontal' jet run in the `dense' CGM (fourth row), where the projected velocities do not exceed $140\;{\rm km\,s^{-1}}$ and $220\;{\rm km\,s^{-1}}$ in the warm and hot phases, respectively. The highest velocities are found in the `high' spin, `horizontal' jet run in the `standard' CGM (third row), which has projected velocities that reach $410\;{\rm km\,s^{-1}}$ and $1380\;{\rm km\,s^{-1}}$ in the warm and hot phases, respectively. It is worth highlighting the fact that the initial conditions for these two simulations differ only in the density of the CGM. The disparity between the velocities in these two runs can be attributed to the fact that in the `dense' CGM case, the outflow is trapped by the higher CGM column densities, meaning that a large amount of the jet energy goes into perturbing the CND rather than escaping vertically (consistent with conclusions made in Sections~\ref{Sec: CNDs}~and~\ref{Sec: Loadings}).

Across these simulations we find a wide range of outflow velocities that encompass those found in observational studies such as \cite{2021Venturi+,2021Speranza+,2019Jarvis+} and in simulations such as those of misaligned jets in \cite{2018MMukherjee+} and of clusters in \cite{2020Ehlert+}. 

In general, the hot gas velocities in the $z$-direction are larger than those in the $y$-direction, which is consistent with the quasi-bipolar nature of these outflows. This is also true of the warm gas velocities in all runs except the `low' power $70^\circ$ jet (second row), although these high velocities in the $y$-direction are likely tracing rotational motion in the disc. This `low' power $70^\circ$ jet run has largely coherent velocities in the majority of the outflow, with signs of turbulence in the outer regions as the jet interacts with the CGM. In all other cases, the velocity structure is more complex with both the warm and cold phases of the outflows hosting multiple components with different kinematics.

From Fig.~\ref{fig: los velocity hot}, it is clear that the velocity structures of our jet-driven outflows are particularly sensitive to the initial jet configuration as well as the surrounding environment. Additionally, we see that the power of the jet can influence the phase distribution of the gas in the outflows. The `low' power `horizontal' jet (first row) drives an outflow that has a more extended warm phase than the analogous `high' power run (third row). In this case, the lower amount of warm gas in the `high' power run is due to the stronger shocks that it is able to drive.

Focusing now on the kinematics of the CND, from the main panels it is clear that the central regions of the CNDs largely contain cold gas, with warm and hot gas found in the outer regions. The inset plots indicate that the kinematics of both the warm and the hot phase can be significantly altered by the launching of the jets. In all cases there is evidence for lateral shocks being driven in the central regions, although the effect of the jet is smaller both at lower powers and when the jet is less inclined. The largest perturbations to the CND are found in the `high' spin, `horizontal' jet run in the `dense' CGM (fourth row), which is consistent with our discussion of the CND profiles in Fig.~\ref{fig: cnd profiles slices thick}.

\subsubsection{Cold phase LoS velocities}
\label{Sec: los velocities cold}
We now turn to the kinematics of the cold gas component of the AGN jet-driven outflows in our simulations. Whilst many of the jet configurations explored in this work do not drive cold outflows beyond a few scale-heights of the CND (see the discussion in Section~\ref{Sec: Outflow composition}), jets with low power that are misaligned to such a degree that they interact with the CND can drive extended cold outflows (see Section~\ref{Sec: Outflow composition}).

Each of the columns in Fig.~\ref{fig: los velocity cold} shows LoS velocity maps of the cold ($T<2\times10^4\,{\rm K}$) gas phase for a simulation in which a `low' power jet is launched from the `thick' CND. All simulations correspond to highly inclined jet configurations ($\theta_{\rm BH,0} = 70^\circ, 80^\circ, 90^\circ$ from left to right, respectively). In this analysis we explore the kinematics of the cold gas when the $z$-extent of the outflow (as determined by the jet tracer) is closest to $3$~kpc rather than when it is closest to $2$~kpc, as was used in the analysis of the hot and warm gas kinematics in Section~\ref{Sec: los velocities warm hot}. We do so as interesting features that differentiate between the properties of the cold outflows driven by jets with different configurations take longer to become apparent due to the fact that the cold gas velocities are lower than those of the warm and hot components (as we will shortly see). The inset plots associated with each of these maps show the residual LoS velocity map, relative to an analogous simulation without a jet and, again, correspond to the central $1\times1\times0.4\;{\rm kpc}$ and $1\times0.4\times1\;{\rm kpc}$ for the projections down the $z$ and $y$ axes, respectively.

The maximum LoS velocities range from $\sim70\;{\rm km\,s^{-1}}$ in the $70^\circ$ case to $\sim90\;{\rm km\,s^{-1}}$ in the $90^\circ$ case which are consistent with those found by \cite{2014GarciaBurillo+} and \cite{2019Alonso-Herrero+}. They are, however, lower than velocities found by \cite{2017Osterloo+, 2015Morganti+} in the cold molecular phase (up to $800\;{\rm km\,s^{-1}}$) and by \cite{2014Tadhunter+} in the warm molecular phase (up to $600\;{\rm km\,s^{-1}}$) for the jet-driven molecular outflow in IC~$5036$, although the black hole in this system is more than an order of magnitude more massive than those considered in this work. 

The masses of cold gas in these outflows\footnote{Here we define cold outflowing gas to be that which has radial velocity $v_r > 50 \, {\rm km \, s^{-1}}$, which is an upper limit to the radial velocities present in the cold gas before the jets were launched.} lie in the range $4\times10^4-2\times10^5 \, {\rm M_\odot}$, with the highest cold gas mass found in the outflow driven by `horizontal' jet and the lowest found in the outflow driven by the $70^\circ$ jet. The cold gas mass in these outflows, however, is subdominant relative to the warm and hot components which have masses that are typically a factor of $10-100$ larger.

The fact that the velocities found in the cold outflow are lower than those in the warm and hot phases is consistent with the picture that this cold outflowing gas originates from the CND and has either been launched from the CND by direct interaction with the jet or has been lifted from the CND surface by jet-induced CGM motions (for example, see the streamlines in Fig.~\ref{fig: cnd profiles slices thincooled} and the accompanying discussion in Section~\ref{Sec: CNDs}).

The relatively low velocities that we find in these cold outflows highlights the difficulty in using velocity cuts to accurately differentiate between cold gas that is outflowing and that which is associated with the rotationally supported CND. This is particularly true for low luminosity systems such as those considered in this work and, indeed, may also be the case in some observational works that show evidence of disturbed kinematics in the cold phase, but are unable to unambiguously associate these motions with molecular outflows \citep[see e.g.][]{2021Runnoe+, 2021BewketuBelete+}. 

In all three simulations shown in Fig.~\ref{fig: los velocity cold}, the cold gas is largely found towards the base of the outflow. As the inclination of the jet increases, the volume filled by the cold gas increases due to the stronger interactions between the jet and the CND. This is particularly clear in the edge-on residual LoS velocity maps of the CND region, where the largest and most wide-spread perturbations to the disc kinematics are found in the $90^\circ$ jet case.

In general, the velocity structure of these cold outflows is not consistent with a that of a simple bipolar outflow. Instead, there is clear evidence that these outflows consist of multiple components with different kinematics. In the $70^\circ$ inclination run, however, the arcs of material with coherent velocities are indicative of cold gas entrainment from the CND by the warm/hot outflow, rather that the expulsion of CND gas seen in the $90^\circ$ inclination run.

\section{Conclusions}
\label{Sec: Conclusions}
In this work we used our sub-grid model for black hole accretion through a (potentially warped) $\alpha$-disc and feedback in the form of a Blandford-Znajek jet to self-consistently explore black hole fuelling and jet-driven outflows in radio loud, Seyfert-like galaxies. Our spin-driven AGN jet model evolves the properties of the black hole on-the-fly during the simulation, is fully interfaced with the surrounding hydro simulation and accurately tracks the evolution of the black hole mass and spin. This information is then used, along with the properties of the accretion flow, to self-consistently determine the power and direction of the jet. Our sub-grid model determines injection of the jet mass, energy and momentum and beyond this, what we observe in the simulations is primarily determined by the surrounding environment. To this end, we performed an extensive suite of simulations in which the jets are launched by black holes embedded in circumnuclear discs (CNDs) that are surrounded by a dilute circumgalactic medium (CGM) and explored a range of densities both for the CGM and for the gas in the CNDs. We also examined a large parameter space of initial black hole spin magnitudes and orientations to gauge their impact on the nature of large-scale outflows. Our main results are:
\begin{itemize}
    
\item{AGN bolometric luminosities in our simulated Seyfert galaxies range from $10^{41}-10^{43} \,{\rm erg \, s^{-1}}$. These are comparable to typical Seyfert nuclear luminosities in the local Universe for black holes with $10^6 \; {\rm M_\odot}$ accreting at a moderate Eddington rate ($f_{\rm Edd} \sim 10^{-3}$).}
    
\item In the absence of a jet, the black hole is fed by steady, coherent axisymmetric gas flows, driven by secular processes in the CND. The launching of a jet generates funnels of high angular momentum, outflowing material, which shock heat the local gas and alter its kinematics. Jet-driven backflows, however, draw low angular momentum gas back towards the centre which replenish the gas supply that feeds the black hole and refuel the jet activity. The balance between these two processes is highly dependent, not only on the jet power, but also on the thermodynamics of the gas in the CND, both of which can significantly affect the variability and mass content of the accretion flow.

\item The complex interplay between the processes that drive inflows and outflows leads to significant evolution in the jet power (by orders of magnitude on Myr timescales). In comparison, the evolution of the black hole spin is relatively modest in all cases. Interestingly, we find that spins can either increase or decrease, due to shifts in the balance between spin-up torques from the accretion of co-rotating gas and spin-down torques due to jet-launching.
  
\item AGN jets are able to significantly alter the thermodynamics and kinematics of the gas in the CND. They drive strong shocks into the disc, directly entrain CND material and alter the kinematics of the gas at the CND-CGM interface, causing the upper strata of the disc to be drawn up into the large-scale outflow. Indeed, we find that the jets can unbind up to $10\%$ of mass of the CND in just a few Myrs, despite their relatively modest power. Highly inclined jets have the most disruptive effect on the CND structure. These jets are, however, unable to advance laterally to a significant distance, but are instead re-directed towards the galactic polar regions which offer the least resistance to outflow propagation. 

\item AGN jets are able to generate a wide variety of large-scale outflows, the morphology and evolution of which are particularly sensitive to the power of the jet, its orientation and the properties of the surrounding environment. High power jets and those launched perpendicular to the plane of the CND drive outflows that are, on the whole, light, hot and fast. On the other hand, low power, misaligned jets launch highly mass-loaded (with mass-loading factors reaching $\sim1000$), multiphase, quasi-bipolar outflows which entrain a significant mass of CND material. In general, enhanced pressure confinement of the CGM leads to outflows that propagate with lower velocities and are more mass-loaded.
    
\item In terms of mass outflow rates and large-scale kinetic powers in warm ionised gas, our simulations are comparable to recent observations of Seyfert galaxies that host black holes of similar masses and bolometric luminosities \citep{2021Venturi+,2021Ruschel-Dutra}. Additionally, line-of-sight (LoS) velocity maps of this warm gas component indicate that our jet-driven outflows have high velocities (up to $\sim 500\;{\rm km\,s^{-1}}$) that are comparable to those found in \cite{2021Venturi+,2021Speranza+,2019Jarvis+}. 

\item Low power jets launched directly into the CND generate the largest amount of cold, outflowing gas (up to $10^5 \, {\rm M_\odot}$) with modest LoS velocities of $\sim90\;{\rm km\,s^{-1}}$ consistent with observations by \cite{2014GarciaBurillo+} and \cite{2019Alonso-Herrero+}. In our simulations, all cold, dense gas in the outflows originates from the CND. We, therefore, expect that inclusion of realistic radiative cooling in outflows will lead to the formation of even more outflowing cold gas moving at higher velocities.  

\end{itemize}

Future observational facilities such as JWST, SKA, ngVLA, Athena, XRISM, and Lynx have the potential to revolutionise our understanding of AGN physics, investigating black hole jet feedback at earlier times, at lower luminosities and with higher spatial and spectral resolution. Additionally, the LISA mission will expand the field of multi-messenger astronomy to the low-frequency range, detecting gravitational waves from the coalescence of supermassive black holes all the way back to cosmic dawn. 

The diversity of outflows, in terms of morphology, thermodynamics and kinematics, that we have shown can be produced solely by AGN jets highlights the importance of applying physically motivated models of AGN feedback to realistic galaxy formation contexts. Indeed, this will be vital if simulations are to provide firm theoretical predictions for, as well as accurately interpret, the wealth of data that these next-generation observational facilities will undoubtedly provide.

\section*{Acknowledgements}
We would like to thank the referee for their very thoughtful and constructive report. We would also like to thank Tiago Costa, George Efstathiou, Martin Haehnelt, Roberto Maiolino and Hannah {\"U}bler for useful discussions and helpful comments during the development of this manuscript. RYT, DS and MAB acknowledge the support from the ERC starting grant 638707 `Black holes and their host galaxies: co-evolution across cosmic time' and the STFC. This work was performed using the Cambridge Service for Data Driven Discovery (CSD3), part of which is operated by the University of Cambridge Research Computing on behalf of the STFC DiRAC HPC Facility (www.dirac.ac.uk). The DiRAC component of CSD3 was funded by BEIS capital funding via STFC capital grants ST/P002307/1 and ST/R002452/1 and STFC operations grant ST/R00689X/1. DiRAC is part of the National e-Infrastructure. This work used the DiRAC@Durham facility managed by the Institute for Computational Cosmology on behalf of the STFC DiRAC HPC Facility (www.dirac.ac.uk). The equipment was funded by BEIS capital funding via STFC capital grants ST/P002293/1 and ST/R002371/1, Durham University and STFC operations grant ST/R000832/1. DiRAC is part of the National e-Infrastructure. This work made significant use of the NumPy \citep{2020Harris+}, SciPy \citep{2020Virtanen+}, and Matplotlib \citep{2007Hunter} Python packages.

\section*{Data availability}
The data underlying this article will be shared upon request to the corresponding author.



\bibliographystyle{mnras}
\bibliography{references} 

\begin{thebibliography}{}
\makeatletter
\relax
\def\mn@urlcharsother{\let\do\@makeother \do\$\do\&\do\#\do\^\do\_\do\%\do\~}
\def\mn@doi{\begingroup\mn@urlcharsother \@ifnextchar [ {\mn@doi@}
  {\mn@doi@[]}}
\def\mn@doi@[#1]#2{\def\@tempa{#1}\ifx\@tempa\@empty \href
  {http://dx.doi.org/#2} {doi:#2}\else \href {http://dx.doi.org/#2} {#1}\fi
  \endgroup}
\def\mn@eprint#1#2{\mn@eprint@#1:#2::\@nil}
\def\mn@eprint@arXiv#1{\href {http://arxiv.org/abs/#1} {{\tt arXiv:#1}}}
\def\mn@eprint@dblp#1{\href {http://dblp.uni-trier.de/rec/bibtex/#1.xml}
  {dblp:#1}}
\def\mn@eprint@#1:#2:#3:#4\@nil{\def\@tempa {#1}\def\@tempb {#2}\def\@tempc
  {#3}\ifx \@tempc \@empty \let \@tempc \@tempb \let \@tempb \@tempa \fi \ifx
  \@tempb \@empty \def\@tempb {arXiv}\fi \@ifundefined
  {mn@eprint@\@tempb}{\@tempb:\@tempc}{\expandafter \expandafter \csname
  mn@eprint@\@tempb\endcsname \expandafter{\@tempc}}}

\bibitem[\protect\citeauthoryear{{Alonso-Herrero} et~al.,}{{Alonso-Herrero}
  et~al.}{2019}]{2019Alonso-Herrero+}
{Alonso-Herrero} A.,  et~al., 2019, \mn@doi [\aap]
  {10.1051/0004-6361/201935431}, \href
  {https://ui.adsabs.harvard.edu/abs/2019A&A...628A..65A} {628, A65}

\bibitem[\protect\citeauthoryear{{Antonuccio-Delogu} \&
  {Silk}}{{Antonuccio-Delogu} \& {Silk}}{2008}]{2008Antonuccio-DeloguSilk}
{Antonuccio-Delogu} V.,  {Silk} J.,  2008, \mn@doi [\mnras]
  {10.1111/j.1365-2966.2008.13663.x}, \href
  {https://ui.adsabs.harvard.edu/abs/2008MNRAS.389.1750A} {389, 1750}

\bibitem[\protect\citeauthoryear{{Antonuccio-Delogu} \&
  {Silk}}{{Antonuccio-Delogu} \& {Silk}}{2010}]{2010Antonuccio-DeloguSilk}
{Antonuccio-Delogu} V.,  {Silk} J.,  2010, \mn@doi [\mnras]
  {10.1111/j.1365-2966.2010.16532.x}, \href
  {https://ui.adsabs.harvard.edu/abs/2010MNRAS.405.1303A} {405, 1303}

\bibitem[\protect\citeauthoryear{{Banda-Barrag{\'a}n}, {Br{\"u}ggen}, {Heesen},
  {Scannapieco}, {Cottle}, {Federrath}  \& {Wagner}}{{Banda-Barrag{\'a}n}
  et~al.}{2021}]{2021Banda-Barragan+}
{Banda-Barrag{\'a}n} W.~E.,  {Br{\"u}ggen} M.,  {Heesen} V.,  {Scannapieco} E.,
   {Cottle} J.,  {Federrath} C.,   {Wagner} A.~Y.,  2021, \mn@doi [\mnras]
  {10.1093/mnras/stab1884}, \href
  {https://ui.adsabs.harvard.edu/abs/2021MNRAS.506.5658B} {506, 5658}

\bibitem[\protect\citeauthoryear{{Bardeen} \& {Petterson}}{{Bardeen} \&
  {Petterson}}{1975}]{1975BardeenPetterson}
{Bardeen} J.~M.,  {Petterson} J.~A.,  1975, \mn@doi [\apj] {10.1086/181711},
  \href {https://ui.adsabs.harvard.edu/abs/1975ApJ...195L..65B} {195, L65}

\bibitem[\protect\citeauthoryear{{Barnes} \& {Hut}}{{Barnes} \&
  {Hut}}{1986}]{1986BarnesHut}
{Barnes} J.,  {Hut} P.,  1986, \mn@doi [\nat] {10.1038/324446a0}, \href
  {https://ui.adsabs.harvard.edu/abs/1986Natur.324..446B} {324, 446}

\bibitem[\protect\citeauthoryear{{Baron} \& {Netzer}}{{Baron} \&
  {Netzer}}{2019}]{2019BaronNetzer}
{Baron} D.,  {Netzer} H.,  2019, \mn@doi [\mnras] {10.1093/mnras/stz1070},
  \href {https://ui.adsabs.harvard.edu/abs/2019MNRAS.486.4290B} {486, 4290}

\bibitem[\protect\citeauthoryear{{Beckmann} et~al.,}{{Beckmann}
  et~al.}{2019}]{2019Beckmann+}
{Beckmann} R.~S.,  et~al., 2019, \mn@doi [\aap] {10.1051/0004-6361/201936188},
  \href {https://ui.adsabs.harvard.edu/abs/2019A&A...631A..60B} {631, A60}

\bibitem[\protect\citeauthoryear{{Best}, {von der Linden}, {Kauffmann},
  {Heckman}  \& {Kaiser}}{{Best} et~al.}{2007}]{2007Best+}
{Best} P.~N.,  {von der Linden} A.,  {Kauffmann} G.,  {Heckman} T.~M.,
  {Kaiser} C.~R.,  2007, \mn@doi [\mnras] {10.1111/j.1365-2966.2007.11937.x},
  \href {https://ui.adsabs.harvard.edu/abs/2007MNRAS.379..894B} {379, 894}

\bibitem[\protect\citeauthoryear{{Bewketu Belete} et~al.,}{{Bewketu Belete}
  et~al.}{2021}]{2021BewketuBelete+}
{Bewketu Belete} A.,  et~al., 2021, arXiv e-prints, \href
  {https://ui.adsabs.harvard.edu/abs/2021arXiv210506867B} {p. arXiv:2105.06867}

\bibitem[\protect\citeauthoryear{{B{\^\i}rzan}, {Rafferty}, {Nulsen},
  {McNamara}, {R{\"o}ttgering}, {Wise}  \& {Mittal}}{{B{\^\i}rzan}
  et~al.}{2012}]{2012Birzan}
{B{\^\i}rzan} L.,  {Rafferty} D.~A.,  {Nulsen} P.~E.~J.,  {McNamara} B.~R.,
  {R{\"o}ttgering} H.~J.~A.,  {Wise} M.~W.,   {Mittal} R.,  2012, \mn@doi
  [\mnras] {10.1111/j.1365-2966.2012.22083.x}, \href
  {https://ui.adsabs.harvard.edu/abs/2012MNRAS.427.3468B} {427, 3468}

\bibitem[\protect\citeauthoryear{{Blandford} \& {Znajek}}{{Blandford} \&
  {Znajek}}{1977}]{1977BlandfordZnajek}
{Blandford} R.~D.,  {Znajek} R.~L.,  1977, \mn@doi [\mnras]
  {10.1093/mnras/179.3.433}, \href
  {https://ui.adsabs.harvard.edu/abs/1977MNRAS.179..433B} {179, 433}

\bibitem[\protect\citeauthoryear{{Blandford}, {Meier}  \&
  {Readhead}}{{Blandford} et~al.}{2019}]{2019Blandford+}
{Blandford} R.,  {Meier} D.,   {Readhead} A.,  2019, \mn@doi [\araa]
  {10.1146/annurev-astro-081817-051948}, \href
  {https://ui.adsabs.harvard.edu/abs/2019ARA&A..57..467B} {57, 467}

\bibitem[\protect\citeauthoryear{{Boroson}}{{Boroson}}{2002}]{2002Boroson}
{Boroson} T.~A.,  2002, \mn@doi [\apj] {10.1086/324486}, \href
  {https://ui.adsabs.harvard.edu/abs/2002ApJ...565...78B} {565, 78}

\bibitem[\protect\citeauthoryear{{Bourne} \& {Sijacki}}{{Bourne} \&
  {Sijacki}}{2017}]{2017BourneSijacki}
{Bourne} M.~A.,  {Sijacki} D.,  2017, \mn@doi [\mnras] {10.1093/mnras/stx2269},
  \href {https://ui.adsabs.harvard.edu/abs/2017MNRAS.472.4707B} {472, 4707}

\bibitem[\protect\citeauthoryear{{Bourne} \& {Sijacki}}{{Bourne} \&
  {Sijacki}}{2021}]{2021BourneSijacki}
{Bourne} M.~A.,  {Sijacki} D.,  2021, \mn@doi [\mnras]
  {10.1093/mnras/stab1662}, \href
  {https://ui.adsabs.harvard.edu/abs/2021MNRAS.506..488B} {506, 488}

\bibitem[\protect\citeauthoryear{{Bourne}, {Sijacki}  \& {Puchwein}}{{Bourne}
  et~al.}{2019}]{2019Bourne+}
{Bourne} M.~A.,  {Sijacki} D.,   {Puchwein} E.,  2019, \mn@doi [\mnras]
  {10.1093/mnras/stz2604}, \href
  {https://ui.adsabs.harvard.edu/abs/2019MNRAS.490..343B} {490, 343}

\bibitem[\protect\citeauthoryear{{Bustamante} \& {Springel}}{{Bustamante} \&
  {Springel}}{2019}]{2019BustamanteSpringel}
{Bustamante} S.,  {Springel} V.,  2019, \mn@doi [\mnras]
  {10.1093/mnras/stz2836}, \href
  {https://ui.adsabs.harvard.edu/abs/2019MNRAS.490.4133B} {490, 4133}

\bibitem[\protect\citeauthoryear{{Chen} et~al.,}{{Chen}
  et~al.}{2020}]{2020Chen+}
{Chen} S.,  et~al., 2020, \mn@doi [\mnras] {10.1093/mnras/staa2373}, \href
  {https://ui.adsabs.harvard.edu/abs/2020MNRAS.498.1278C} {498, 1278}

\bibitem[\protect\citeauthoryear{{Cielo}, {Antonuccio-Delogu}, {Macci{\`o}},
  {Romeo}  \& {Silk}}{{Cielo} et~al.}{2014}]{2014Cielo+}
{Cielo} S.,  {Antonuccio-Delogu} V.,  {Macci{\`o}} A.~V.,  {Romeo} A.~D.,
  {Silk} J.,  2014, \mn@doi [\mnras] {10.1093/mnras/stu161}, \href
  {https://ui.adsabs.harvard.edu/abs/2014MNRAS.439.2903C} {439, 2903}

\bibitem[\protect\citeauthoryear{{Cielo}, {Babul}, {Antonuccio-Delogu}, {Silk}
  \& {Volonteri}}{{Cielo} et~al.}{2018}]{2018Cielo+}
{Cielo} S.,  {Babul} A.,  {Antonuccio-Delogu} V.,  {Silk} J.,   {Volonteri} M.,
   2018, \mn@doi [\aap] {10.1051/0004-6361/201832582}, \href
  {https://ui.adsabs.harvard.edu/abs/2018A&A...617A..58C} {617, A58}

\bibitem[\protect\citeauthoryear{{Clarke}, {Kinney}  \& {Pringle}}{{Clarke}
  et~al.}{1998}]{1998Clarke+}
{Clarke} C.~J.,  {Kinney} A.~L.,   {Pringle} J.~E.,  1998, \mn@doi [\apj]
  {10.1086/305285}, \href
  {https://ui.adsabs.harvard.edu/abs/1998ApJ...495..189C} {495, 189}

\bibitem[\protect\citeauthoryear{{Combes} et~al.,}{{Combes}
  et~al.}{2013}]{2013Combes+}
{Combes} F.,  et~al., 2013, \mn@doi [\aap] {10.1051/0004-6361/201322288}, \href
  {https://ui.adsabs.harvard.edu/abs/2013A&A...558A.124C} {558, A124}

\bibitem[\protect\citeauthoryear{{Congiu} et~al.,}{{Congiu}
  et~al.}{2020}]{2020Congiu+}
{Congiu} E.,  et~al., 2020, \mn@doi [\mnras] {10.1093/mnras/staa3024}, \href
  {https://ui.adsabs.harvard.edu/abs/2020MNRAS.499.3149C} {499, 3149}

\bibitem[\protect\citeauthoryear{{Costa}, {Sijacki}  \& {Haehnelt}}{{Costa}
  et~al.}{2014}]{2014Costa+}
{Costa} T.,  {Sijacki} D.,   {Haehnelt} M.~G.,  2014, \mn@doi [\mnras]
  {10.1093/mnras/stu1632}, \href
  {https://ui.adsabs.harvard.edu/abs/2014MNRAS.444.2355C} {444, 2355}

\bibitem[\protect\citeauthoryear{{Costa}, {Pakmor}  \& {Springel}}{{Costa}
  et~al.}{2020}]{2020Costa+}
{Costa} T.,  {Pakmor} R.,   {Springel} V.,  2020, \mn@doi [\mnras]
  {10.1093/mnras/staa2321}, \href
  {https://ui.adsabs.harvard.edu/abs/2020MNRAS.497.5229C} {497, 5229}

\bibitem[\protect\citeauthoryear{{Couto}, {Storchi-Bergmann}, {Axon},
  {Robinson}, {Kharb}  \& {Riffel}}{{Couto} et~al.}{2013}]{2013Couto+}
{Couto} G.~S.,  {Storchi-Bergmann} T.,  {Axon} D.~J.,  {Robinson} A.,  {Kharb}
  P.,   {Riffel} R.~A.,  2013, \mn@doi [\mnras] {10.1093/mnras/stt1491}, \href
  {https://ui.adsabs.harvard.edu/abs/2013MNRAS.435.2982C} {435, 2982}

\bibitem[\protect\citeauthoryear{{Crain} et~al.,}{{Crain}
  et~al.}{2015}]{2015Crain+}
{Crain} R.~A.,  et~al., 2015, \mn@doi [\mnras] {10.1093/mnras/stv725}, \href
  {https://ui.adsabs.harvard.edu/abs/2015MNRAS.450.1937C} {450, 1937}

\bibitem[\protect\citeauthoryear{{Cresci} et~al.,}{{Cresci}
  et~al.}{2015}]{2015Cresci+}
{Cresci} G.,  et~al., 2015, \mn@doi [\aap] {10.1051/0004-6361/201526581}, \href
  {https://ui.adsabs.harvard.edu/abs/2015A&A...582A..63C} {582, A63}

\bibitem[\protect\citeauthoryear{{Curtis} \& {Sijacki}}{{Curtis} \&
  {Sijacki}}{2015}]{2015CurtisSijacki}
{Curtis} M.,  {Sijacki} D.,  2015, \mn@doi [\mnras] {10.1093/mnras/stv2246},
  \href {https://ui.adsabs.harvard.edu/abs/2015MNRAS.454.3445C} {454, 3445}

\bibitem[\protect\citeauthoryear{{Curtis} \& {Sijacki}}{{Curtis} \&
  {Sijacki}}{2016}]{2016CurtisSijacki}
{Curtis} M.,  {Sijacki} D.,  2016, \mn@doi [\mnras] {10.1093/mnrasl/slv199},
  \href {https://ui.adsabs.harvard.edu/abs/2016MNRAS.457L..34C} {457, L34}

\bibitem[\protect\citeauthoryear{{Das}, {Crenshaw}, {Kraemer}  \& {Deo}}{{Das}
  et~al.}{2006}]{2006Das+}
{Das} V.,  {Crenshaw} D.~M.,  {Kraemer} S.~B.,   {Deo} R.~P.,  2006, \mn@doi
  [\aj] {10.1086/504899}, \href
  {https://ui.adsabs.harvard.edu/abs/2006AJ....132..620D} {132, 620}

\bibitem[\protect\citeauthoryear{{Dav{\'e}}, {Angl{\'e}s-Alc{\'a}zar},
  {Narayanan}, {Li}, {Rafieferantsoa}  \& {Appleby}}{{Dav{\'e}}
  et~al.}{2019}]{2019Dave+}
{Dav{\'e}} R.,  {Angl{\'e}s-Alc{\'a}zar} D.,  {Narayanan} D.,  {Li} Q.,
  {Rafieferantsoa} M.~H.,   {Appleby} S.,  2019, \mn@doi [\mnras]
  {10.1093/mnras/stz937}, \href
  {https://ui.adsabs.harvard.edu/abs/2019MNRAS.486.2827D} {486, 2827}

\bibitem[\protect\citeauthoryear{{Davies} et~al.,}{{Davies}
  et~al.}{2020}]{2020Davies+}
{Davies} R.,  et~al., 2020, \mn@doi [\mnras] {10.1093/mnras/staa2413}, \href
  {https://ui.adsabs.harvard.edu/abs/2020MNRAS.498.4150D} {498, 4150}

\bibitem[\protect\citeauthoryear{{Dehnen} \& {King}}{{Dehnen} \&
  {King}}{2013}]{2013DehnenKing}
{Dehnen} W.,  {King} A.,  2013, \mn@doi [\apjl] {10.1088/2041-8205/777/2/L28},
  \href {https://ui.adsabs.harvard.edu/abs/2013ApJ...777L..28D} {777, L28}

\bibitem[\protect\citeauthoryear{{Dubois}, {Devriendt}, {Slyz}  \&
  {Teyssier}}{{Dubois} et~al.}{2012}]{2012Dubois+}
{Dubois} Y.,  {Devriendt} J.,  {Slyz} A.,   {Teyssier} R.,  2012, \mn@doi
  [\mnras] {10.1111/j.1365-2966.2011.20236.x}, \href
  {https://ui.adsabs.harvard.edu/abs/2012MNRAS.420.2662D} {420, 2662}

\bibitem[\protect\citeauthoryear{{Dubois}, {Volonteri}, {Silk}, {Devriendt}  \&
  {Slyz}}{{Dubois} et~al.}{2014}]{2014Dubois+}
{Dubois} Y.,  {Volonteri} M.,  {Silk} J.,  {Devriendt} J.,   {Slyz} A.,  2014,
  \mn@doi [\mnras] {10.1093/mnras/stu425}, \href
  {https://ui.adsabs.harvard.edu/abs/2014MNRAS.440.2333D} {440, 2333}

\bibitem[\protect\citeauthoryear{{Ehlert}, {Weinberger}, {Pfrommer}, {Pakmor}
  \& {Springel}}{{Ehlert} et~al.}{2018}]{2018Ehlert+}
{Ehlert} K.,  {Weinberger} R.,  {Pfrommer} C.,  {Pakmor} R.,   {Springel} V.,
  2018, \mn@doi [\mnras] {10.1093/mnras/sty2397}, \href
  {https://ui.adsabs.harvard.edu/abs/2018MNRAS.481.2878E} {481, 2878}

\bibitem[\protect\citeauthoryear{{Ehlert}, {Weinberger}, {Pfrommer}  \&
  {Springel}}{{Ehlert} et~al.}{2021}]{2020Ehlert+}
{Ehlert} K.,  {Weinberger} R.,  {Pfrommer} C.,   {Springel} V.,  2021, \mn@doi
  [\mnras] {10.1093/mnras/stab551}, \href
  {https://ui.adsabs.harvard.edu/abs/2021MNRAS.503.1327E} {503, 1327}

\bibitem[\protect\citeauthoryear{{Fabian}}{{Fabian}}{2012}]{2012Fabian}
{Fabian} A.~C.,  2012, \mn@doi [\araa] {10.1146/annurev-astro-081811-125521},
  \href {https://ui.adsabs.harvard.edu/abs/2012ARA&A..50..455F} {50, 455}

\bibitem[\protect\citeauthoryear{{Fiacconi}, {Sijacki}  \&
  {Pringle}}{{Fiacconi} et~al.}{2018}]{2018Fiacconi+}
{Fiacconi} D.,  {Sijacki} D.,   {Pringle} J.~E.,  2018, \mn@doi [\mnras]
  {10.1093/mnras/sty893}, \href
  {https://ui.adsabs.harvard.edu/abs/2018MNRAS.477.3807F} {477, 3807}

\bibitem[\protect\citeauthoryear{{Fiore} et~al.,}{{Fiore}
  et~al.}{2017}]{2017Fiore+}
{Fiore} F.,  et~al., 2017, \mn@doi [\aap] {10.1051/0004-6361/201629478}, \href
  {https://ui.adsabs.harvard.edu/abs/2017A&A...601A.143F} {601, A143}

\bibitem[\protect\citeauthoryear{{Fluetsch} et~al.,}{{Fluetsch}
  et~al.}{2021}]{2021Fluetsch+}
{Fluetsch} A.,  et~al., 2021, \mn@doi [\mnras] {10.1093/mnras/stab1666}, \href
  {https://ui.adsabs.harvard.edu/abs/2021MNRAS.505.5753F} {505, 5753}

\bibitem[\protect\citeauthoryear{{Foschini} et~al.,}{{Foschini}
  et~al.}{2015}]{2015Foschini+}
{Foschini} L.,  et~al., 2015, \mn@doi [\aap] {10.1051/0004-6361/201424972},
  \href {https://ui.adsabs.harvard.edu/abs/2015A&A...575A..13F} {575, A13}

\bibitem[\protect\citeauthoryear{{Gaibler}, {Khochfar}  \& {Krause}}{{Gaibler}
  et~al.}{2011}]{2011Gaibler+}
{Gaibler} V.,  {Khochfar} S.,   {Krause} M.,  2011, \mn@doi [\mnras]
  {10.1111/j.1365-2966.2010.17674.x}, \href
  {https://ui.adsabs.harvard.edu/abs/2011MNRAS.411..155G} {411, 155}

\bibitem[\protect\citeauthoryear{{Gaibler}, {Khochfar}, {Krause}  \&
  {Silk}}{{Gaibler} et~al.}{2012}]{2012Gaibler+}
{Gaibler} V.,  {Khochfar} S.,  {Krause} M.,   {Silk} J.,  2012, \mn@doi
  [\mnras] {10.1111/j.1365-2966.2012.21479.x}, \href
  {https://ui.adsabs.harvard.edu/abs/2012MNRAS.425..438G} {425, 438}

\bibitem[\protect\citeauthoryear{{Gallimore}, {Axon}, {O'Dea}, {Baum}  \&
  {Pedlar}}{{Gallimore} et~al.}{2006}]{2006Gallimore+}
{Gallimore} J.~F.,  {Axon} D.~J.,  {O'Dea} C.~P.,  {Baum} S.~A.,   {Pedlar} A.,
   2006, \mn@doi [\aj] {10.1086/504593}, \href
  {https://ui.adsabs.harvard.edu/abs/2006AJ....132..546G} {132, 546}

\bibitem[\protect\citeauthoryear{{Gammie}}{{Gammie}}{2001}]{2001Gammie}
{Gammie} C.~F.,  2001, \mn@doi [\apj] {10.1086/320631}, \href
  {https://ui.adsabs.harvard.edu/abs/2001ApJ...553..174G} {553, 174}

\bibitem[\protect\citeauthoryear{{Garc{\'\i}a-Burillo}
  et~al.,}{{Garc{\'\i}a-Burillo} et~al.}{2014}]{2014GarciaBurillo+}
{Garc{\'\i}a-Burillo} S.,  et~al., 2014, \mn@doi [\aap]
  {10.1051/0004-6361/201423843}, \href
  {https://ui.adsabs.harvard.edu/abs/2014A&A...567A.125G} {567, A125}

\bibitem[\protect\citeauthoryear{{Grupe}, {Komossa}, {Leighly}  \&
  {Page}}{{Grupe} et~al.}{2010}]{2010Grupe+}
{Grupe} D.,  {Komossa} S.,  {Leighly} K.~M.,   {Page} K.~L.,  2010, \mn@doi
  [\apjs] {10.1088/0067-0049/187/1/64}, \href
  {https://ui.adsabs.harvard.edu/abs/2010ApJS..187...64G} {187, 64}

\bibitem[\protect\citeauthoryear{Harris et~al.,}{Harris
  et~al.}{2020}]{2020Harris+}
Harris C.~R.,  et~al., 2020, \mn@doi [Nature] {10.1038/s41586-020-2649-2}, 585,
  357–362

\bibitem[\protect\citeauthoryear{{Harrison}, {Thomson}, {Alexander}, {Bauer},
  {Edge}, {Hogan}, {Mullaney}  \& {Swinbank}}{{Harrison}
  et~al.}{2015}]{2015Harrison+}
{Harrison} C.~M.,  {Thomson} A.~P.,  {Alexander} D.~M.,  {Bauer} F.~E.,  {Edge}
  A.~C.,  {Hogan} M.~T.,  {Mullaney} J.~R.,   {Swinbank} A.~M.,  2015, \mn@doi
  [\apj] {10.1088/0004-637X/800/1/45}, \href
  {https://ui.adsabs.harvard.edu/abs/2015ApJ...800...45H} {800, 45}

\bibitem[\protect\citeauthoryear{{Henden}, {Puchwein}, {Shen}  \&
  {Sijacki}}{{Henden} et~al.}{2018}]{2018Henden+}
{Henden} N.~A.,  {Puchwein} E.,  {Shen} S.,   {Sijacki} D.,  2018, \mn@doi
  [\mnras] {10.1093/mnras/sty1780}, \href
  {https://ui.adsabs.harvard.edu/abs/2018MNRAS.479.5385H} {479, 5385}

\bibitem[\protect\citeauthoryear{{Holt}, {Tadhunter}  \& {Morganti}}{{Holt}
  et~al.}{2008}]{2008Holt+}
{Holt} J.,  {Tadhunter} C.~N.,   {Morganti} R.,  2008, \mn@doi [\mnras]
  {10.1111/j.1365-2966.2008.13089.x}, \href
  {https://ui.adsabs.harvard.edu/abs/2008MNRAS.387..639H} {387, 639}

\bibitem[\protect\citeauthoryear{{Hota} \& {Saikia}}{{Hota} \&
  {Saikia}}{2006}]{2006HotaSaikia}
{Hota} A.,  {Saikia} D.~J.,  2006, \mn@doi [\mnras]
  {10.1111/j.1365-2966.2006.10738.x}, \href
  {https://ui.adsabs.harvard.edu/abs/2006MNRAS.371..945H} {371, 945}

\bibitem[\protect\citeauthoryear{Hunter}{Hunter}{2007}]{2007Hunter}
Hunter J.~D.,  2007, \mn@doi [Computing in Science Engineering]
  {10.1109/MCSE.2007.55}, 9, 90

\bibitem[\protect\citeauthoryear{{Jarvis} et~al.,}{{Jarvis}
  et~al.}{2019}]{2019Jarvis+}
{Jarvis} M.~E.,  et~al., 2019, \mn@doi [\mnras] {10.1093/mnras/stz556}, \href
  {https://ui.adsabs.harvard.edu/abs/2019MNRAS.485.2710J} {485, 2710}

\bibitem[\protect\citeauthoryear{{Kinney}, {Schmitt}, {Clarke}, {Pringle},
  {Ulvestad}  \& {Antonucci}}{{Kinney} et~al.}{2000}]{2000Kinney+}
{Kinney} A.~L.,  {Schmitt} H.~R.,  {Clarke} C.~J.,  {Pringle} J.~E.,
  {Ulvestad} J.~S.,   {Antonucci} R.~R.~J.,  2000, \mn@doi [\apj]
  {10.1086/309016}, \href
  {https://ui.adsabs.harvard.edu/abs/2000ApJ...537..152K} {537, 152}

\bibitem[\protect\citeauthoryear{{Komossa}}{{Komossa}}{2018}]{2018Komossa}
{Komossa} S.,  2018, in Revisiting Narrow-Line Seyfert 1 Galaxies and their
  Place in the Universe. p.~15 (\mn@eprint {arXiv} {1807.03666})

\bibitem[\protect\citeauthoryear{{Kormendy} \& {Ho}}{{Kormendy} \&
  {Ho}}{2013}]{2013KormendyHo}
{Kormendy} J.,  {Ho} L.~C.,  2013, \mn@doi [\araa]
  {10.1146/annurev-astro-082708-101811}, \href
  {https://ui.adsabs.harvard.edu/abs/2013ARA&A..51..511K} {51, 511}

\bibitem[\protect\citeauthoryear{{Krause}, {Alexander}, {Riley}  \&
  {Hopton}}{{Krause} et~al.}{2012}]{2012Krause+}
{Krause} M.,  {Alexander} P.,  {Riley} J.,   {Hopton} D.,  2012, \mn@doi
  [\mnras] {10.1111/j.1365-2966.2012.21645.x}, \href
  {https://ui.adsabs.harvard.edu/abs/2012MNRAS.427.3196K} {427, 3196}

\bibitem[\protect\citeauthoryear{{Li} \& {Bryan}}{{Li} \&
  {Bryan}}{2014}]{2014LiBryan}
{Li} Y.,  {Bryan} G.~L.,  2014, \mn@doi [\apj] {10.1088/0004-637X/789/2/153},
  \href {https://ui.adsabs.harvard.edu/abs/2014ApJ...789..153L} {789, 153}

\bibitem[\protect\citeauthoryear{{Liska}, {Hesp}, {Tchekhovskoy}, {Ingram},
  {van der Klis}  \& {Markoff}}{{Liska} et~al.}{2018}]{2018Liska+}
{Liska} M.,  {Hesp} C.,  {Tchekhovskoy} A.,  {Ingram} A.,  {van der Klis} M.,
  {Markoff} S.,  2018, \mn@doi [\mnras] {10.1093/mnrasl/slx174}, \href
  {https://ui.adsabs.harvard.edu/abs/2018MNRAS.474L..81L} {474, L81}

\bibitem[\protect\citeauthoryear{{McNamara} \& {Nulsen}}{{McNamara} \&
  {Nulsen}}{2007}]{2007McNamaraNulsen}
{McNamara} B.~R.,  {Nulsen} P.~E.~J.,  2007, \mn@doi [\araa]
  {10.1146/annurev.astro.45.051806.110625}, \href
  {https://ui.adsabs.harvard.edu/abs/2007ARA&A..45..117M} {45, 117}

\bibitem[\protect\citeauthoryear{{Mellema}, {Kurk}  \&
  {R{\"o}ttgering}}{{Mellema} et~al.}{2002}]{2002Mellema+}
{Mellema} G.,  {Kurk} J.~D.,   {R{\"o}ttgering} H.~J.~A.,  2002, \mn@doi [\aap]
  {10.1051/0004-6361:20021408}, \href
  {https://ui.adsabs.harvard.edu/abs/2002A&A...395L..13M} {395, L13}

\bibitem[\protect\citeauthoryear{{Mezcua}, {Suh}  \& {Civano}}{{Mezcua}
  et~al.}{2019}]{2019Mezcua+}
{Mezcua} M.,  {Suh} H.,   {Civano} F.,  2019, \mn@doi [\mnras]
  {10.1093/mnras/stz1760}, \href
  {https://ui.adsabs.harvard.edu/abs/2019MNRAS.488..685M} {488, 685}

\bibitem[\protect\citeauthoryear{{Mingo}, {Hardcastle}, {Croston}, {Evans},
  {Hota}, {Kharb}  \& {Kraft}}{{Mingo} et~al.}{2011}]{2011Mingo+}
{Mingo} B.,  {Hardcastle} M.~J.,  {Croston} J.~H.,  {Evans} D.~A.,  {Hota} A.,
  {Kharb} P.,   {Kraft} R.~P.,  2011, \mn@doi [\apj]
  {10.1088/0004-637X/731/1/21}, \href
  {https://ui.adsabs.harvard.edu/abs/2011ApJ...731...21M} {731, 21}

\bibitem[\protect\citeauthoryear{{Mingo}, {Hardcastle}, {Croston}, {Evans},
  {Kharb}, {Kraft}  \& {Lenc}}{{Mingo} et~al.}{2012}]{2012Mingo+}
{Mingo} B.,  {Hardcastle} M.~J.,  {Croston} J.~H.,  {Evans} D.~A.,  {Kharb} P.,
   {Kraft} R.~P.,   {Lenc} E.,  2012, \mn@doi [\apj]
  {10.1088/0004-637X/758/2/95}, \href
  {https://ui.adsabs.harvard.edu/abs/2012ApJ...758...95M} {758, 95}

\bibitem[\protect\citeauthoryear{{Mittal}, {Hudson}, {Reiprich}  \&
  {Clarke}}{{Mittal} et~al.}{2009}]{2009Mittal+}
{Mittal} R.,  {Hudson} D.~S.,  {Reiprich} T.~H.,   {Clarke} T.,  2009, \mn@doi
  [\aap] {10.1051/0004-6361/200810836}, \href
  {https://ui.adsabs.harvard.edu/abs/2009A&A...501..835M} {501, 835}

\bibitem[\protect\citeauthoryear{{Molyneux}, {Harrison}  \&
  {Jarvis}}{{Molyneux} et~al.}{2019}]{2019Molyneux+}
{Molyneux} S.~J.,  {Harrison} C.~M.,   {Jarvis} M.~E.,  2019, \mn@doi [\aap]
  {10.1051/0004-6361/201936408}, \href
  {https://ui.adsabs.harvard.edu/abs/2019A&A...631A.132M} {631, A132}

\bibitem[\protect\citeauthoryear{{Morganti}, {Killeen}, {Ekers}  \&
  {Oosterloo}}{{Morganti} et~al.}{1999}]{1999Morganti+}
{Morganti} R.,  {Killeen} N.~E.~B.,  {Ekers} R.~D.,   {Oosterloo} T.~A.,  1999,
  \mn@doi [\mnras] {10.1046/j.1365-8711.1999.02622.x}, \href
  {https://ui.adsabs.harvard.edu/abs/1999MNRAS.307..750M} {307, 750}

\bibitem[\protect\citeauthoryear{{Morganti}, {Oosterloo}, {Oonk}, {Frieswijk}
  \& {Tadhunter}}{{Morganti} et~al.}{2015}]{2015Morganti+}
{Morganti} R.,  {Oosterloo} T.,  {Oonk} J.~B.~R.,  {Frieswijk} W.,
  {Tadhunter} C.,  2015, \mn@doi [\aap] {10.1051/0004-6361/201525860}, \href
  {https://ui.adsabs.harvard.edu/abs/2015A&A...580A...1M} {580, A1}

\bibitem[\protect\citeauthoryear{{Morganti}, {Oosterloo}, {Schulz}, {Tadhunter}
   \& {Oonk}}{{Morganti} et~al.}{2018}]{2018Morganti+}
{Morganti} R.,  {Oosterloo} T.,  {Schulz} R.,  {Tadhunter} C.,   {Oonk}
  J.~B.~R.,  2018, arXiv e-prints, \href
  {https://ui.adsabs.harvard.edu/abs/2018arXiv180707245M} {p. arXiv:1807.07245}

\bibitem[\protect\citeauthoryear{{Morganti}, {Oosterloo}, {Tadhunter},
  {Bernhard}  \& {Oonk}}{{Morganti} et~al.}{2021}]{2021Morganti+}
{Morganti} R.,  {Oosterloo} T.,  {Tadhunter} C.,  {Bernhard} E.~P.,   {Oonk}
  J.~B.~R.,  2021, arXiv e-prints, \href
  {https://ui.adsabs.harvard.edu/abs/2021arXiv210913516M} {p. arXiv:2109.13516}

\bibitem[\protect\citeauthoryear{{Mukherjee}, {Bicknell}, {Sutherland}  \&
  {Wagner}}{{Mukherjee} et~al.}{2016}]{2016Mukherjee+}
{Mukherjee} D.,  {Bicknell} G.~V.,  {Sutherland} R.,   {Wagner} A.,  2016,
  \mn@doi [\mnras] {10.1093/mnras/stw1368}, \href
  {https://ui.adsabs.harvard.edu/abs/2016MNRAS.461..967M} {461, 967}

\bibitem[\protect\citeauthoryear{{Mukherjee}, {Bicknell}, {Wagner},
  {Sutherland}  \& {Silk}}{{Mukherjee} et~al.}{2018}]{2018MMukherjee+}
{Mukherjee} D.,  {Bicknell} G.~V.,  {Wagner} A.~Y.,  {Sutherland} R.~S.,
  {Silk} J.,  2018, \mn@doi [\mnras] {10.1093/mnras/sty1776}, \href
  {https://ui.adsabs.harvard.edu/abs/2018MNRAS.479.5544M} {479, 5544}

\bibitem[\protect\citeauthoryear{{Nagar} \& {Wilson}}{{Nagar} \&
  {Wilson}}{1999}]{1999NagarWilson}
{Nagar} N.~M.,  {Wilson} A.~S.,  1999, \mn@doi [\apj] {10.1086/307109}, \href
  {https://ui.adsabs.harvard.edu/abs/1999ApJ...516...97N} {516, 97}

\bibitem[\protect\citeauthoryear{{Nayakshin}, {Cuadra}  \&
  {Springel}}{{Nayakshin} et~al.}{2007}]{2007Nayakshin+}
{Nayakshin} S.,  {Cuadra} J.,   {Springel} V.,  2007, \mn@doi [\mnras]
  {10.1111/j.1365-2966.2007.11938.x}, \href
  {https://ui.adsabs.harvard.edu/abs/2007MNRAS.379...21N} {379, 21}

\bibitem[\protect\citeauthoryear{{Nelson} et~al.,}{{Nelson}
  et~al.}{2019}]{2019Nelson+}
{Nelson} D.,  et~al., 2019, \mn@doi [\mnras] {10.1093/mnras/stz2306}, \href
  {https://ui.adsabs.harvard.edu/abs/2019MNRAS.490.3234N} {490, 3234}

\bibitem[\protect\citeauthoryear{{Nesvadba}, {Lehnert}, {De Breuck}, {Gilbert}
  \& {van Breugel}}{{Nesvadba} et~al.}{2008}]{2008Nevsbada+}
{Nesvadba} N.~P.~H.,  {Lehnert} M.~D.,  {De Breuck} C.,  {Gilbert} A.~M.,
  {van Breugel} W.,  2008, \mn@doi [\aap] {10.1051/0004-6361:200810346}, \href
  {https://ui.adsabs.harvard.edu/abs/2008A&A...491..407N} {491, 407}

\bibitem[\protect\citeauthoryear{{Oosterloo}, {Raymond Oonk}, {Morganti},
  {Combes}, {Dasyra}, {Salom{\'e}}, {Vlahakis}  \& {Tadhunter}}{{Oosterloo}
  et~al.}{2017}]{2017Osterloo+}
{Oosterloo} T.,  {Raymond Oonk} J.~B.,  {Morganti} R.,  {Combes} F.,  {Dasyra}
  K.,  {Salom{\'e}} P.,  {Vlahakis} N.,   {Tadhunter} C.,  2017, \mn@doi [\aap]
  {10.1051/0004-6361/201731781}, \href
  {https://ui.adsabs.harvard.edu/abs/2017A&A...608A..38O} {608, A38}

\bibitem[\protect\citeauthoryear{{Pakmor}, {Springel}, {Bauer}, {Mocz},
  {Munoz}, {Ohlmann}, {Schaal}  \& {Zhu}}{{Pakmor} et~al.}{2016}]{2016Pakmor+}
{Pakmor} R.,  {Springel} V.,  {Bauer} A.,  {Mocz} P.,  {Munoz} D.~J.,
  {Ohlmann} S.~T.,  {Schaal} K.,   {Zhu} C.,  2016, \mn@doi [\mnras]
  {10.1093/mnras/stv2380}, \href
  {https://ui.adsabs.harvard.edu/abs/2016MNRAS.455.1134P} {455, 1134}

\bibitem[\protect\citeauthoryear{{Reines} \& {Deller}}{{Reines} \&
  {Deller}}{2012}]{2012ReinesDeller}
{Reines} A.~E.,  {Deller} A.~T.,  2012, \mn@doi [\apjl]
  {10.1088/2041-8205/750/1/L24}, \href
  {https://ui.adsabs.harvard.edu/abs/2012ApJ...750L..24R} {750, L24}

\bibitem[\protect\citeauthoryear{Reynolds}{Reynolds}{2021}]{2021Reynolds}
Reynolds C.~S.,  2021, \mn@doi [Annual Review of Astronomy and Astrophysics]
  {10.1146/annurev-astro-112420-035022}, 59, 117

\bibitem[\protect\citeauthoryear{{Rice}, {Armitage}, {Bonnell}, {Bate},
  {Jeffers}  \& {Vine}}{{Rice} et~al.}{2003}]{2003Rice+}
{Rice} W.~K.~M.,  {Armitage} P.~J.,  {Bonnell} I.~A.,  {Bate} M.~R.,  {Jeffers}
  S.~V.,   {Vine} S.~G.,  2003, \mn@doi [\mnras]
  {10.1111/j.1365-2966.2003.07317.x}, \href
  {https://ui.adsabs.harvard.edu/abs/2003MNRAS.346L..36R} {346, L36}

\bibitem[\protect\citeauthoryear{{Rice}, {Lodato}  \& {Armitage}}{{Rice}
  et~al.}{2005}]{2005Rice+}
{Rice} W.~K.~M.,  {Lodato} G.,   {Armitage} P.~J.,  2005, \mn@doi [\mnras]
  {10.1111/j.1745-3933.2005.00105.x}, \href
  {https://ui.adsabs.harvard.edu/abs/2005MNRAS.364L..56R} {364, L56}

\bibitem[\protect\citeauthoryear{{Riffel}, {Storchi-Bergmann}  \&
  {Riffel}}{{Riffel} et~al.}{2014}]{2014Riffel+}
{Riffel} R.~A.,  {Storchi-Bergmann} T.,   {Riffel} R.,  2014, \mn@doi [\apjl]
  {10.1088/2041-8205/780/2/L24}, \href
  {https://ui.adsabs.harvard.edu/abs/2014ApJ...780L..24R} {780, L24}

\bibitem[\protect\citeauthoryear{{Riffel}, {Storchi-Bergmann}  \&
  {Riffel}}{{Riffel} et~al.}{2015}]{2015Riffel+}
{Riffel} R.~A.,  {Storchi-Bergmann} T.,   {Riffel} R.,  2015, \mn@doi [\mnras]
  {10.1093/mnras/stv1129}, \href
  {https://ui.adsabs.harvard.edu/abs/2015MNRAS.451.3587R} {451, 3587}

\bibitem[\protect\citeauthoryear{{Roy} et~al.,}{{Roy} et~al.}{2021}]{2021Roy+}
{Roy} N.,  et~al., 2021, arXiv e-prints, \href
  {https://ui.adsabs.harvard.edu/abs/2021arXiv210902609R} {p. arXiv:2109.02609}

\bibitem[\protect\citeauthoryear{{Runnoe}, {G{\"u}ltekin}, {Rupke}  \&
  {L{\'o}pez-Sepulcre}}{{Runnoe} et~al.}{2021}]{2021Runnoe+}
{Runnoe} J.~C.,  {G{\"u}ltekin} K.,  {Rupke} D.,   {L{\'o}pez-Sepulcre} A.,
  2021, \mn@doi [\mnras] {10.1093/mnras/stab1579}, \href
  {https://ui.adsabs.harvard.edu/abs/2021MNRAS.505.6017R} {505, 6017}

\bibitem[\protect\citeauthoryear{{Ruschel-Dutra} et~al.,}{{Ruschel-Dutra}
  et~al.}{2021}]{2021Ruschel-Dutra}
{Ruschel-Dutra} D.,  et~al., 2021, arXiv e-prints, \href
  {https://ui.adsabs.harvard.edu/abs/2021arXiv210707635R} {p. arXiv:2107.07635}

\bibitem[\protect\citeauthoryear{{Russell} et~al.,}{{Russell}
  et~al.}{2017}]{2017Russell+}
{Russell} H.~R.,  et~al., 2017, \mn@doi [\apj] {10.3847/1538-4357/836/1/130},
  \href {https://ui.adsabs.harvard.edu/abs/2017ApJ...836..130R} {836, 130}

\bibitem[\protect\citeauthoryear{{Santoro}, {Tadhunter}, {Baron}, {Morganti}
  \& {Holt}}{{Santoro} et~al.}{2020}]{2020Santoro+}
{Santoro} F.,  {Tadhunter} C.,  {Baron} D.,  {Morganti} R.,   {Holt} J.,  2020,
  \mn@doi [\aap] {10.1051/0004-6361/202039077}, \href
  {https://ui.adsabs.harvard.edu/abs/2020A&A...644A..54S} {644, A54}

\bibitem[\protect\citeauthoryear{{Schmitt}, {Kinney}, {Storchi-Bergmann},
  {Antonucci}  \& {Robert}}{{Schmitt} et~al.}{1997}]{1997Schmitt+}
{Schmitt} H.~R.,  {Kinney} A.~L.,  {Storchi-Bergmann} T.,  {Antonucci}
  {Robert} 1997, \mn@doi [\apj] {10.1086/303744}, \href
  {https://ui.adsabs.harvard.edu/abs/1997ApJ...477..623S} {477, 623}

\bibitem[\protect\citeauthoryear{{Shakura} \& {Sunyaev}}{{Shakura} \&
  {Sunyaev}}{1973}]{1973ShakuraSunyaev}
{Shakura} N.~I.,  {Sunyaev} R.~A.,  1973, \aap, \href
  {https://ui.adsabs.harvard.edu/abs/1973A%26A....24..337S} {24, 337}

\bibitem[\protect\citeauthoryear{{Shimizu} et~al.,}{{Shimizu}
  et~al.}{2019}]{2019Shimizu+}
{Shimizu} T.~T.,  et~al., 2019, \mn@doi [\mnras] {10.1093/mnras/stz2802}, \href
  {https://ui.adsabs.harvard.edu/abs/2019MNRAS.490.5860S} {490, 5860}

\bibitem[\protect\citeauthoryear{{Sijacki}, {Springel}, {Di Matteo}  \&
  {Hernquist}}{{Sijacki} et~al.}{2007}]{2007Sijacki+}
{Sijacki} D.,  {Springel} V.,  {Di Matteo} T.,   {Hernquist} L.,  2007, \mn@doi
  [\mnras] {10.1111/j.1365-2966.2007.12153.x}, \href
  {https://ui.adsabs.harvard.edu/abs/2007MNRAS.380..877S} {380, 877}

\bibitem[\protect\citeauthoryear{{Sijacki}, {Vogelsberger}, {Genel},
  {Springel}, {Torrey}, {Snyder}, {Nelson}  \& {Hernquist}}{{Sijacki}
  et~al.}{2015}]{2015Sijacki+}
{Sijacki} D.,  {Vogelsberger} M.,  {Genel} S.,  {Springel} V.,  {Torrey} P.,
  {Snyder} G.~F.,  {Nelson} D.,   {Hernquist} L.,  2015, \mn@doi [\mnras]
  {10.1093/mnras/stv1340}, \href
  {https://ui.adsabs.harvard.edu/abs/2015MNRAS.452..575S} {452, 575}

\bibitem[\protect\citeauthoryear{{Simionescu}, {Tremblay}, {Werner}, {Canning},
  {Allen}  \& {Oonk}}{{Simionescu} et~al.}{2018}]{2018Simionescu+}
{Simionescu} A.,  {Tremblay} G.,  {Werner} N.,  {Canning} R.~E.~A.,  {Allen}
  S.~W.,   {Oonk} J.~B.~R.,  2018, \mn@doi [\mnras] {10.1093/mnras/sty047},
  \href {https://ui.adsabs.harvard.edu/abs/2018MNRAS.475.3004S} {475, 3004}

\bibitem[\protect\citeauthoryear{{Speranza} et~al.,}{{Speranza}
  et~al.}{2021}]{2021Speranza+}
{Speranza} G.,  et~al., 2021, \mn@doi [\aap] {10.1051/0004-6361/202140686},
  \href {https://ui.adsabs.harvard.edu/abs/2021A&A...653A.150S} {653, A150}

\bibitem[\protect\citeauthoryear{{Springel}}{{Springel}}{2005}]{2005Springel}
{Springel} V.,  2005, \mn@doi [\mnras] {10.1111/j.1365-2966.2005.09655.x},
  \href {https://ui.adsabs.harvard.edu/abs/2005MNRAS.364.1105S} {364, 1105}

\bibitem[\protect\citeauthoryear{{Springel}}{{Springel}}{2010}]{2010Springel}
{Springel} V.,  2010, \mn@doi [\mnras] {10.1111/j.1365-2966.2009.15715.x},
  \href {https://ui.adsabs.harvard.edu/abs/2010MNRAS.401..791S} {401, 791}

\bibitem[\protect\citeauthoryear{{Su} et~al.,}{{Su} et~al.}{2021}]{2021Su+}
{Su} K.-Y.,  et~al., 2021, \mn@doi [\mnras] {10.1093/mnras/stab2021}, \href
  {https://ui.adsabs.harvard.edu/abs/2021MNRAS.507..175S} {507, 175}

\bibitem[\protect\citeauthoryear{{Tadhunter}, {Morganti}, {Rose}, {Oonk}  \&
  {Oosterloo}}{{Tadhunter} et~al.}{2014}]{2014Tadhunter+}
{Tadhunter} C.,  {Morganti} R.,  {Rose} M.,  {Oonk} J.~B.~R.,   {Oosterloo} T.,
   2014, \mn@doi [\nat] {10.1038/nature13520}, \href
  {https://ui.adsabs.harvard.edu/abs/2014Natur.511..440T} {511, 440}

\bibitem[\protect\citeauthoryear{{Talbot}, {Bourne}  \& {Sijacki}}{{Talbot}
  et~al.}{2021}]{2021Talbot+}
{Talbot} R.~Y.,  {Bourne} M.~A.,   {Sijacki} D.,  2021, \mn@doi [\mnras]
  {10.1093/mnras/stab804}, \href
  {https://ui.adsabs.harvard.edu/abs/2021MNRAS.504.3619T} {504, 3619}

\bibitem[\protect\citeauthoryear{{Tchekhovskoy}, {Narayan}  \&
  {McKinney}}{{Tchekhovskoy} et~al.}{2010}]{2010Tchekhovskoy+}
{Tchekhovskoy} A.,  {Narayan} R.,   {McKinney} J.~C.,  2010, \mn@doi [\apj]
  {10.1088/0004-637X/711/1/50}, \href
  {https://ui.adsabs.harvard.edu/abs/2010ApJ...711...50T} {711, 50}

\bibitem[\protect\citeauthoryear{{Tchekhovskoy}, {McKinney}  \&
  {Narayan}}{{Tchekhovskoy} et~al.}{2012}]{2012Tchekhovskoy}
{Tchekhovskoy} A.,  {McKinney} J.~C.,   {Narayan} R.,  2012, in Journal of
  Physics Conference Series. p. 012040 (\mn@eprint {arXiv} {1202.2864}),
  \mn@doi{10.1088/1742-6596/372/1/012040}

\bibitem[\protect\citeauthoryear{{Tremblay} et~al.,}{{Tremblay}
  et~al.}{2018}]{2018Tremblay+}
{Tremblay} G.~R.,  et~al., 2018, \mn@doi [\apj] {10.3847/1538-4357/aad6dd},
  \href {https://ui.adsabs.harvard.edu/abs/2018ApJ...865...13T} {865, 13}

\bibitem[\protect\citeauthoryear{{Vayner}, {Wright}, {Murray}, {Armus},
  {Larkin}  \& {Mieda}}{{Vayner} et~al.}{2017}]{2017Vayner+}
{Vayner} A.,  {Wright} S.~A.,  {Murray} N.,  {Armus} L.,  {Larkin} J.~E.,
  {Mieda} E.,  2017, \mn@doi [\apj] {10.3847/1538-4357/aa9c42}, \href
  {https://ui.adsabs.harvard.edu/abs/2017ApJ...851..126V} {851, 126}

\bibitem[\protect\citeauthoryear{{Venturi} et~al.,}{{Venturi}
  et~al.}{2021}]{2021Venturi+}
{Venturi} G.,  et~al., 2021, \mn@doi [\aap] {10.1051/0004-6361/202039869},
  \href {https://ui.adsabs.harvard.edu/abs/2021A&A...648A..17V} {648, A17}

\bibitem[\protect\citeauthoryear{Virtanen et~al.,}{Virtanen
  et~al.}{2020}]{2020Virtanen+}
Virtanen P.,  et~al., 2020, \mn@doi [Nature Methods]
  {10.1038/s41592-019-0686-2}, \href {https://rdcu.be/b08Wh} {17, 261}

\bibitem[\protect\citeauthoryear{{Wagner} \& {Bicknell}}{{Wagner} \&
  {Bicknell}}{2011}]{2011WagnerBicknell}
{Wagner} A.~Y.,  {Bicknell} G.~V.,  2011, \mn@doi [\apj]
  {10.1088/0004-637X/728/1/29}, \href
  {https://ui.adsabs.harvard.edu/abs/2011ApJ...728...29W} {728, 29}

\bibitem[\protect\citeauthoryear{{Wagner}, {Bicknell}, {Umemura}, {Sutherland}
  \& {Silk}}{{Wagner} et~al.}{2016}]{2016Wagner+}
{Wagner} A.~Y.,  {Bicknell} G.~V.,  {Umemura} M.,  {Sutherland} R.~S.,   {Silk}
  J.,  2016, \mn@doi [Astronomische Nachrichten] {10.1002/asna.201512287},
  \href {https://ui.adsabs.harvard.edu/abs/2016AN....337..167W} {337, 167}

\bibitem[\protect\citeauthoryear{{Weinberger}, {Ehlert}, {Pfrommer}, {Pakmor}
  \& {Springel}}{{Weinberger} et~al.}{2017}]{2017Weinberger+}
{Weinberger} R.,  {Ehlert} K.,  {Pfrommer} C.,  {Pakmor} R.,   {Springel} V.,
  2017, \mn@doi [\mnras] {10.1093/mnras/stx1409}, \href
  {https://ui.adsabs.harvard.edu/abs/2017MNRAS.470.4530W} {470, 4530}

\bibitem[\protect\citeauthoryear{{Weinberger} et~al.,}{{Weinberger}
  et~al.}{2018}]{2018Weinberger+}
{Weinberger} R.,  et~al., 2018, \mn@doi [\mnras] {10.1093/mnras/sty1733}, \href
  {https://ui.adsabs.harvard.edu/abs/2018MNRAS.479.4056W} {479, 4056}

\bibitem[\protect\citeauthoryear{{Weinberger}, {Springel}  \&
  {Pakmor}}{{Weinberger} et~al.}{2020}]{2020Weinberger+}
{Weinberger} R.,  {Springel} V.,   {Pakmor} R.,  2020, \mn@doi [\apjs]
  {10.3847/1538-4365/ab908c}, \href
  {https://ui.adsabs.harvard.edu/abs/2020ApJS..248...32W} {248, 32}

\bibitem[\protect\citeauthoryear{{Williams} et~al.,}{{Williams}
  et~al.}{2017}]{2017Williams+}
{Williams} D.~R.~A.,  et~al., 2017, \mn@doi [\mnras] {10.1093/mnras/stx2205},
  \href {https://ui.adsabs.harvard.edu/abs/2017MNRAS.472.3842W} {472, 3842}

\bibitem[\protect\citeauthoryear{{Xu}, {Komossa}, {Zhou}, {Lu}, {Li}, {Grupe},
  {Wang}  \& {Yuan}}{{Xu} et~al.}{2012}]{2012Xu+}
{Xu} D.,  {Komossa} S.,  {Zhou} H.,  {Lu} H.,  {Li} C.,  {Grupe} D.,  {Wang}
  J.,   {Yuan} W.,  2012, \mn@doi [\aj] {10.1088/0004-6256/143/4/83}, \href
  {https://ui.adsabs.harvard.edu/abs/2012AJ....143...83X} {143, 83}

\bibitem[\protect\citeauthoryear{{Yang}, {Gurvits}, {Paragi}, {Frey}, {Conway},
  {Liu}  \& {Cui}}{{Yang} et~al.}{2020}]{2020Yang+}
{Yang} J.,  {Gurvits} L.~I.,  {Paragi} Z.,  {Frey} S.,  {Conway} J.~E.,  {Liu}
  X.,   {Cui} L.,  2020, \mn@doi [\mnras] {10.1093/mnrasl/slaa052}, \href
  {https://ui.adsabs.harvard.edu/abs/2020MNRAS.495L..71Y} {495, L71}

\makeatother
\end{thebibliography}



\appendix
\section{Angular momentum direction evolution}
\label{App: bh dir evolution}

In this appendix we explore the processes that lead to evolution in the direction of the angular momentum of the black hole and $\alpha$-disc. In all simulations presented in the main body of the paper, the initial angular momentum direction of the $\alpha$-disc is aligned with the vertical (i.e. $\theta_{\rm d} \equiv \cos^{-1}(\boldsymbol{j}_{\rm d}\cdot\boldsymbol{\hat{z}}) = 0$). In this appendix we, therefore, additionally explore the evolution the black hole-disc system when the angular momenta of {\it both} are inclined with respect to the vertical.

\subsection{Jet direction evolution}
\begin{figure}
    \centering
    \includegraphics[width=0.9\columnwidth]{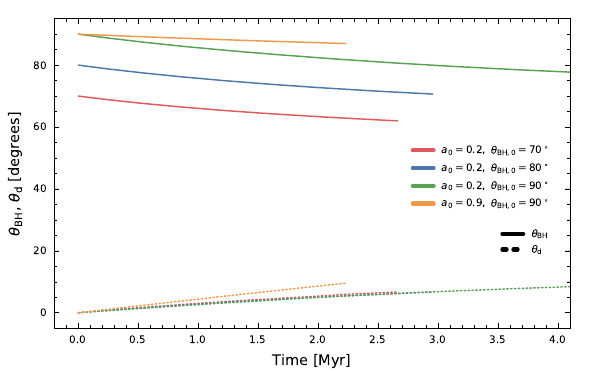}
    \caption{Solid lines show the evolution of the angle between the black hole spin direction and the vertical, $\theta_{\rm BH} = \cos^{-1}(\boldsymbol{j}_{\rm BH}\cdot\boldsymbol{\hat{z}})$, and dotted lines show the angle between the $\alpha$-disc angular momentum and the vertical, $\theta_{\rm d} = \cos^{-1}(\boldsymbol{j}_{\rm d}\cdot\boldsymbol{\hat{z}})$. These are shown for a selection of the `high' resolution runs in the `thick' CND with the colour of the line indicating the initial jet configuration (see the legends).}
    \label{fig: subgrid direction}
\end{figure}

\begin{figure*}
    \centering
    \includegraphics[width=0.8\textwidth]{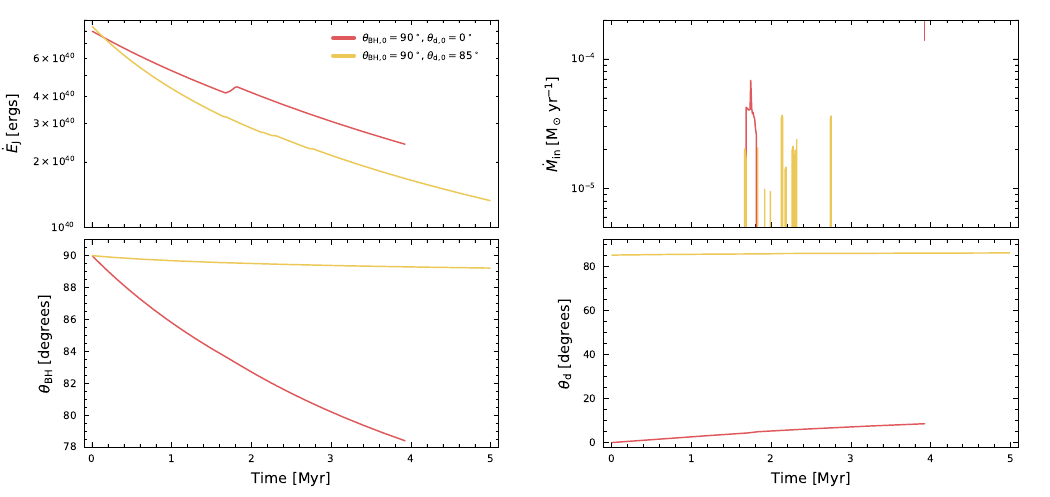}
    \caption{Evolution of the jet power (top-left), the mass inflow rate (top-right), the angle between the black hole spin direction and the vertical (bottom-left) and the angle between the $\alpha$-disc angular momentum direction and the vertical (bottom right) for two `low' resolution, `low' power runs in the `thick' CND. In both runs, the black hole is initially directed into the CND ($\theta_{\rm BH,0} = 90^\circ$). In one run, the angular momentum of the $\alpha$-disc is aligned with the vertical ($\theta_{\rm d,0} = 0^\circ$, red line) and in the other, it is highly inclined to the vertical, but still misaligned with that of the black hole ($\theta_{\rm d,0} = 85^\circ$, gold line).}
    \label{fig: small misalignment}
\end{figure*}

Accretion from the ISCO of the $\alpha$-disc and the launching of the Blandford-Znajek jet alter only the magnitude of the black hole angular momentum. Bardeen-Petterson torques, however, will alter the direction of the black hole spin (and therefore the direction of the jet). A necessary condition for changes in the jet direction is therefore that the black hole spin and $\alpha$-disc angular momentum are misaligned.

If the angular momenta of the black hole and $\alpha$-disc are aligned (as in the `vertical' jet cases) then the Bardeen-Petterson torques vanish and neither the black hole angular momentum direction, nor that of the $\alpha$-disc angular momentum can evolve. Change to the direction of `vertical' jets, therefore, requires inflows of gas with angular momentum that is not aligned with that of the $\alpha$-disc.

Fig.~\ref{fig: subgrid direction} shows the evolution of the angle between the black hole spin direction and the vertical, $\theta_{\rm BH} \equiv \cos^{-1}(\boldsymbol{j}_{\rm BH}\cdot\boldsymbol{\hat{z}})$, and that of the angle between the $\alpha$-disc angular momentum and the vertical, $\theta_{\rm d} \equiv \cos^{-1}(\boldsymbol{j}_{\rm d}\cdot\boldsymbol{\hat{z}})$, for a selection of the `low' and `high' power runs in the `thick' CND. We do not show the `vertical' jet cases as neither the jet direction nor the $\alpha$-disc angular momentum directions undergo appreciable evolution when aligned, given that no Bardeen-Petterson torques act and they are both aligned with the large-scale angular momentum of the CND \citep{2018Fiacconi+}. 

From Fig.~\ref{fig: subgrid direction} it is clear that all of these jets undergo non-negligible Bardeen-Petterson torquing, with both the black hole and $\alpha$-disc directions evolving as they align with the total angular momentum of the black hole-$\alpha$-disc system. 

We also see that the rate at which the angular momenta of `low' spin black holes and their $\alpha$-discs are torqued is not significantly affected by changes to the initial spin inclination. Comparing the `low' and `high' power `horizontal' jets (yellow and green lines), however, it is apparent that the angular momentum direction of the `high' spin black hole is torqued slower than that of the `low' spin black hole, while the $\alpha$-disc angular momentum in the `high' spin case is torqued faster than the `low' spin case. To understand this, recall that Bardeen-Petterson torques act to align the black hole and $\alpha$-disc with their total angular momentum $\boldsymbol{J}_{\rm tot}=\boldsymbol{J}_{\rm BH} + \boldsymbol{J}_{\rm d}$. For fixed $\alpha$-disc angular momentum, the angle between $\boldsymbol{J}_{\rm tot}$ and $\boldsymbol{J}_{\rm BH}$ will decrease as the spin of the black hole increases. The timescale on which the the black hole aligns with the total angular momentum is the same as that on which the $\alpha$-disc aligns with the total angular momentum which explains why we observe the black hole/disc inclination angle changing more slowly/quickly in the `high' spin run compared to the `low' spin run.

\subsection{Inclination comparison}

Fig.~\ref{fig: small misalignment} shows the evolution of the jet power (top-left), the mass inflow rate (top-right), the angle between the black hole spin direction and the vertical (bottom-left) and the angle between the $\alpha$-disc angular momentum direction and the vertical (bottom right) for two `low' resolution, `low' power jets that are launched from the `thick' CND. In both runs, the black hole spin is initially `horizontal' (i.e. $\theta_{\rm BH,0} = 90^\circ$). In one of the runs, the angular momentum of the $\alpha$-disc is aligned with the vertical ($\theta_{\rm d,0} = 0^\circ$, red line) and in the other, the $\alpha$-disc angular momentum is highly inclined to the vertical but still misaligned with that of the black hole ($\theta_{\rm d,0} = 85^\circ$, gold line).

Comparing these two runs, it is clear that both the black hole and $\alpha$-disc angular momentum directions are torqued faster when the misalignment between the two is greater, as the Bardeen-Petterson torques that act in this scenario are stronger. This is due to the dependence of the Bardeen-Petterson alignment rate on the inclination angle between the black hole and $\alpha$-disc angular momenta via the $\boldsymbol{j}_{\rm BH} \times \boldsymbol{j}_{\rm d}$ term in equations~(\ref{eq: jdotbh})~and~(\ref{eq: jdotd}). 

Whilst the differences in the evolution of the black hole and $\alpha$-disc angular momentum directions are fairly large, the differences in the evolution of the jet powers and mass inflow rates are less considerable. The jet power drops faster in the case where the the $\alpha$-disc is misaligned, this occurs for the same reason that the power of `vertical' jets drops faster than that of `horizontal' jets, which is discussed in detail in Section~\ref{Sec: Jet power evolution}. Mass inflow restarts at similar times in both cases, although in the `vertical' disc case, the inflow has a shorter duration but leads to a noticeable jump in the power of the jet, whereas in the `misaligned' disc case, inflow is much more bursty and leads to a more modest increase in jet power.

While it is encouraging that the evolution of the jet power and inflow rate are not significantly affected by the initial orientation of the black hole and $\alpha$-disc, it is worth reiterating that our sub-grid model is only guaranteed to capture their evolution when the misalignment between them is small (see section~$2.2$ of \cite{2021Talbot+}). To analyse in detail the evolution of the black hole-$\alpha$-disc system when their angular momenta are significantly misaligned requires a theoretical framework with capabilities beyond that of our sub-grid model.

\section{Resolution study}
\label{app: res test}

\begin{figure}
    \centering
    \includegraphics[width=\columnwidth]{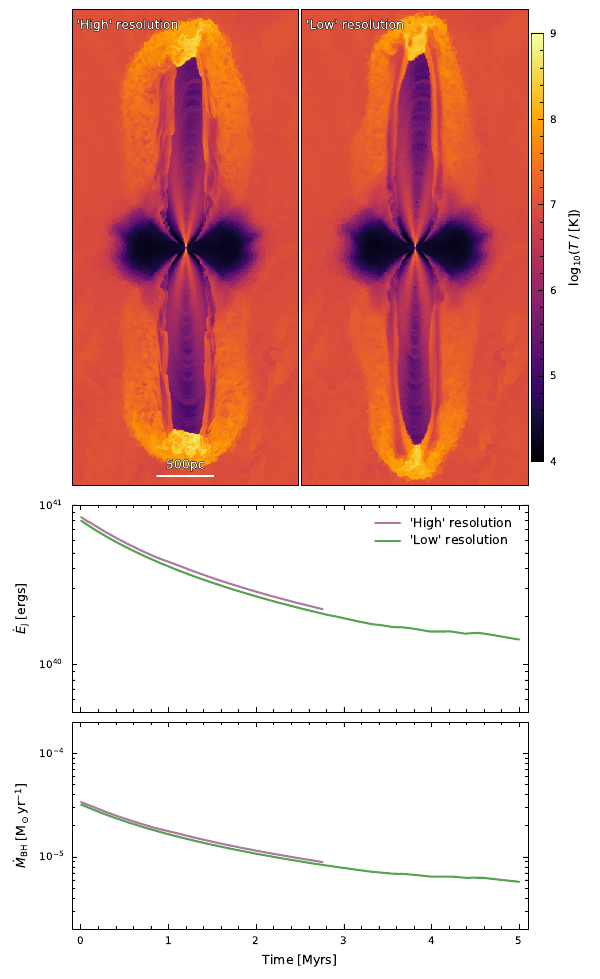}
    \caption{A comparison between different resolution simulations. The two panels in the first row show temperature slices in the $x$-$z$ plane, centred on the black hole, for `low' and `high' resolution simulations in which the `low' power `vertical' jet is launched from the `thick' CND into the `standard' CGM. These are shown at the time when the length of the `high' resolution jet is closest to $2$~kpc. The panels in the second and third row show the evolution of the jet power and black hole growth rate for these two cases.}
    \label{fig: resolution comparison}
\end{figure}

In this work, we used a suite of `high' resolution simulations to perform analysis that requires accurately resolving the jet lobe evolution and interaction with the CGM. In sections where we were focusing on the evolution of the black hole and $\alpha$-disc, we used `low' resolution simulations in which the spatial resolution of the jet lobes was reduced (for a description of the difference between these `high' and `low' resolution simulations, see Section~\ref{Sec: Resolution} and Table~\ref{tab: all runs}).

In this appendix, we show that the `low' resolution simulations are still able to accurately capture the evolution of the black hole-$\alpha$-disc system, despite having less well resolved jet lobes. 

In Fig.~\ref{fig: resolution comparison}, the panel in the second/third row shows the evolution of the jet power/black hole growth rate for the `low' power `vertical' jet launched from the `thick' CND into the `standard' CGM. From these panels, it is evident that changing the resolution in the jet lobes has a negligible impact on the evolution of both the jet power and black hole growth rate. We have verified that this result holds in general, both for other relevant black hole and $\alpha$-disc properties and also for the other simulations presented in this work.

The two slices in Fig.~\ref{fig: resolution comparison} show the temperature field of the gas at the time when the length of the `high' resolution jet is closest to $2$~kpc. Whilst the jet morphology shown in these two slices is largely similar, it is clear that the instabilities along the jet lobe-CGM interface are not as well resolved in the `low' resolution simulation. Similarly, the shocks at the head of the jet are narrower and the cocoon of shocked CGM gas is less extended in the `low' resolution case. 

We found that these differences were sufficient to justify the use of the `high' resolution simulations when considering the jet lobe evolution, however, we opted to use the `low' resolution data to analyse the evolution of the black hole and $\alpha$-disc properties so that we could make use of the additional temporal data that these simulations provided.

\bsp	
\label{lastpage}
\end{document}